\documentclass[a4paper]{article}
\pdfoutput=1
\usepackage{jheppub}

\usepackage{amsmath}

\usepackage{amssymb}
\usepackage{graphicx,color}

\usepackage{amsmath,amsthm}
\usepackage{txfonts}

\usepackage{mathrsfs}

\usepackage{bm}

\newcommand{\Ref}[1]{(\ref{#1})}

\newtheorem{Theorem}{Theorem}[section]

\newtheorem{Lemma}[Theorem]{Lemma}

\newcommand{\Z}{\mathbb{Z}}
\newcommand{\R}{\mathbb{R}}
\newcommand{\C}{\mathbb{C}}

\newcommand{\half}{\frac{1}{2}}

\newcommand{\ccirc}{\kern0.2ex\vcenter{\hbox{$\scriptstyle\circ$}}\kern0.2ex}

\newcommand{\Tr}{\text{Tr}}

\newcommand{\Slc}{\mathrm{SL}(2,\mathbb{C})}

\newcommand{\Su}{\mathrm{SU}(2)}

\def\be{\begin{eqnarray}}
\def\ee{\end{eqnarray}}

\newcommand{\cd}{\mathcal D}
\newcommand{\ce}{\mathcal E}

\newcommand{\ch}{\mathcal H}

\newcommand{\cp}{\mathcal P}

\newcommand{\calr}{\mathcal R}
\newcommand{\cs}{\mathcal S}

\newcommand{\fa}{\mathfrak{a}}

\renewcommand{\a}{\alpha}
\renewcommand{\b}{\beta}
\newcommand{\g}{\gamma}

\newcommand{\eps}{\varepsilon}

\newcommand{\sig}{\sigma}

\renewcommand{\l}{\lambda}
\renewcommand{\L }{\Lambda}

\renewcommand{\t}{\tau}

\newcommand{\rmd}{\mathrm d}

\newcommand{\lt}{\left}
\newcommand{\rt}{\right}

\newcommand{\lag}{\left\langle}
\newcommand{\rag}{\right\rangle}

\newcommand{\tr}{\mathrm{tr}}

\title{Effective Dynamics from Coherent State Path Integral of Full Loop Quantum Gravity}

\author[1,2]{Muxin Han}  

\affiliation[1]{Department of Physics, Florida Atlantic University, 777 Glades Road, Boca Raton, FL 33431-0991, USA}

\affiliation[2]{Institut f\"ur Quantengravitation, Universit\"at Erlangen-N\"urnberg, Staudtstr. 7/B2, 91058 Erlangen, Germany}

\author[3,4,5]{\ Hongguang Liu}

\affiliation[3]{Aix Marseille Univ, Universite de Toulon, CNRS, CPT, Marseille, France}

\affiliation[4]{Laboratoire de Math\'ematiques et Physique Th\'eorique (UMR CNRS 7350), Universit\'e Francois Rabelais, Parc de Grandmont, 37200 Tours, France}

\affiliation[5]{Laboratoire Astroparticule et Cosmologie, Universit\'e Denis Diderot Paris 7, 75013 Paris, France}

\emailAdd{hanm(At)fau.edu}
\emailAdd{liu.hongguang(At)cpt.univ-mrs.fr}

\abstract{A new routine is proposed to relate Loop Quant Cosmology (LQC) to Loop Quantum Gravity (LQG) from the perspective of effective dynamics. We derive the big-bang singularity resolution and big bounce from the first principle of full canonical LQG. Our results are obtained in the framework of the reduced phase space quantization of LQG. As a key step in our work, we derive with coherent states a new discrete path integral formula of the transition amplitude generated by the physical Hamiltonian. The semiclassical approximation of the path integral formula gives an interesting set of effective equations of motion (EOMs) for full LQG. When solving the EOMs with homogeneous and isotropic ansatz, we reproduce the LQC effective dynamics in $\mu_0$-scheme. The solution replaces the big-bang singularity by a big bounce. In the end, we comment on the possible relation between the $\bar{\mu}$-scheme of effective dynamics and the continuum limit of the path integral formula.

}


\begin{document}

\maketitle

\section{Introduction}

Loop Quantum Gravity (LQG) is a promising attempt toward a non-perturbative and background independent theory of quantum gravity (see e.g. \cite{book, review, review1} for reviews). Among many important achievements of LQG, one of the most profound physical predictions is the resolution of singularities by quantum effects e.g. \cite{Bojowald:2001xe,Ashtekar:2006wn,Ashtekar:2006rx,Singh:2009mz,Assanioussi:2019iye,Ashtekar:2018cay,Ashtekar:2018lag,Assanioussi:2019twp,Gambini:2013hna, BenAchour:2018khr, Bojowald:2018xxu, Bodendorfer:2019cyv, Rovelli:2014cta,Han:2016fgh}. It is well-known that the classical theory of Einstein gravity breaks down at singularities, while the purpose of quantum gravity is to extend the gravity theory to describe the physics of singularities.    

The first concrete example of resolving singularity in LQG was made in Loop Quantum Cosmology (LQC) \cite{Bojowald:2001xe,Ashtekar:2006wn,Ashtekar:2006rx} (see e.g. \cite{Bojowald:2006da,Ashtekar:2008zu,Agullo:2016tjh} for reviews on LQC). LQC applies the quantization procedure of LQG to the homogeneous and isotropic sector of gravity. The classical symmetry-reduction to the homogeneous and isotropic sector reduces from infinitely many degrees of freedom (DOFs) of gravity to a single DOF (the scale factor). Then LQC quantizes the 2-dimensional phase space of cosmology by LQG method and imposes Hamiltonian constraint at the quantum level. The quantum dynamics of LQC have been studied extensively. It has been shown that the quantum evolution of coherent state in LQC gives rise to an effective equation, which reduces to classical Friedmann equation at low energy density regiem, while modifying Friedmann equation at high energy density near the big-bang singularity \cite{Taveras:2008ke}. The solution of effective equation demonstrates that the big-bang singularity is resolved and replaced by a big bounce, where the curvature is finite and Planckian. The time evolution of cosmology governed by the LQC effective equation is often called the effective dynamics. 

However LQC, from quantization after classical symmetry-reduction, only takes into account DOFs that are homogeneous and isotropic, thus fails to predict for instance quantum fluctuations beyond the homogeneous and isotropic sector \cite{Bojowald:2018sgf}. Moreover LQC has been suffered from the long-standing issue on the relation with the full theory of LQG (see e.g. \cite{Alesci:2013xd, Bodendorfer:2014vea, Bodendorfer:2015hwl, Alesci:2016rmn,Dapor:2017rwv,Engle:2007zz,2016arXiv160105531H,Fleischhack:2010zt,Rovelli:2008aa} for some earlier works on this perspective). It has not been clarified if the singularity resolution and big bounce can be derived from the first principle of full LQG. Symmetry-reduced models of loop quantum black holes share the same issues as LQC.

Our present work partially addresses the above issues of LQC. We propose a new routine to relate LQC to LQG from the perspective of effective dynamics, and demonstrate the singularity resolution and big bounce from the first principle of the full canonical LQG. Our results are obtained in the framework of the reduced phase space quantization of LQG. As a key step in our work, we derive with coherent states a new discrete path integral formula of the transition amplitude generated by the physical Hamiltonian. The semiclassical approximation of the path integral formula gives an interesting set of effective equations of motion (EOMs) of full LQG. When solving the EOMs with homogeneous and isotropic ansatz, we reproduce the solution of LQC effective dynamics in $\mu_0$-scheme. The solution replaces the big-bang singularity by a big bounce. 

Since our framework embeds the cosmological effective dynamics in the path integral formulation of full LQG, all quantum fluctuations within and beyond the homogenous and isotropic sector are in principle computable by e.g. perturbative expansion or other non-perturbative methods.   

Our work takes advantage of the reduced phase space quantization of LQG \cite{Giesel:2007wn,Giesel:2012rb} and studies the quantum dynamics given by the full LQG physical Hamiltonian $\hat{\bf H}$ defined on the physical Hilbert space $\ch$ of Dirac observables. The reduced phase space quantization is based on classical deparametrized models of gravity such as gravity coupled to dust fields or scalars. Our work takes into account three different scenarios: gravity coupled to the Brown-Kucha\v{r} dust, gaussian dust, and massless scalar field \cite{Brown:1994py,Kuchar:1990vy,Giesel:2007wn,Giesel:2012rb,Domagala:2010bm,Rovelli:1993bm}. In the quantum theory, we set the Hilbert space $\ch=\ch_\g$ to be based on a fixed graph $\g$ which is a finite cubic lattice partitioning 3-torus \footnote{The result is also valid in the case of an infinite space.}, and define $\hat{\bf H}$ to be the non-graph-changing Hamiltonian. 

There are two popular Hamiltonians of LQG, based on Giesel-Thiemann's construction \cite{QSD,Giesel:2007wn}, and the construction by Alesci-Assanioussi-Lewandowski-Makinen (AALM) \cite{Alesci:2015wla,Assanioussi:2015gka} using scalar curvature operator \cite{Alesci:2014aza}. Our work analyzes both possibilities.  

The seminal works \cite{Thiemann:2000bw,Thiemann:2000ca,Thiemann:2000bx} tell us that the Hilbert space $\ch$ has an over-complete basis given by the complexifier coherent states $\psi^t_g$, where at each edge $e\in E(\g)$\footnote{$E(\g)$ and $V(\g)$ denote the set of edges and vertices in $\g$.}, the coherent state label $g=\{g(e)\}_{e\in E(\g)}$ with $g(e)=e^{-ip^a(e)\t^a/2}h(e)=e^{-ip^a(e)\t^a/2}e^{\theta^a(e)\t^a/2}\in\Slc$ is the holomorphic coordinate on the LQG phase space of holonomies $h(e)$ and gauge covariant fluxes $p^a(e)$. The dimensionless semiclassicality parameter $t=\ell_P^2/a^2$ where $\ell_P^2=\hbar\kappa$ is the Planck scale, and $a$ is a length unit e.g. 1cm. The semiclassical limit is given by $t\to0$ or $\ell_P\ll a$. Thanks to the overcompleteness and semiclassical properties of coherent states, and by the standard discretization and coherent state path integral method, the transition amplitude between gauge invariant coherent states $\langle\Psi^t_{[g]}|\,\exp[-\frac{i}{\hbar}\hat{\bf H} T]\, |\Psi^t_{[g']}\rangle$ can be written as a discrete path integral formula (see Section \ref{CSPI} for details)
\be
\int\rmd h\prod_{i=1}^{N+1}\rmd g_i\,\nu[g]\, e^{S[{g},h]/t}, \label{LQGPI}
\ee 
where the integral is over $N+1$ intermediate states $g_i\in\Slc^{|E(\g)|}$ and $h=\{h_v\}_{v\in V(\g)}\in \Su^{|V(\g)|}$ (for ensuring SU(2) gauge invariance). $\nu[g]$ is a path integral measure. $S[{g},h]$ is the effective action for LQG extracted from the path integral:
\be
S[g,h]&=&\sum_{i=0}^{N+1} K\left(g_{i+1}, g_{i}\right)-\frac{i \kappa}{a^{2}} \sum_{i=1}^{N} \Delta \tau\frac{\left\langle\psi_{g_{i+1}}^{t}|\hat{\mathbf{H}}| \psi_{g_{i}}^{t}\right\rangle}{\left\langle\psi_{g_{i+1}}^{t} | \psi_{g_{i}}^{t}\right\rangle}\\
K\left(g_{i+1}, g_{i}\right)&=&\sum_{e \in E(\gamma)}\left[z_{i+1, i}(e)^{2}-\frac{1}{2} p_{i+1}(e)^{2}-\frac{1}{2} p_{i}(e)^{2}\right]
\ee
The discretization step $\Delta\t=T/N$ is arbitrarily small while $N$ is arbitrarily large. Here $z_{i+1,i}(e)$ relates to $g_i(e),g_{i+1}(e)$ by $z_{i+1,i}(e)=\mathrm{arccosh}(\half\mathrm{tr}[g_{i+1}(e)^\dagger g_i(e)])$. $S[g,h]$ depends on $h$ through $g_0=g'^h=h_{s(e)}^{-1}\{g'(e)h_{t(e)}\}_{e\in E(\g)}$ where $g'$ is the initial data. $g_{N+2}=g$ is the final data.

To our knowledge, this path integral formula Eq.\Ref{LQGPI} of the full LQG hasn't been derived before, although it has been motivated in e.g. \cite{link,Han:2009bb}. It should shed light on the relation between canonical LQG and Spinfoam formulation. Interestingly in the scenario of gravity coupled to scalar field and AALM Hamiltonian, this path integral formula should be the coherent state representation of the spinfoam model in \cite{Kisielowski:2018oiv} if their model was defined with a non-graph-changing Hamiltonian. 

The path integral formula in the context of LQC is studied in \cite{Ashtekar:2009dn,Henderson:2010qd,Craig:2016iuw,Craig:2016lxm}. The coherent state path integral is applied to LQC in \cite{Qin:2012xh}.

The discrete path integral Eq.\Ref{LQGPI} is well-defined as a finite integral as far as $N$ is finite (although arbitrarily large). The semiclassical limit $t\to0$ motivates us to apply the stationary phase approximation. The integral is dominated by contributions from solutions of the EOM obtained from the variational principle $\delta S[g,h]=0$. We propose that the cosmological effective dynamics should emerge from solutions of EOM given by the path integral of full LQG. Here the EOM may be ($a$) the classical EOM obtained from the stationary phase approximation, or ($b$) the quantum EOM derived from the quantum effective action which includes all quantum corrections. Our work focuses on the proposal $(a)$.

The EOMs from the semiclassical approximation of the path integral Eq.\Ref{LQGPI} is derived in Section \ref{SAEEQN}: In the continuous time approximation $\Delta\t\to0$, we obtain

\begin{itemize}
 \item For $i=1,\cdots,N$, at every edge $e\in E(\g)$,
   \be
   \frac{z_{i+1,i}(e)\,\tr\lt[\t^a g_{i+1}(e)^\dagger g_i(e)\rt]}{\sqrt{x_{i+1,i}(e)-1}\sqrt{x_{i+1,i}(e)+1}}-\frac{p_i(e)\,\tr\lt[\t^a g_{i}(e)^\dagger g_i(e)\rt]}{\sinh(p_i(e))}
   =\frac{i\kappa \Delta\t}{a^2}\frac{\partial\,{\mathbf{H}\lt[g_i^\eps\rt]}}{\partial{\eps^a(e)}}\Bigg|_{\vec{\eps}=0}\label{eoms1I}
   \ee
   
   \item For $i=2,\cdots,N+1$, at every edge $e\in E(\g)$,
   \be
   \frac{z_{i,i-1}(e)\,\tr\lt[\t^a g_{i}(e)^\dagger g_{i-1}(e)\rt]}{\sqrt{x_{i,i-1}(e)-1}\sqrt{x_{i,i-1}(e)+1}}-\frac{p_i(e)\,\tr\lt[\t^a g_{i}(e)^\dagger g_i(e)\rt]}{\sinh(p_i(e))}
   =-\frac{i\kappa \Delta\t}{a^2}\frac{\partial\,{\mathbf{H}\lt[g_i^\eps\rt]}}{\partial{\bar{\eps}^a(e)}}\Bigg|_{\vec{\eps}=0}.\label{eoms2I}
   \ee
   
   \item The closure condition at every vertex $v\in V(\g)$ for initial data:
   \be
   -\sum_{e, s(e)=v}p_1^a(e)+\sum_{e, t(e)=v}\L^a_{\ b}\lt(\vec{\theta}_1(e)\rt)\,p_1^b(e)=0.\label{closure0I}
   \ee
\end{itemize}
\noindent   
In the above, $z_{i+1,i}(e)$ and $x_{i+1,i}(e)$ are given by $z_{i+1,i}(e)= \mathrm{arccosh}\lt(x_{i+1,i}(e)\rt)$, $x_{i+1,i}(e)=\half\tr\lt[g_{i+1}(e)^\dagger g_{i}(e)\rt]$. The initial and final conditions are given by $g_{1}=g'{}^h$ and $g_{N+1}=g$. ${\bf H}[g^\eps]$ is the classical physical Hamiltonian discretized on the lattice $\g$, and evaluated at the phase space point $g^\eps(e)=g(e)e^{\eps^a(e)\t^a}$. $\L^a_{\ b}(\vec{\theta})\in\mathrm{SO(3)}$ is given by $e^{\theta^a\t^a/2}\t^a e^{-\theta^a\t^a/2}=\L^a_{\ b}(\vec{\theta})\t^b$.

We may call Eqs.\Ref{eoms1I}, \Ref{eoms2I}, and \Ref{closure0I} effective equations if we call $S[g,h]$ the effective action. These effective equations are interesting because 
\begin{itemize}
   
   \item They govern the semiclassical dynamics of full LQG, and are valid for all dynamical scenarios/spacetimes.
   
   \item All ingredients of the equations are explicitly computable. The equations are solvable analytically or numerically. In particular, Eqs.\Ref{eoms1I} and \Ref{eoms2I} are dynamical evolution equations with natural discretization implied by LQG. They are ready for numerically simulation and can relate to numerical relativity.
   
   \item Quantum correction can be computed (by perturbative expansion or other non-perturbative methods) since these equations are derived from path integral formula. In contrast of the LQC effective dynamics, we can take into account all quantum fluctuations in our framework.  
\end{itemize}

In Sections \ref{HICOS} and \ref{CEE}, we apply the above effective EOMs to cosmology. To search for the cosmological solution, we insert the following ansatz respecting the homogeneous and isotropic symmetries
\be
g_i(e_I(v))=e^{\lt(\theta_i-{i}p_i\rt)\frac{\t^I}{2}},
\label{nenI}
\ee 
where $\theta_i$ and $p_i$ are constants on every spatial slice, and relates to holonomies and fluxes by  
\be
h_i(e_I(v))\equiv h_i(I)=e^{\theta_i\t_I/2}
\quad p^a_i(e_I(v))\equiv p_i^a(I)=p_i\,\delta^a_I.
\ee
EOMs \Ref{eoms1I}, \Ref{eoms2I}, and \Ref{closure0I} are simplified by the above ansatz, and interestingly reproduces the LQC effective equations in $\mu_0$-scheme. The solution resolves big-bang singularity in all scenarios of gravity coupled to the Brown-Kucha\v{r} dust, Gaussian dust, and massless scalar. The big-bang singularity is replaced by an (time-refection) unsymmetric bounce in the case of the Giesel-Thiemann's Hamiltonian. The bounce is symmetric in the case of Alesci-Assanioussi-Lewandowski-Makinen's Hamiltonian.  

The above summarizes our strategy to derive the cosmological effective dynamics and singularity resolution in full LQG from top to down. We would like to mention that our work is inspired by the work by Dapor and Liegener \cite{Dapor:2017rwv} where they propose the LQC effective dynamics to be generated by the cosmological coherent state expectation value of $\hat{H}$ of full LQG, and demonstrate the unsymmetric bounce (their effective equation coincides with the one proposed earlier by Ding, Ma, and Yang in \cite{Yang:2009fp}). However in their scheme, the relation with the full LQG relies on the conjecture of existing dynamically stabled coherent state. But this conjecture is difficult to verify. It is usually difficult to find dynamically stable coherent state in interacting quantum field theories. However lessons from quantum field theory indicate that path integral should be the most powerful tool for studying effective theories. Our work suggests to use path integral for deriving effective dynamics in LQG and bypassing the difficulty of finding dynamically stable coherent state. 

In Section \ref{Semiclassical amplitude}, we compute the contribution of the cosmological solutions to the transition amplitude in the semiclassical approximation, and show that in a formal continuum limit of the lattice $\g$, the contribution is independent of the choice of Hamiltonian. 

The cosmological effective dynamics obtained here correspond to the $\mu_0$-scheme of LQC. It is known that in LQC, the $\mu_0$-scheme suffers the problem that the bounce may happen at low critical density (or large critical volume), and the better scheme is the $\bar{\mu}$-scheme where the critical density is Plankian (see e.g. \cite{Ashtekar:2006wn,Assanioussi:2019iye}). In Section \ref{Onmubar}, we comment on the problem of $\mu_0$-scheme from the full LQG point of view, and suggest that if we take into account of the continuum limit of the lattice $\g$, the critical density should be always high in our framework. However the continuum limit requires to take non-perturbative quantum corrections into consideration, and is beyond the scope of this paper (see some recent work e.g. \cite{Dittrich:2014ala,Bahr:2016hwc,Lang:2017beo,Han:2018fmu} on the continuum limit in LQG). We propose that the $\bar{\mu}$-scheme should relate to the \emph{quantum} effective EOM, which may be derived from the \emph{quantum} effective action of our path integral.\\

The architecture of this paper is the following: Firstly in Section \ref{Preliminaries} we review briefly some key results and tools in the reduced phase space quantization of LQG and complexifier coherent states. Section \ref{CSPI} derives the path integral formula from the transition amplitude of full LQG. Section \ref{SAEEQN} derives the effective EOMs of full LQG from the variational principle. Section \ref{HICOS} studies the effective EOMs with the homogeneous and isotropic ansatz. Section \ref{CEE} derives the cosmological effective equations with different choices of Hamiltonians. Section \ref{BOUNCE} obtains the solution with singularity resolution and unsymmetric/symmetric bounce. Section \ref{Semiclassical amplitude} computes the on-shell action and the semiclassical approximation of the transition amplitude. Section \ref{Onmubar} discusses the relation between the continuum limit and effective dynamics in $\bar{\mu}$-scheme, as well as a few other future perspectives.

\section{Preliminaries}\label{Preliminaries}

\subsection{Reduced phase space quantization of LQG}

Several models have been proposed to introduce scalar material reference frame and deparametrize GR into systems with physical Hamiltonians (see e.g. \cite{Kuchar:1990vy,Brown:1994py,Giesel:2007wn,Alesci:2015wla,Giesel:2012rb,Husain:2011tk}). The deparametrization procedure solves Hamiltonian and diffeomorphism constraints classically and construct models in terms of Dirac observables. These models have physical time evolution generated by physical Hamiltonian. LQG quantization of these models bypass the standard difficulty of solving quantum Hamiltonian constraint, because Hilbert spaces and operators are resulting from quantizing Dirac observables.

Among models of deparametrized GR, our analysis takes into account 3 different scenarios: the Brown-Kucha\v{r} dust, Gaussian dust, and massless scalar field: Firstly, we denote by $S_{BKD}$ and $S_{GD}$ the dust actions of Brown-Kucha\v{r} and Gaussian dust models \cite{Brown:1994py,Kuchar:1990vy,Giesel:2007wn,Giesel:2012rb}:
\be
S_{BKD}[\rho,g_{\mu\nu},T,S^j,W_j]&=& -\frac{1}{2}\int\rmd^4x\ \sqrt{|\det(g)|}\ \rho\ [g^{\mu\nu}U_\mu U_\nu+1],\quad U_\mu=-\partial_\mu T+W_j\partial_\mu S^j,\\
S_{GD}[\rho,g_{\mu\nu},T,S^j,W_j]&=&-\int\rmd^4x\ \sqrt{|\det(g)|}\lt[ \frac{\rho}{2}\ \lt(g^{\mu\nu}\partial_\mu T\partial_\nu T+1\rt)+g^{\mu\nu}\partial_\mu T \lt(W_j\partial_\nu S^j\rt)\rt],\label{dustaction}
\ee
where $T, S^{j=1,2,3}$ form the dust reference frame, and $\rho,\ W_j$ are Lagrangian multipliers. When we couple either $S_{BKD}$ or $S_{GD}$ to Einstein gravity and carry out the Hamiltonian analysis \cite{Giesel:2012rb}, the total Hamiltonian and diffeomorphism constraints can be cast in the following form ($\a=1,2,3$)
\be
C^{tot}&=&P+h(p,q,\partial_\a T),\label{Hsolve}\\
C^{tot}_j&=&P_j+S^\a_j\lt[C_\a(p,q)+P\partial_\a T\rt],\label{csolve}
\ee
which are strongly Poisson commutative. $S^\a_j$ is the inverse matrix of $\partial_\a S^j$ ($\a=1,2,3$), $P,P_j$ are momenta conjugate to $T,S^j$, and $p,q$ denote canonical variables of gravity. $C_\a$ are the diffeomorphism constraint of gravity. The classical constraints $C^{tot}=C^{tot}_j=0$ are solved by $P=-h$ and $P_j=-S^\a_j\lt[C_\a(p,q)+P\partial_\a T\rt]$. Gauge invariant Dirac observables are constructed relationally by parametrizing gravity variables with values of dust fields $T(x)\equiv\t,S^j(x)\equiv\sig^j$, e.g. $p(\sig,\t)=p(x)|_{T(x)\equiv\t,\,S^j(x)\equiv\sig^j}$ and $q(\sig,\t)=q(x)|_{T(x)\equiv\t,\,S^j(x)\equiv\sig^j}$, where $\sig,\t$ are physical space and time coordinates in the dust frame. 

We choose gravity canonical variables as su(2) Ashtekar-Barbero connection $A_j^a(\sig,\t)$ and densitized triad $E^j_a(\sig,\t)$. Both $A_j^a(\sig,\t)$ and $E^j_a(\sig,\t)$ are Dirac observables ($a=1,2,3$ is the su(2) index and $j=1,2,3$ is the coordinate index in $\cs$). They satisfy the Poisson bracket $\{E^i_a(\sig,\t),A_j^b(\sig',\t)\}=\frac{1}{2}\kappa \b\delta^{i}_j\delta^b_a\delta^{3}(\sig,\sig')$ where $\b$ is the Barbero-Immirzi parameter and $\kappa=16\pi G$. The phase space $\cp$ of $A_j^a(\sig,\t),E^j_a(\sig,\t)$ is free of Hamiltonian and diffeomorphism constraints, and all phase space functions are Dirac observables.

As an important Dirac observable, the physical Hamiltonian is given by integrating $h$ on the constant $T=\t$ slice $\cs$: $\mathbf{H}=\int_\cs\rmd^3\sig\, h|_{\partial_\a T=0}$ \cite{Giesel:2007wn,Giesel:2012rb},
\be
\text{Brown-Kucha\v{r} dust:}&&\mathbf{H}=\int_\cs\rmd^3\sig\, \sqrt{C(\sig,\t)^2-\frac{1}{4}\sum_{j=1}^3C_j(\sig,\t)C_j(\sig,\t)},\label{ham1}\\
\text{Gaussian dust:}&&\mathbf{H}=\int_\cs\rmd^3\sig\, C(\sig,\t).\label{ham2}
\ee 
The constant $\t$ slice $\cs$ is coordinated by the value of dust scalars $S^j=\sig^j$ thus is often referred to as the dust space \cite{Giesel:2007wn,Giesel:2012rb}. The physical Hamiltonian $\mathbf{H}$ generates the $\t$-time evolution:
\be
\frac{\rmd f}{\rmd\t}=\lt\{\mathbf{H}, f\rt\},
\ee
for all phase space function $f$ of $A_j^a(\sig,\t)$ and $E^j_a(\sig,\t)$. In either the Brown-Kucha\v{r} or the Gaussian dust model, $\mathbf{H}$ relates to the Hamiltonian and diffeomorphism constraints $C(\sig,\t),C_j(\sig,\t)$ evaluated on $\cs$. $C(\sig,\t),C_j(\sig,\t)$ are in terms of $A_j^a(\sig,\t)$ and $E^j_a(\sig,\t)$: 
\be
C&=&-\frac{2}{\kappa\sqrt{\det(q)}}\tr\lt(F_{jk}\lt[E^j,E^k\rt]\rt)+\frac{2({1-s\b^2})}{\kappa\sqrt{\det(q)}}\tr\lt(\lt[K_j,K_k\rt]\lt[E^j,E^k\rt]\rt),\label{Csigmatau}\\
C_j&=&-\frac{2}{\kappa\sqrt{\det(q)}}\tr\lt(\t_j F_{kl}\lt[E^k,E^l\rt]\rt)\label{Cjsigmatau}.
\ee  
where $E^j=E^j_a\t^a/2$, $K_j=K_j^a\t^a/2$ is the extrinsic curvature, and $F_{jk}=F_{jk}^a\t^a/2$ is the curvature of $A_j=A_j^a\t^a/2$. $\t^a=-i(\text{Pauli matrix})^a$. $s=1$ or $-1$ corresponds respectively to the Euclidean or Lorentzian signature. 

The gravity-dust models only deparametrize the Hamiltonian and diffeomorphism constraints, while the SU(2) Gauss constraint $D_j E^j=0$ still has to be imposed to the classical phase space. In addition, we impose some non-holonomic constraints to the phase space. In the Brown-Kucha\v{r} dust model, solving $C^{tot}=C^{tot}_j=0$ restricts $C(\sig,\t)^2-\frac{1}{4}\sum_{j=1}^3C_j(\sig,\t)C_j(\sig,\t)\geq 0$. In the Gaussian dust model, we split the phase space $\cp$ into two sector $C(\sig,\t)\geq 0$ and $C(\sig,\t)< 0$, where we can write $\mathbf{H} =\int_\cs\rmd^3\sig\,| C(\sig,\t)|$ and $\mathbf{H} =-\int_\cs\rmd^3\sig\,| C(\sig,\t)|$ respectively. Note that since we consider the dust is the only matter coupling to gravity, we work with physical Brown-Kucha\v{r} dust with positive energy density in order to fulfill the energy condition. We also choose the physical time $\t$ to flow backward in order to make ${\bf H}$ positive (see \cite{Giesel:2007wi} for details).

A real massless scalar field coupling to gravity is another interesting deparametrized model for reduced phase space quantization. We minimally couple the following scalar field action to gravity:
\be
S_\phi[g_{\mu\nu},\phi]=-\frac{1}{2}\int\rmd^4x\ \sqrt{|\det(g)|}\, g^{\mu\nu}\partial_\mu\phi\,\partial_\nu\phi.
\ee
Similar to $T$ in the above dust models, the value of the real scalar $\phi(x)=\t$ plays the role of the physical time variable. Canonical variables of gravity are parametrized by the physcical time $\t$: $p(\vec{x},\t)=p(x)|_{\phi(x)\equiv\t}$ and $q(\vec{x},\t)=q(x)|_{\phi(x)\equiv\t}$. The total Hamiltonian and diffeomorphism constraints can be written as
\be
C^{tot}&=&\pi-h(p,q),\quad h(p,q)=\sqrt{s\sqrt{\det(q)}\,C+\sqrt{\det(q)}\sqrt{C^2-q^{\a\b}C_\a C_\b}}\nonumber\\
C^{tot}_\a&=&C_\a+\pi\partial_\a\phi,
\ee
where $\pi$ is the momentum conjugate to $\phi$. The Hamiltonian constriant is solved simply by $\pi=h(p,q)$. We obtain the physical Hamiltonian \cite{Domagala:2010bm,Rovelli:1993bm}:
\be
\text{Gravity-scalar}:&& {\bf H}=-\int_\cs\rmd^3x \sqrt{s\sqrt{\det(q)}\,C+\sqrt{\det(q)}\sqrt{C^2-q^{\a\b}C_\a C_\b}}.\label{physHamScalar}
\ee
where $\cs$ is again the constant $\t$ slice. There is no analog of $S^j$ to solve the diffeomorphism constraint since we only have a single scalar field. Therefore we have to impose the diffeomorphism constraint in addition to the Gauss constraint to the phase space $\cp$ and solve them at the quantum level. We have to imposes again the non-holonomic constraint that the quantities under the square-roots are non-negative. We remove the overall minus sign by a refection of time $\t\to-\t$.

The above classical theory can be quantized in the LQG framework. In this work, we assume topologically $\cs\simeq T^3$ and define the quantum theory on a cubic lattice $\g$ as a partition of $\cs$ (sets of edges and vertices are denoted by $E(\g)$ and $V(\g)$). Edges in $\g$ are oriented such that every vertex $v$ connects to 3 outgoing edges $e_{I}(v)$ (oriented toward the $I=1,2,3$ spatial directions) and 3 incoming edges $e_{I}(v-\hat{I})$. The physical Hamiltonian is going to be quantized as a non-graph-changing operator $\hat{\bf H}$. In LQG, we use the following holonomy and gauge covariant flux variables at every $e\in E(\g)$:
\be
h(e):=\cp \exp \int_{e}A,\quad
p^a(e):=-\frac{1}{2\b a^2}\tr\lt[\t^a\int_{S_e}\eps_{ijk}\rmd \sig^i\wedge\rmd \sig^j\ h\lt(\rho_e(\sig)\rt)\, E_b^k(\sig)\t^b\, h\lt(\rho_e(\sig)\rt)^{-1}\rt],\label{hpvari}
\ee 
where $S_e$ is a 2-face in the dual lattice $\g^*$, and $\rho_e(\sig)\subset S_e$ is a path starting at the begin point of $e$ and traveling along $e$ until $e\cap S_e$, then running in $S_e$ until $x$. $a$ is a length unit for making $p^a(e)$ dimensionless. 

In the quantum theory, the Hilbert space $\ch_\g$ is spanned by gauge invariant functions of all $h(e)$'s on $\g$, and is a proper subspace of $\ch_\g^0=\otimes_e L^2(\Su)$. $\hat{h}(e)$ becomes multiplication operators on functions, while $\hat{p}^a(e)=i t\,\hat{R}_e^a/2$ where $\hat{R}_e^a$ is the right invariant vector field on SU(2): $R^a f(h)=\frac{\rmd}{\rmd \eps}\big|_{\eps=0} f(e^{\eps\t^a}h)$. $t=\ell^2_p/a^2$ is a dimensionless semiclassicality parameter ($\ell^2_p=\hbar\kappa$). They satisfy the commutation relation:
\be
\lt[\hat{h}(e),\hat{h}(e')\rt] &=&0\nonumber\\
\lt[\hat{p}^a(e),\hat{h}(e')\rt] &=&i t \delta_{e,e'} \frac{\t^a}{2} {h}(e')\nonumber\\
\lt[\hat{p}^a(e),\hat{p}^b(e')\rt]&=&-it \delta_{e,e'} \eps_{abc} {p}^c(e'). \label{ph}
\ee

The (non-graph-changing) physical Hamiltonian operators $\hat{\bf H}$ of dust models are given by \cite{Giesel:2007wn}:
\be
\hat{\mathbf{H}}&=&\sum_{v\in V(\g)}\lt[\hat{M}_-^\dagger(v) \hat{M}_-(v)\rt]^{1/4}\quad \hat{M}_-(v)=\hat{C}_{v}^{\ \dagger}\hat{C}_{v}-\frac{\a}{4}\hat{C}_{j,v}^{\ \dagger}\hat{C}_{j,v},\quad \a=\begin{cases}
1,&\text{Brown-Kucha\v{r} dust,}\\
0,&\text{Gaussian dust.}\label{physHam}
\end{cases}
\ee 
where $\hat{C}_{v},\hat{C}_{j,v}$ quantize $C(\sig,\t),C_j(\sig,\t)$ and preserve the graph $\g$. The Hamiltonian operator $\hat{\mathbf{H}}$ is positive semi-definite and self-adjoint because $\hat{M}_-^\dagger(v) \hat{M}_-(v)$ is manifestly positive semi-definite and Hermitian, therefore admits a self-adjoint extension. $\hat{\mathbf{H}}$ quantizes the classical Hamiltonian $\mathbf{H}$ in Eqs.\Ref{ham1} in the phase space with the non-holonomic constraint $C(\sig,\t)^2-\frac{1}{4}C_j(\sig,\t)C_j(\sig,\t)\geq 0$ for the Brown-Kucha\v{r} dust model. For the Gaussian dust model, $\hat{\bf H}$ quantizes $\int_\cs\rmd^3\sig|C(\sig,\t)|$ and removes the overall minus sign in the phase space sector $C(\sig,\t)<0$ by flipping $\t\to-\t$.

For the deparametrized model with scalar field, we have to solve the diffeomophism constraint at the quantum level by averaging over the diffeomorphism group on $\cs$ \cite{Ashtekar:1995zh,book,review1,review}. In this context, we promote $\ch_\g$ to $\ch_{Diff,\g}$ of diffeomorphism invariant states
\be
\eta_{Diff}(\psi_\g)=\frac{1}{| GS_\g|}\sum_{\varphi_1\in Diff/Diff_\g}\sum_{\varphi_2\in GS_\g}U_{\varphi_1}U_{\varphi_2} \psi_\g.
\ee
where $\psi_\g\in\ch_\g$ is a cylindrical function, and $U_\varphi$ is the unitary operator representing diffeomorphisms. $Diff_\g$ denotes the subgroup of diffeomorphisms leaving $\g$ invariant, while $GS_\g\subset Diff_\g$ is the graph symmetry group. The resulting diffeomorphism invariant state $\eta_{Diff}(\psi_\g)$ depends on the equivalence class of $\g$ generated by diffeomorphisms acting on the cubic lattice. The inner product of $\ch_{Diff,\g}$ between diffeomorphism invariant states is defined by
\be
\lag\eta_{Diff}(\psi_\g),\eta_{Diff}(\psi_{\g}')\rag_{Diff}=\frac{1}{| GS_\g|}\sum_{\varphi_1\in Diff/Diff_\g}\sum_{\varphi_2\in GS_\g}\langle U_{\varphi_1}U_{\varphi_2} \psi_\g|\psi'_{\g}\rangle_{\ch_\g}.
\ee
In quantizing the physical Hamiltonian Eq.\Ref{physHamScalar}, we set $C_\a=0$ since $\hat{\bf H}$ is defined on diffeomorphism invariant states. The Hamiltonian operator is given by
\be
\text{Gravity-scalar:}&&\hat{\bf H}=\sum_{v\in V(\g)}\lt(\hat{V}_v \hat{C}_v^\dagger\hat{C}_v\hat{V}_v \rt)^{1/4},
\ee
which preserves the graph $\g$ and is defined on $\ch_{Diff,\g}$: $\hat{\bf H}\,\eta_{Diff}(\psi_\g)=\eta_{Diff}(\hat{\bf H}\psi_\g)$ by the diffeomorphism invariance $U_\varphi \hat{\bf H} U^{-1}_\varphi=\hat{\bf H}$. $\hat{V}_v \hat{C}_v^\dagger\hat{C}_v\hat{V}_v$ is manifestly positive semi-definite and Hermitian, therefore admits a self-adjoint extension and guarantees the self-adjointness of $\hat{\bf H}$. The above $\hat{\bf H}$ applies a different operator-ordering comparing to \cite{Domagala:2010bm,Giesel:2016gxq}, and doesn't require $\hat{C}_v$ to be self-adjoint.


In Sections \ref{CSPI}, \ref{SAEEQN}, and \ref{HICOS}, we are going to derive path integral formula and effective equations of motion in full LQG, with the application to homogeneous and isotropic cosmology. Our discussions there are in general and valid for all choices of $\hat{\bf H}$ (from different deparametrized models and from different choices of regularization, operator ordering, etc), while we only require that $\hat{\bf H}$ is self-adjoint and its coherent state expectation value gives the correct semiclassical limit. In Section \ref{CEE}, we will insert concrete expressions of $\hat{\bf{H}}$ and analyze their implications in cosmology.

 \subsection{Coherent states}

 The complexifier coherent state in LQG \cite{Thiemann:2000bw,Sahlmann:2001nv,Bianchi:2009ky} provides a useful tool for the semiclassical analysis of the quantum theory, and will be employed in Section \ref{CSPI} to derive a discrete path integral formula from the unitary time-evolution by the Hamiltonian $\hat{\mathbf{H}}$. In the path integral method, the most important property of the coherent states is the over-compeleteness relation. In addition, they have the following useful properties: 1) the coherent state label parametrizes LQG phase space, 2) The overlap function behaves as a sharply peaked Gaussian in phase space.  These useful facts of the coherent state is briefly reviewed in this section.

The coherent state associated to a single edge $e\in E(\g)$ can be written as a function of $h(e)\in\Su$:
\be
\psi^{t}_{g(e)}\lt(h(e)\rt)
&=&\sum_{j_e\in\Z_+/2\cup\{0\}}(2j_e+1)\ e^{-tj_e(j_e+1)/2}\chi_{j_e}\lt(g(e)h(e)^{-1}\rt)\label{coherent}
\ee
where $\chi_j$ is the SU(2) character at spin-$j$. The coherent state label $g(e)\in\Slc$ is the complexified holonomy 
\be
g(e)=
e^{-ip^a(e)\t^a/2}e^{\theta^a(e)\t^a/2}, \quad p^a(e),\ \theta^a(e)\in\R^3.\label{gthetap}
\ee
where $e^{\theta^a(e)\t^a/2}$ parametrizes the classical holonomy variable in Eq.\Ref{hpvari}. $g(e)$ and $\bar{g}(e)$ are complex coordinates on the LQG phase space of holonomies and fluxes. The coherent state $\psi^{t}_{g(e)}(h(e))$ depends on $g(e)$ holomorphically.

The above coherent state is not normalized, the normalized coherent state is denoted by
\be
\tilde{\psi}^t_{g(e)}=\frac{\psi^{t}_{g(e)}}{||\psi^{t}_{g(e)}||}.
\ee
It is useful to review the overlap amplitude of $\tilde{\psi}^t_{g(e)}$ \cite{book,Thiemann:2000ca}:
\be
\langle \tilde{\psi}^t_{g_{2}(e)}|\tilde{\psi}^t_{g_{1}(e)}\rangle&\simeq& \lt[\frac{\sinh\lt(p_1(e)\rt)\sinh\lt(p_2(e)\rt)}{p_1(e)p_2(e)}\rt]^{1/2}\frac{z_{21}(e)}{\sinh(z_{21}(e))} \, e^{K\lt(g_2(e),g_1(e)\rt)/t},\label{overlap}\\
 K\lt(g_2(e),g_1(e)\rt)&=&z_{21}(e)^2-\half p_2(e)^2-\half p_1(e)^2, \quad z_{21}(e)=\mathrm{arccosh}\lt( \half\tr\lt[g_2(e)^\dagger g_1(e)\rt]\rt)\label{K12}
\ee
where ``$\simeq$'' neglects $O(t^\infty)$ and $p_{1,2}(e)=\sqrt{p_{1,2}^a(e)p_{1,2}^a(e)}= \text{arccosh}(\frac{1}{2}\Tr[g_{1,2}(e)^{\dagger} g_{1,2}(e)])$.  $|\langle \tilde{\psi}^t_{g_{2}(e)}|\tilde{\psi}^t_{g_{1}(e)}\rangle|$ behaves as a Gaussian sharply peaked at $g_1(e)=g_2(e)$ a width given by $| p^a_1(e)-p^a_2(e)|\sim| \theta^a_1(e)-\theta^a_2(e)|\sim\sqrt{t}$. Note that $\langle \tilde{\psi}^t_{g_{2}(e)}|\tilde{\psi}^t_{g_{1}(e)}\rangle$ is clearly invariant under $z_{21}\mapsto -z_{21}$ which relates to the Weyl refection of SU(2). We fix the sign ambiguity of $z_{21}$ by using the inverse hyperbolic cosine function, so $\mathrm{Re}(z_{21})\geq 0$. Our convention for the inverse hyperbolic cosine is $\mathrm{arccosh}(x)=\ln(x+\sqrt{x+1}\sqrt{x-1})$.

It is important that at every edge $e$, the normalized coherent states form an over-complete basis in $\ch_e=L^2(\Su)$ \cite{Thiemann:2000ca}:
\be
\int_{G^\mathbb{C}}\rmd g(e)\ |\tilde{\psi}^{t}_{g(e)}\rangle\langle\tilde{\psi}^{t}_{g(e)}|=1_{\ch_e},\quad \rmd g(e)=\frac{c}{t^3}\rmd\mu_H(h(e))\,\rmd^3p(e),\quad c=\frac{2}{\pi}+o(t^\infty)\label{overcomplete}
\ee
where $\rmd\mu_H(h)$ is the Haar measure on SU(2).

The matrix elements of operators $\hat{p}_a(e)$ and $\hat{h}(e)$ with respect to coherent states are computed in \cite{Thiemann:2000bx} and shown to satisfy the anticipated semiclassical properties:
\be
\lt|\langle \tilde{\psi}^t_{g_{2}(e)}|\hat{p}^a(e)|\tilde{\psi}^t_{g_{1}(e)}\rangle-p^a_1(e)\langle \tilde{\psi}^t_{g_{2}(e)}|\tilde{\psi}^t_{g_{1}(e)}\rangle\rt|&\leq& t f_p\lt(\vec{p}_1(e),\vec{p}_2(e)\rt)\lt|\langle \tilde{\psi}^t_{g_{2}(e)}|\tilde{\psi}^t_{g_{1}(e)}\rangle\rt|\nonumber\\
\lt|\langle \tilde{\psi}^t_{g_{2}(e)}|\hat{h}(e)|\tilde{\psi}^t_{g_{1}(e)}\rangle-h_1(e)\langle \tilde{\psi}^t_{g_{2}(e)}|\tilde{\psi}^t_{g_{1}(e)}\rangle\rt|&\leq& t f_h\lt(\vec{p}_1(e),\vec{p}_2(e)\rt)\lt|\langle \tilde{\psi}^t_{g_{2}(e)}|\tilde{\psi}^t_{g_{1}(e)}\rangle\rt|,\label{hpbound}
\ee
where the bounds are proportional to the semiclassicality parameter $t$, and $f_h(\vec{p}_1,\vec{p}_2),f_p(\vec{p}_1,\vec{p}_2)$ are functions which grow no faster than exponentials as $\vec{p}_{1,2}$ going large.

We make the tensor product over all $e\in E(\g)$ to define coherent states in $\ch^0_\g$,
\be
{\psi}^t_g=\bigotimes_{e\in E(\g)}{\psi}^t_{g(e)}.
\ee
The (unnormalized) gauge invariant coherent state $\Psi^t_{[g]}$ is defined by group averaging the SU(2) gauge transformations \cite{Thiemann:2000bw}:
\be
\Psi^t_{[g]}(h)&=&P_{\mathrm{SU(2)}}\,{\psi}^t_g(h)=\int_{\mathrm{SU(2)}^{|V(\g)|}}\prod_{v\in V(\g)}\rmd\mu_H(h_v)\prod_{e\in E(\g)}{\psi}^t_{g(e)}\lt(h_{s(e)}h(e)h_{t(e)}^{-1}\rt)\nonumber\\
&=&\int_{\mathrm{SU(2)}^{|V(\g)|}}\prod_{v\in V(\g)}\rmd\mu_H(h_v)\prod_{e\in E(\g)}{\psi}^t_{h_{s(e)}^{-1}g(e)h_{t(e)}}\lt(h(e)\rt)\label{gaugeinv}
\ee
where $P_{\mathrm{SU(2)}}$ is the projection from $\ch_\g^0$ onto the Hilbert space $\ch_\g$ of gauge invariant states. $s(e)$ and $t(e)$ are source and target of the edge $e$. $\Psi^t_{[g]}(h)$ is labelled by the equivalence class $[g]$ generated by gauge transformations $g(e)\mapsto g^h(e)\equiv h_{s(e)}^{-1}g(e)h_{t(e)}$. We may write Eq.\Ref{gaugeinv} as
\be
|\Psi^t_{[g]}\rangle=\int\rmd h\, |{\psi}^t_{g^h}\rangle,\quad \text{where}\quad \rmd h=\prod_{v\in V(\g)}\rmd\mu_H(h_v),\quad \lt|{\psi}^t_{g^h}\rag=\bigotimes_{e\in E(\g)} \lt|{\psi}^t_{h_{s(e)}^{-1}g(e)h_{t(e)}}\rag.\label{ginvcoh}
\ee

\section{Coherent state path integral}\label{CSPI}

The unitary time-evolution by the physical Hamiltonian defines the transition amplitude $A_{[g],[g']}$ between a pair of gauge invariant coherent states ${\Psi}^t_{[g]},{\Psi}^t_{[g']}$: 
\be
A_{[g],[g']}:=\lag\Psi^t_{[g]}\rt|U(\tau)\lt|{\Psi}^t_{[g']}\rag_{\ch_\g},\quad U(\t):=\exp\lt( -\frac{i}{\hbar} \t \hat{\mathbf{H}}\rt).\label{Agg}
 \ee
In the context of deparametrized model with scalar field, the amplitude is defined in $\ch_{Diff,\g}$. The above $\Psi^t_{[g]}$ has to be replaced by the diffeomorphism average:
\be
A_{Diff;[g],[g']}&:=&\frac{1}{| GS_\g|}\sum_{\varphi_1\in Diff/Diff_\g}\sum_{\varphi_2\in GS_\g}\lag U_{\varphi_1}U_{\varphi_2}\Psi^t_{[g]}\rt|U(\tau)\lt|{\Psi}^t_{[g']}\rag_{\ch_\g}\nonumber\\
&=&\frac{1}{| GS_\g|}\sum_{\varphi_2\in GS_\g}\lag U_{\varphi_2}\Psi^t_{[g]}\rt|U(\tau)\lt|{\Psi}^t_{[g']}\rag_{\ch_\g}\label{Aggdiff}
\ee
In the above average, only trivial $\varphi_1$ give nonzero term in the sum since the Hamiltonian is non-graph-changing. Thus $A_{Diff;[g],[g']}$ is a finite sum over $GS_\g$ (graph symmetry group of the finite cubic lattice) of $A_{[g],[g']}$ in Eq.\Ref{Agg}.

We focus on computing $A_{[g],[g']}$. The gauge invariance of $\hat{\bf H}$ implies $[P_{\Su},U(\tau)]=0$. We can write $A_{[g],[g']}$ in terms of non-gauge-invariant coherent states $\psi^t_{g}$, $\psi^t_{g^h}$ and the inner product $\lag\cdot|\cdot\rag$ of $\ch_\g^0$:
\be
A_{[g],[g']}=\lag\psi^t_{g}\rt|P_{\Su}U(\tau)P_{\Su}\lt|{\psi}^t_{{g'}}\rag=\int \rmd h\lag\psi^t_{g}\rt|U(\tau)\lt|{\psi}^t_{g'{}^{h}}\rag
\ee 
where we have used $P_{\Su}^2=P_{\Su}$ and Eq.\Ref{ginvcoh}. The above integrand can be computed by discretizing the time $\t$ into $N$ steps, where $N$ can be arbitrarily large:
\be
&&\lag\psi^t_{g}\rt|U(\tau)\lt|{\psi}^t_{{g'}^{h}}\rag=\lag\psi^t_{g}\rt|\lt[e^{ -\frac{i}{\hbar}\Delta\t \hat{\mathbf{H}}}\rt]^N\lt|{\psi}^t_{{g'}^{h}}\rag,\quad\quad\text{where}\quad \Delta\t=T/N,\nonumber\\
&=&\int\mathrm{d}g_{N+1}\cdots\mathrm{d}g_1\langle\psi^t_{g}|\tilde{\psi}^t_{g_{N+1}}\rangle\langle \tilde{\psi}^t_{g_{N+1}}\big|e^{ -\frac{i}{\hbar}\Delta\t \hat{\mathbf{H}}}\big|\tilde{\psi}^t_{g_{N}}\rangle
\langle \tilde{\psi}^t_{g_{N}}\big|e^{ -\frac{i}{\hbar}\Delta\t \hat{\mathbf{H}}}\big|\tilde{\psi}^t_{g_{N-1}}\rangle\cdots
\langle \tilde{\psi}^t_{g_2}\big|e^{ -\frac{i}{\hbar}\Delta\t \hat{\mathbf{H}}}\big|\tilde{\psi}^t_{g_1}\rangle\langle\tilde{\psi}^t_{g_1}|{\psi}^t_{g'{}^{h}}\rangle 
\end{eqnarray}
where we have inserted $N+1$ overcompleteness relations in $\ch^0_\g$:
\be
\int\rmd g_i\ |\tilde{\psi}^{t}_{g_i}\rangle\langle\tilde{\psi}^{t}_{g_i}|=1_{\ch_\g^0},\quad \rmd g_i=\lt(\frac{c}{t^3}\rt)^{|E(\g)|}\prod_{e\in E(\g)}\rmd\mu_H(h_i(e))\,\rmd^3p_i(e),\quad i=1,\cdots,N-1.
\ee

Following the standard coherent state functional integral method, we let $N$ arbitrarily large thus $\Delta\t$ arbitrarily small. $U(\Delta\t)$ is a strongly continuous unitary group and ${[U(\Delta\t)|\psi\rangle-|\psi\rangle]}/{\Delta\t}\to -\frac{i}{\hbar} \hat{\mathbf{H}}\, |\psi\rangle$, so $\hat{\eps}\lt(\frac{\Delta\t}{\hbar}\rt):=\frac{\hbar}{\Delta\t}[U(\Delta\t)-1+\frac{i {\Delta \tau} }{\hbar} \hat{\mathbf{H}}]$ satisfies the strong limit $\hat{\eps}\lt(\frac{\Delta\t}{\hbar}\rt)|\psi\rangle\to0$ as $\Delta\t\to0$ for all $\psi$ in the domain of $\hat{\mathbf{H}}$. The coherent state $\tilde{\psi}^t_{g}$ belongs to the domain of $\hat{\mathbf{H}}$, thus $\eps_{i+1,i}\lt(\frac{\Delta \t}{\hbar}\rt)=\langle \tilde{\psi}^t_{g_{i+1}}|\hat{\eps}\lt(\frac{\Delta\t}{\hbar}\rt)| \tilde{\psi}^t_{g_{i}}\rangle$ satisfies $\lim\limits_{\Delta\t\to0}\eps_{i+1,i}\lt(\frac{\Delta \t}{\hbar}\rt)=0$. We obtain the following relation
\be
&&\langle \tilde{\psi}^t_{g_{i+1}}\big|\exp \lt( -\frac{i}{\hbar}\Delta\t \hat{\mathbf{H}}\rt)\big|\tilde{\psi}^t_{g_{i}}\rangle
=\langle \tilde{\psi}^t_{g_{i+1}}\big|1- \frac{i\Delta\t}{\hbar} \hat{\mathbf{H}} \big|\tilde{\psi}^t_{g_{i}}\rangle+\frac{\Delta \t}{\hbar}\eps_{i+1,i}\lt(\frac{\Delta \t}{\hbar}\rt)\nonumber\\
&=&\langle \tilde{\psi}^t_{g_{i+1}}\big|\tilde{\psi}^t_{g_{i}}\rangle\lt[1-\frac{i\Delta\t}{\hbar}\frac{\langle \tilde{\psi}^t_{g_{i+1}}\big| \hat{\mathbf{H}}\big|\tilde{\psi}^t_{g_{i}}\rangle}{\langle \tilde{\psi}^t_{g_{i+1}}\big|\tilde{\psi}^t_{g_{i}}\rangle}\rt] +\frac{\Delta \t}{\hbar}\eps_{i+1,i}\lt(\frac{\Delta \t}{\hbar}\rt)=\langle \tilde{\psi}^t_{g_{i+1}}\big|\tilde{\psi}^t_{g_{i}}\rangle\,e^{-\frac{i\Delta\t}{\hbar}\frac{\langle \tilde{\psi}^t_{g_{i+1}}\lt| \hat{\mathbf{H}}\rt|\tilde{\psi}^t_{g_{i}}\rangle}{\langle \tilde{\psi}^t_{g_{i+1}}|\tilde{\psi}^t_{g_{i}}\rangle} +\frac{\Delta \t}{\hbar}\tilde{\eps}_{i+1,i}\lt(\frac{\Delta \t}{\hbar}\rt)}\label{smallstep}
\ee
where 
\be
\frac{\Delta \t}{\hbar}\tilde{\eps}_{i+1,i}\lt(\frac{\Delta \t}{\hbar}\rt)=\ln\lt[1-\frac{i\Delta\t}{\hbar}\frac{\langle \tilde{\psi}^t_{g_{i+1}}\big| \hat{\mathbf{H}}\big|\tilde{\psi}^t_{g_{i}}\rangle}{\langle \tilde{\psi}^t_{g_{i+1}}\big|\tilde{\psi}^t_{g_{i}}\rangle} +\frac{\Delta \t}{\hbar}\frac{\eps_{i+1,i}\lt({\Delta \t}/{\hbar}\rt)}{\langle \tilde{\psi}^t_{g_{i+1}}\big|\tilde{\psi}^t_{g_{i}}\rangle}\rt]+\frac{i\Delta\t}{\hbar}\frac{\langle \tilde{\psi}^t_{g_{i+1}}\lt| \hat{\mathbf{H}}\rt|\tilde{\psi}^t_{g_{i}}\rangle}{\langle \tilde{\psi}^t_{g_{i+1}}|\tilde{\psi}^t_{g_{i}}\rangle}
\ee
and satisfy $\lim\limits_{\Delta\t\to0}\tilde{\eps}_{i+1,i}\lt(\frac{\Delta \t}{\hbar}\rt)=0$. 

By Eq.\Ref{smallstep}, the amplitude $A_{[g],[g']}$ can be written as a discrete path integral with an effective action $S[g,h]$ if we neglect $O(t^\infty)$ in the integrand:
\be
A_{[g],[g']}&=&||\psi^t_{g}||\,||{\psi}^t_{g'}||\int\rmd h\prod_{i=1}^{N+1}\rmd g_i\, e^{-\frac{i }{\hbar}\sum_{i=1}^{N}\Delta\t\lt[\frac{\langle \tilde{\psi}^t_{g_{i+1}}\lt| \hat{\mathbf{H}}\rt|\tilde{\psi}^t_{g_{i}}\rangle}{\langle \tilde{\psi}^t_{g_{i+1}}|\tilde{\psi}^t_{g_{i}}\rangle} +i\tilde{\eps}_{i+1,i}\lt(\frac{\Delta \t}{\hbar}\rt)\rt]}\,\langle\tilde{\psi}^t_{g}|\tilde{\psi}^t_{g_{N+1}}\rangle \left( \prod_{i=1}^N\langle \tilde{\psi}^t_{g_{i+1}}\big|\tilde{\psi}^t_{g_{i}}\rangle \right)\langle\tilde{\psi}^t_{g_1}|\tilde{\psi}^t_{g'{}^{h}}\rangle\nonumber\\
&=&||\psi^t_{g}||\,||{\psi}^t_{g'}||\int\rmd h\prod_{i=1}^{N+1}\rmd g_i\,\nu[g]\, e^{S[{g},h]/t},\label{integral}
\ee
where
\be
&& S[{g},h]=\sum_{i=0}^{N+1}K(g_{i+1},g_i)-\frac{i \kappa}{a^2}\sum_{i=1}^{N}\Delta\t\lt[\frac{\langle {\psi}^t_{g_{i+1}}\lt| \hat{\mathbf{H}}\rt|{\psi}^t_{g_{i}}\rangle}{\langle {\psi}^t_{g_{i+1}}|{\psi}^t_{g_{i}}\rangle} +i\tilde{\eps}_{i+1,i}\lt(\frac{\Delta \t}{\hbar}\rt)\rt],\quad g_0\equiv g'{^h},\ g_{N+2}\equiv g,\label{action}\\
&&\nu[g]=\prod_{i=0}^{N+1}\prod_{e\in E(\g)}\lt[\frac{\sinh(p_{i+1}(e))}{p_{i+1}(e)}\frac{\sinh(p_{i}(e))}{p_{i}(e)}\rt]^{1/2}\frac{z_{{i+1},i}(e)}{\sinh(z_{{i+1},i}(e))}.
\ee
In the above derivation, we have used the formula Eq.\Ref{overlap} of $\langle \tilde{\psi}^t_{g_{i+1}}|\tilde{\psi}^t_{g_{i}}\rangle$ with
\be
K(g_{i+1},g_i)&=&\sum_{e\in E(\g)}\lt[z_{i+1,i}(e)^2-\half p_{i+1}(e)^2-\half p_i(e)^2\rt]\\ 
z_{i+1,i}(e)&=& \mathrm{arccosh}\lt(x_{i+1,i}(e)\rt),\quad x_{i+1,i}(e)=\half\tr\lt[g_{i+1}(e)^\dagger g_{i}(e)\rt]
\ee
and the fact that $||{\psi}^t_{g'{}^{h}}||=||{\psi}^t_{g'}||$ does not depend on $h$.
The effective action $S[g,h]$ has been expressed in terms of unnormalized coherent states since
\be
\frac{\langle \tilde{\psi}^t_{g_{i+1}}\lt| \hat{\mathbf{H}}\rt|\tilde{\psi}^t_{g_{i}}\rangle}{\langle \tilde{\psi}^t_{g_{i+1}}|\tilde{\psi}^t_{g_{i}}\rangle}=\frac{\langle {\psi}^t_{g_{i+1}}\lt| \hat{\mathbf{H}}\rt|{\psi}^t_{g_{i}}\rangle}{\langle {\psi}^t_{g_{i+1}}|{\psi}^t_{g_{i}}\rangle},
\ee
which will be useful in deriving equations of motion (EOMs) in the following. Here $\nu[g]$ does not depend on $t$, so does not involve in the derivation of EOMs.

We remark that by Eq.\Ref{smallstep}, $\langle \tilde{\psi}^t_{g_{i+1}}|\exp \lt( -\frac{i}{\hbar}\Delta\t \hat{\mathbf{H}}\rt)|\tilde{\psi}^t_{g_{i}}\rangle$ is dominated by the overlap amplitude $\langle \tilde{\psi}^t_{g_{i+1}}|\tilde{\psi}^t_{g_{i}}\rangle$ as $\Delta\t$ is arbitrarily small. $\langle \tilde{\psi}^t_{g_{i+1}}|\tilde{\psi}^t_{g_{i}}\rangle$ decays exponentially fast to zero unless $g_{i+1}$ is within a small neighborhood at $g_i$ of radius $\sqrt{t}$. Therefore for sufficiently large $N$, the dominant contribution to $A_{[g],[g']}$ comes from the integral in Eq.\Ref{integral} over a neighborhood where all $g_{i+1}$ are close to $g_i$ ($i=0,\cdots,N$). In the following we focus on this dominant contribution by assuming $|\vec{p}_i-\vec{p}_{i+1}|$ and $|\vec{\theta}_i-\vec{\theta}_{i+1}|$ to be small and of $O(\sqrt{t})$.

\section{Semiclassical approximation and effective equations}\label{SAEEQN}

Given the integral representation Eq.\Ref{integral}, we apply the stationary phase approximation to Eq.\Ref{integral} to study the semiclassical behavior of the transition amplitude $A_{[g],[g']}$ as $t\to 0$.

To derive EOMs from the path integral, we apply the variational principle $\delta S[g,h]=0$ for all integration variables $h_v$, $g_i(e)$ ($i=1,\cdots,N+1$). We will send $\delta \tau \to 0$ and neglect arbitraily small $\tilde{\eps}$ contributions to get the cotinuum limit. First of all, we deform $h_v\to h_v^\eta=h_v\,e^{\eta^a_v\t^a}$ with $\eta^a_v\in\R$. Since only $g_0=g'{}^h$ involves $h_v$, $\delta_h S=0$ gives the following equation at every $v\in V(g)$
\be
-\sum_{e, s(e)=v}\frac{z_{10}(e)\,\tr\lt[g_1(e)^\dagger \t^a g'{}^h(e)\rt]}{\sqrt{x_{10}(e)-1}\sqrt{x_{10}(e)+1}}+\sum_{e, t(e)=v}\frac{z_{10}(e)\,\tr\lt[\t^a g_1(e)^\dagger g'{}^h(e)\rt]}{\sqrt{x_{10}(e)-1}\sqrt{x_{10}(e)+1}}=0.\label{closure}
\ee

To compute $\delta_g S=0$, we deform 
\be
g_i(e)\mapsto g_i^\eps(e)=g_i(e)\,e^{\eps_i^a(e)\t^a}
\ee 
with $\eps_i^a(e)\in\C, i=1,\cdots,N+1$. The variational principle gives the following EOMs at every $e\in E(\g)$ for $i=2,\cdots,N$:
\be
\lt[\frac{z_{i+1,i}(e)\,\tr\lt(\t^a g_{i+1}(e)^\dagger g_i(e)\rt)}{\sqrt{x_{i+1,i}(e)-1}\sqrt{x_{i+1,i}(e)+1}}-\frac{p_i(e)\,\tr\lt(\t^a g_{i}(e)^\dagger g_i(e)\rt)}{\sinh(p_i(e))}\rt]&=&\frac{i\kappa {\Delta\t}}{a^2}\frac{\partial}{\partial{\eps_i^a(e)}}\lt[\frac{\langle {\psi}^t_{g_{i+1}}\lt| \hat{\mathbf{H}}\rt|{\psi}^t_{g^\eps_{i}}\rangle}{\langle {\psi}^t_{g_{i+1}}|{\psi}^t_{g^\eps_{i}}\rangle} \rt]_{\vec{\eps}=0},\label{eom1}\\
-\lt[\frac{z_{i,i-1}(e)\,\tr\lt(\t^a g_{i}(e)^\dagger g_{i-1}(e)\rt)}{\sqrt{x_{i,i-1}(e)-1}\sqrt{x_{i,i-1}(e)+1}}-\frac{p_i(e)\,\tr\lt(\t^a g_{i}(e)^\dagger g_i(e)\rt)}{\sinh(p_i(e))}\rt]&=&\frac{i\kappa {\Delta\t}}{a^2}\frac{\partial}{\partial{\bar{\eps}_i^a(e)}}\lt[\frac{\langle {\psi}^t_{g^\eps_{i}}\lt| \hat{\mathbf{H}}\rt|{\psi}^t_{g_{i-1}}\rangle}{\langle {\psi}^t_{g^\eps_{i}}| {\psi}^t_{g_{i-1}}\rangle} \rt]_{\vec{\eps}=0}\label{eom2}.
\ee
where the left-hand sides are $\partial_{\eps_i^a(e)}\lt[K(g_{i+1},g^\eps_i)+K(g^\eps_{i},g_{i-1})\rt]_{\vec{\eps}=0}$ and $\partial_{\bar{\eps}^a(e)}\lt[K(g_{i+1},g^\eps_i)+K(g^\eps_{i},g_{i-1})\rt]$.  We have neglect arbitrarily small contributions from $\tilde{\eps}_{i+1,i}\lt({\Delta \t}/{\hbar}\rt)$. Note that ${\langle {\psi}^t_{g_{i+1}}\lt| \hat{\mathbf{H}}\rt|{\psi}^t_{g^\eps_{i}}\rangle}$ and ${\langle {\psi}^t_{g_{i+1}}|{\psi}^t_{g^\eps_{i}}\rangle}$ depend on $\eps^a(e)$ holomorphically, while ${\langle {\psi}^t_{g^\eps_{i}}\lt| \hat{\mathbf{H}}\rt|{\psi}^t_{g_{i-1}}\rangle}$ and ${\langle {\psi}^t_{g^\eps_{i}}|{\psi}^t_{g_{i-1}}\rangle}$ depend on $\eps^a(e)$ anti-holomorphically. 

The variation $\delta_{g_1} S=0$ gives
\be
\lt[\frac{z_{21}(e)\,\tr\lt(\t^a g_{2}(e)^\dagger g_1(e)\rt)}{\sqrt{x_{21}(e)-1}\sqrt{x_{21}(e)+1}}-\frac{p_1(e)\,\tr\lt(\t^a g_{1}(e)^\dagger g_1(e)\rt)}{\sinh(p_1(e))}\rt]&=&\frac{i\kappa {\Delta\t}}{a^2}\frac{\partial}{\partial{\eps_1^a(e)}}\lt[\frac{\langle {\psi}^t_{g_{2}}\lt| \hat{\mathbf{H}}\rt| {\psi}^t_{g^\eps_{1}}\rangle}{\langle {\psi}^t_{g_{2}}|{\psi}^t_{g^\eps_{1}}\rangle}\rt]_{\vec{\eps}=0},\label{eom11}\\
\frac{z_{10}(e)\,\tr\lt(\t^a g_{1}(e)^\dagger g'{}^h(e)\rt)}{\sqrt{x_{10}(e)-1}\sqrt{x_{10}(e)+1}}-\frac{p_1(e)\,\tr\lt(\t^a g_{1}(e)^\dagger g_1(e)\rt)}{\sinh(p_1(e))}&=&0\label{eom21}.
\ee
Similarly $\delta_{g_{N+1}} S=0$ gives
\be
\frac{z_{N+2,N+1}(e)\,\tr\lt(\t^a g(e)^\dagger g_{N+1}(e)\rt)}{\sqrt{x_{N+2,N+1}(e)-1}\sqrt{x_{N+2,N+1}(e)+1}}-\frac{p_{N+1}(e)\,\tr\lt(\t^a g_{N+1}(e)^\dagger g_{N+1}(e)\rt)}{\sinh(p_{N+1}(e))}&=&0,\label{eom1N+1}\\
-\lt[\frac{z_{N+1,N}(e)\,\tr\lt(\t^a g_{N+1}(e)^\dagger g_{N}(e)\rt)}{\sqrt{x_{N+1,N}(e)-1}\sqrt{x_{N+1,N}(e)+1}}-\frac{p_{N+1}(e)\,\tr\lt(\t^a g_{N+1}(e)^\dagger g_{N+1}(e)\rt)}{\sinh(p_{N+1}(e))}\rt]
&=&\frac{i\kappa {\Delta\t}}{a^2}\frac{\partial}{\partial{\bar{\eps}_{N+1}^a(e)}}\lt[\frac{\langle {\psi}^t_{g^\eps_{N+1}}\lt| \hat{\mathbf{H}}\rt|{\psi}^t_{g_{N}}\rangle}{\langle {\psi}^t_{g^\eps_{N+1}}|{\psi}^t_{g_{N}}\rangle} \rt]_{\vec{\eps}=0}\label{eom2N+1}.
\ee
Detailed steps of deriving Eqs.\Ref{closure} - \Ref{eom2N+1} are given in Appendix \ref{VariationofS}.

Right-hand sides of Eqs.\Ref{eom1} - \Ref{eom2N+1} are all infinitesimal or zero since $\Delta\t$ is arbitrarily small\footnote{$\partial_{\eps^a_i(e)}\langle {\psi}^t_{g_{i+1}}\lt| \hat{\mathbf{H}}\rt|{\psi}^t_{g^\eps_{i}}\rangle=\partial_{\eps^a_i(e)}\int\rmd h\,\overline{(\hat{\mathbf{H}}^\dagger {\psi}^t_{g_{i+1}})(h)}\,{\psi}^t_{g^\eps_{i}}(h)=\int\rmd h\,\overline{(\hat{\mathbf{H}}^\dagger {\psi}^t_{g_{i+1}})(h)}\,\partial_{\eps^a_i(e)}{\psi}^t_{g^\eps_{i}}(h)$ since the integral is over a compact space. Moreover since $\psi^t_{g^\eps(e)}(h(e))=\sum_{j_e\in\Z_+/2\cup\{0\}}(2j_e+1)\ e^{-tj_e(j_e+1)/2}\chi_{j_e}\lt(g(e)\,e^{\eps^a(e)\t^a}h(e)^{-1}\rt)$, we have $\partial_{\eps^a_i(e)}{\psi}^t_{g^\eps_{i}}(h)|_{\eps=0}=-\hat{L}^a_e\,\psi^t_{g(e)}\lt(h\rt)$
where $\hat{L}^a_e$ is the left invariant vector field on SU(2) $L^a f(h)=\frac{\rmd}{\rmd \eps}\big|_{\eps=0}f(h\,e^{\eps\t^a})$. $\hat{L}^a_e$ is an anti-hermitian operator on $L^2(\mathrm{SU(2)})$. We obtain
\be
\frac{i\kappa }{a^2}\frac{\partial}{\partial{\eps^a_i(e)}}\lt[\frac{\langle {\psi}^t_{g_{i+1}}\lt| \hat{\mathbf{H}}\rt|{\psi}^t_{g^\eps_{i}}\rangle}{\langle {\psi}^t_{g_{i+1}}|{\psi}^t_{g^\eps_{i}}\rangle}\rt]_{\vec{\eps}=0}
&=&-\frac{i\kappa }{a^2}\lt[\frac{\langle  {\psi}^t_{g_{i+1}}\lt| \hat{\mathbf{H}}\hat{L}_e^a\rt| {\psi}^t_{g_{i}}\rangle}{\langle  {\psi}^t_{g_{i+1}}| {\psi}^t_{g_{i}}\rangle}
-\frac{\langle  {\psi}^t_{g_{i+1}}\lt|\hat{L}_e^a\rt| {\psi}^t_{g_{i}}\rangle\langle  {\psi}^t_{g_{i+1}}\lt| \hat{\mathbf{H}}\rt| {\psi}^t_{g_{i}}\rangle}{\langle  {\psi}^t_{g_{i+1}}| {\psi}^t_{g_{i}}\rangle^2}\rt],\label{HL}\\
\frac{i\kappa }{a^2}\frac{\partial}{\partial{\bar{\eps}_i^a(e)}}\lt[\frac{\langle  {\psi}^t_{g^\eps_{i}}\lt| \hat{\mathbf{H}}\rt| {\psi}^t_{g_{i-1}}\rangle}{\langle  {\psi}^t_{g^\eps_{i}}| {\psi}^t_{g_{i-1}}\rangle} \rt]_{\vec{\eps}=0}&=&\frac{i\kappa }{a^2}\lt[\frac{\langle  {\psi}^t_{g_{i}}\lt|\hat{L}_e^a \hat{\mathbf{H}}\rt| {\psi}^t_{g_{i-1}}\rangle}{\langle  {\psi}^t_{g_{i}}| {\psi}^t_{g_{i-1}}\rangle}
-\frac{\langle  {\psi}^t_{g_{i}}\lt|\hat{L}_e^a\rt| {\psi}^t_{g_{i-1}}\rangle\langle  {\psi}^t_{g_{i}}\lt| \hat{\mathbf{H}}\rt| {\psi}^t_{g_{i-1}}\rangle}{\langle  {\psi}^t_{g_{i}}| {\psi}^t_{g_{i-1}}\rangle^2}\rt]\label{LH}.
\ee
which are clearly finite for arbitrary $g_{i+1},g_i,g_{i-1}$. So the above quantities times $\Delta\t$ are arbitrarily small.}, so left-hand sides must also be infinitesimal or zero for all solutions of these equations. The following quantities:
\be
D_1^a(g_i,g_j)=\frac{z_{ij}}{\sqrt{x_{ij}-1}\sqrt{x_{ij}+1}}\tr\lt(\t^a g_{i}^\dagger g_{j}\rt)-\frac{p_{j}}{\sinh(p_{j})}\tr\lt(\t^a g_{j}^\dagger g_{j}\rt)\\
D_2^a(g_i,g_j)=\frac{z_{ij}}{\sqrt{x_{ij}-1}\sqrt{x_{ij}+1}}\tr\lt(\t^a g_{i}^\dagger g_{j}\rt)-\frac{p_{i}}{\sinh(p_{i})}\tr\lt(\t^a g_{i}^\dagger g_{i}\rt)
\ee
must approach to zero when $\Delta\t\to0$. And in Eqs.\Ref{eom21} and \Ref{eom1N+1}, $D_2(g_1,g'{}^h)$ and $D_1(g,g_{N+1})$ strictly vanish.


\begin{Lemma}\label{isolatezero}

In a neighborhood where $g_{i},g_j$ are sufficiently close (with their distence of $O(\sqrt{t})$), $D^a_1(g_i,g_j)=0$ or $D^a_2(g_i,g_j)=0$ if and only if $g_i=g_j$.

\end{Lemma}

\textbf{Proof:} In the neighborhood, we expand $g_{i}=g_j[1+\Delta\phi^a\t^a+O(\Delta\phi^2)]$ in $D^a_1(g_i,g_j)$ and $g_{j}=g_i[1-\Delta\phi^a\t^a+O(\Delta\phi^2)]$ in $D^a_2(g_i,g_j)$. We obtain (see Appendix \ref{DL4.1} for details)
\be
D_1^a(g_i,g_j)&=&M_1{}^{a}_{\ b}(g_j)\, \Delta\bar{\phi}^b+O(\Delta\phi^2),\quad
D_2^a(g_i,g_j)=M_2{}^{a}_{\ b}(g_i)\, \Delta{\phi}^b+O(\Delta\phi^2)\\
M_{1}{}^a_{\ b}(g_j)&=&2\L^a_{\ c}(\vec{\theta}_j)\L^{b}_{\ d}(\vec{\theta}_j)\lt[\frac{p_j^c}{p_j}\frac{p_j^d}{p_j}-i\eps^{cde}p^e_j+\frac{p_j\cosh(p_j)}{\sinh(p_j)}\lt(\delta^{cd}-\frac{p_j^c}{p_j}\frac{p_j^d}{p_j}\rt)\rt],\\
M_{2}{}^a_{\ b}(g_i)&=&2\L^a_{\ c}(\vec{\theta}_i)\L^{b}_{\ d}(\vec{\theta}_i)\lt[\frac{p_i^c}{p_i}\frac{p_i^d}{p_i}+i\eps^{cde}p^e_i+\frac{p_i\cosh(p_i)}{\sinh(p_i)}\lt(\delta^{cd}-\frac{p_i^c}{p_i}\frac{p_i^d}{p_i}\rt)\rt],
\ee
where $g_j=e^{-i p_j^a\t^a/2}e^{\theta_j^a\t^a/2}$ and $\L^a_{\ b}(\vec{\theta})\in\mathrm{SO(3)}$ is given by $e^{\theta_j^a\t^a/2}\t^a e^{-\theta_j^a\t^a/2}=\L^a_{\ b}(\vec{\theta}_j)\t^b$. The matrix $M_{1,2}{}^a_{\ b}(\vec{p}_j)$ is nondegenerate since
\be
\det\lt(M_{1,2}(g)\rt)=\frac{\sinh^2(p)}{p^2}\neq 0.
\ee
Therefore $g_i=g_j$ is an isolated root for $D^a_1(g_i,g_j)=0$ or $D^a_2(g_i,g_j)=0$.

$\Box$

Therefore when we send $\Delta\t\to0$, all solutions of EOMs Eqs.\Ref{eom1} - \Ref{eom2N+1} must satisfy $g_i\to g_{i+1}$ and $|p^a_i-p^a_{i+1}|\sim|\theta^a_i-\theta_{i+1}^a|\sim \Delta\t $ for all solutions of Eqs.\Ref{eom1} - \Ref{eom2N+1}. Moreover Eqs.\Ref{eom21} and \Ref{eom1N+1} implies
\be
 g_{N+1}=g,\quad \text{and}\quad g_{1}=g'{}^h.\label{bc}
\ee
which set the final and initial conditions for EOMs. The initial condition reduces Eq.\Ref{closure} to the closure condition at $v$ (see Appendix \ref{VariationofS} for derivation):
\be
-\sum_{e, s(e)=v}p_1^a(e)+\sum_{e, t(e)=v}\L^a_{\ b}\lt(\vec{\theta}_1(e)\rt)\,p_1^b(e)=0.\label{closure0}
\ee
where $\L^a_{\ b}(\vec{\theta})\in\mathrm{SO(3)}$ is given by $e^{\theta_j^a\t^a/2}\t^a e^{-\theta_j^a\t^a/2}=\L^a_{\ b}(\vec{\theta}_j)\t^b$. $\vec{p}_1(e)$ is a vector located at the source $s(e)$ of $e$, and $\L^a_{\ b}\lt(\vec{\theta}_1(e)\rt)$ parallel transports $\vec{p}_1(e)$ from the source $s(e)$ to target $t(e)$.

Letting $\Delta\t$ arbitrarily small, we approximate solutions of EOMs by the continuum limit (in the time direction) $\Delta\t\to 0$, which leads to $g_i\to g_{i+1}$. In this limit, matrix elements of $\hat{\bf H}$ in right-hand sides in Eqs.\Ref{eom1} - \Ref{eom11} and \Ref{eom2N+1} reduces to expectation values of $\hat{\bf H}$:

\begin{Lemma}\label{expectation}
\be
\lim_{g_i\to g_{i+1}\equiv g}\frac{\partial}{\partial{\eps_i^a(e)}}\lt[\frac{\langle {\psi}^t_{g_{i+1}}\lt| \hat{\mathbf{H}}\rt|{\psi}^t_{g^\eps_{i}}\rangle}{\langle {\psi}^t_{g_{i+1}}|{\psi}^t_{g^\eps_{i}}\rangle} \rt]_{\vec{\eps}=0}
=\frac{\partial{\langle \tilde{\psi}^t_{g^\eps}\lt| \hat{\mathbf{H}}\rt|\tilde{\psi}^t_{g^\eps}\rangle}}{\partial{\eps^a(e)}}\Bigg|_{\vec{\eps}=0} \label{expectation1}\\
\lim_{g_{i-1}\to g_{i}\equiv g}\frac{\partial}{\partial{\bar{\eps}_i^a(e)}}\lt[\frac{\langle {\psi}^t_{g^\eps_{i}}\lt| \hat{\mathbf{H}}\rt|{\psi}^t_{g_{i-1}}\rangle}{\langle {\psi}^t_{g^\eps_{i}}| {\psi}^t_{g_{i-1}}\rangle} \rt]_{\vec{\eps}=0}=\frac{\partial{\langle \tilde{\psi}^t_{g^\eps}\lt| \hat{\mathbf{H}}\rt|\tilde{\psi}^t_{g^\eps}\rangle}}{\partial{\bar{\eps}^a(e)}}\Bigg|_{\vec{\eps}=0}.\label{expectation2}
\ee

\end{Lemma}

\textbf{Proof:} Firstly we have 
\be
\frac{\partial}{\partial{\eps^a(e)}}\langle {\psi}^t_{g^\eps}\lt| \hat{\mathbf{H}}\rt|{\psi}^t_{g^\eps}\rangle=\frac{\partial}{\partial{\eps^a(e)}}\int\rmd h\,\overline{(\hat{\mathbf{H}}^\dagger {\psi}^t_{g^\eps})(h)}\,{\psi}^t_{g^\eps}(h)=\int\rmd h\,\overline{(\hat{\mathbf{H}}^\dagger {\psi}^t_{g^\eps})(h)}\,\frac{\partial}{\partial{\eps^a(e)}}{\psi}^t_{g^\eps}(h)
\ee 
since the integral is over a compact space and $\overline{(\hat{\mathbf{H}}^\dagger {\psi}^t_{g^\eps})(h)}$ depends on $\eps^a(e)$ anti-holomorphically. By $\psi^t_{g^\eps(e)}(h(e))=\sum_{j_e\in\Z_+/2\cup\{0\}}(2j_e+1)\ e^{-tj_e(j_e+1)/2}\chi_{j_e}\lt(g(e)\,e^{\eps^a(e)\t^a}h(e)^{-1}\rt)$, we have 
\be
\frac{\partial}{\partial{\eps^a(e)}}{\psi}^t_{g^\eps}(h)\bigg|_{\eps=0}=-\hat{L}^a_e\,\psi^t_{g(e)}\lt(h\rt)
\ee
where $\hat{L}^a_e$ is the left invariant vector field on SU(2): $L^a f(h)=\frac{\rmd}{\rmd \eps}\big|_{\eps=0}f(h\,e^{\eps\t^a})$. $\hat{L}^a_e$ is an anti-hermitian operator on $L^2(\mathrm{SU(2)})$. Computing right-hand sides of Eqs.\Ref{expectation1} and \Ref{expectation2} is similar to Eqs.\Ref{HL} and \Ref{LH}, 
\be
\frac{\partial}{\partial{\eps^a(e)}}\frac{\langle {\psi}^t_{g^\eps}\lt| \hat{\mathbf{H}}\rt|{\psi}^t_{g^\eps}\rangle}{\langle {\psi}^t_{g^\eps}|{\psi}^t_{g^\eps}\rangle}\Bigg|_{\vec{\eps}=0}=-\frac{\langle {\psi}^t_{g}\lt| \hat{\mathbf{H}}\hat{L}_e^a\rt|{\psi}^t_{g}\rangle}{\langle {\psi}^t_{g}| {\psi}^t_{g}\rangle}
+\frac{\langle {\psi}^t_{g}\lt|\hat{L}_e^a\rt| {\psi}^t_{g}\rangle\langle  {\psi}^t_{g}\lt| \hat{\mathbf{H}}\rt| {\psi}^t_{g}\rangle}{\langle {\psi}^t_{g}| {\psi}^t_{g}\rangle^2}.\\
\frac{\partial}{\partial{\bar{\eps}^a(e)}}\frac{\langle {\psi}^t_{g^\eps}\lt| \hat{\mathbf{H}}\rt|{\psi}^t_{g^\eps}\rangle}{\langle {\psi}^t_{g^\eps}|{\psi}^t_{g^\eps}\rangle}\Bigg|_{\vec{\eps}=0}=\frac{\langle {\psi}^t_{g}\lt|\hat{L}_e^a \hat{\mathbf{H}}\rt|{\psi}^t_{g}\rangle}{\langle {\psi}^t_{g}| {\psi}^t_{g}\rangle}
-\frac{\langle {\psi}^t_{g}\lt|\hat{L}_e^a\rt| {\psi}^t_{g}\rangle\langle  {\psi}^t_{g}\lt| \hat{\mathbf{H}}\rt| {\psi}^t_{g}\rangle}{\langle {\psi}^t_{g}| {\psi}^t_{g}\rangle^2}.
\ee
These results coincide with Eqs.\Ref{HL} and \Ref{LH} when taking the limit $g_i\to g_{i+1}\equiv g$ and $g_{i-1}\to g_{i}\equiv g$.

$\Box$

The non-polynomial nature of Hamiltonian operator $\hat{\bf H}$ in LQG makes its matrix elements difficult to compute. Fortunately by Lemma \ref{expectation}, the EOMs with continuous time $\t$ only involve the expectation value of $\hat{\bf H}$. We require that $\hat{\bf H}$ must have the correct semiclassical limit, in the sense that $\langle \tilde{\psi}^t_{g^\eps}\lt| \hat{\mathbf{H}}\rt|\tilde{\psi}^t_{g^\eps}\rangle$ must recover the classical discrete Hamiltonian ${\bf H}$ as a function on the LQG phase space in the semiclassical limit:
\be
\langle \tilde{\psi}^t_{g^\eps}\lt| \hat{\mathbf{H}}\rt|\tilde{\psi}^t_{g^\eps}\rangle={\bf H}\lt[g^\eps\rt]+O(\hbar)
\ee 
Thiemann's (non-graph-changing) Hamiltonian has been shown to satisfy our requirement \cite{Giesel:2007wn,Giesel:2006um}. Moreover, classical EOMs derived from $\hbar\to0$ and stationary phase approximation are not sensible to $O(\hbar)$ corrections. Therefore all expectation values $\langle \tilde{\psi}^t_{g^\eps}\lt| \hat{\mathbf{H}}\rt|\tilde{\psi}^t_{g^\eps}\rangle$ can be replaced by the classical Hamiltonian ${\bf H}[g^\eps]$ in EOMs.

By the above discussion, EOMs derived from the variational principle $\delta S=0$ reduce to the following form:

\begin{itemize}

\item For $i=1,\cdots,N$, at every edge $e\in E(\g)$,
\be
\frac{z_{i+1,i}(e)\,\tr\lt[\t^a g_{i+1}(e)^\dagger g_i(e)\rt]}{\sqrt{x_{i+1,i}(e)-1}\sqrt{x_{i+1,i}(e)+1}}-\frac{p_i(e)\,\tr\lt[\t^a g_{i}(e)^\dagger g_i(e)\rt]}{\sinh(p_i(e))}
=\frac{i\kappa \Delta\t}{a^2}\frac{\partial\,{\mathbf{H}\lt[g_i^\eps\rt]}}{\partial{\eps^a(e)}}\Bigg|_{\vec{\eps}=0}\label{eoms1}
\ee

\item For $i=2,\cdots,N+1$, at every edge $e\in E(\g)$,
\be
\frac{z_{i,i-1}(e)\,\tr\lt[\t^a g_{i}(e)^\dagger g_{i-1}(e)\rt]}{\sqrt{x_{i,i-1}(e)-1}\sqrt{x_{i,i-1}(e)+1}}-\frac{p_i(e)\,\tr\lt[\t^a g_{i}(e)^\dagger g_i(e)\rt]}{\sinh(p_i(e))}
=-\frac{i\kappa \Delta\t}{a^2}\frac{\partial\,{\mathbf{H}\lt[g_i^\eps\rt]}}{\partial{\bar{\eps}^a(e)}}\Bigg|_{\vec{\eps}=0}.\label{eoms2}
\ee

\item The closure condition at every vertex $v\in V(\g)$ for initial data:
\be
-\sum_{e, s(e)=v}p_1^a(e)+\sum_{e, t(e)=v}\L^a_{\ b}\lt(\vec{\theta}_1(e)\rt)\,p_1^b(e)=0.\label{closure0}
\ee

\end{itemize}
\noindent
In the above, $z_{i+1,i}(e)$ and $x_{i+1,i}(e)$ are given by 
\be
z_{i+1,i}(e)&=& \mathrm{arccosh}\lt(x_{i+1,i}(e)\rt),\quad x_{i+1,i}(e)=\half\tr\lt[g_{i+1}(e)^\dagger g_{i}(e)\rt]
\ee
The initial and final conditions are given by $g_{1}=g'{}^h$ and $g_{N+1}=g$. Eqs.\Ref{eoms1}, \Ref{eoms2}, and \Ref{closure0} are referred to as the effective equations since they are derived from the effective action $S[g,h]$. These effective equations govern the semiclassical dynamics of full LQG.

\section{Application to cosmology}\label{HICOS}

\subsection{Homogeneous and isotropic ansatz}

We apply the effective equations from full LQG to cosmology and extract solutions that are homogeneous and isotropic on every spatial slice. We orient edges in the cubic lattice $\g$ such that every vertex $v$ connects to 3 outgoing edges $e_{I}(v)$ (oriented toward the $I=1,2,3$ spatial directions) and 3 incoming edges $e_{I}(v-\hat{I})$. We apply the following homogeneous and isotropic ansatz to the effective equations: At every edge $e_I(v)$, $g_i(e_I(v))$ assumes to be independent of $v$ and have the following expression 
\be
g_i(e_I(v))\equiv g_i(I)=e^{\lt(\theta_i-{i}p_i\rt)\frac{\t^I}{2}}=n_I\,e^{\lt(\theta_i-{i} p_i\rt)\frac{\t^3}{2}}\,n_I^\dagger,
\label{nen}
\ee 
where $n_{I=1,2}\in\Su$ rotates $(0,0,1)$ to $(1,0,0)$ or $(0,1,0)$ while $n_{I=3}$ is the identity. $i=0,\cdots,N+2$ labels time steps as above. $\theta_i$ and $p_i$ are constants on every spatial slice, and relates to holonomies and fluxes in Eq.\Ref{hpvari} by  
\be
h_i(e_I(v))\equiv h_i(I)=e^{\theta_i\t_I/2}
\quad p^a_i(e_I(v))\equiv p_i^a(I)=p_i\,\delta^a_I
\ee
in the context of homogeneous and isotropic cosmology. $p_i$ is always positive. Changing orientation of $e_I\to e_I^{-1}$ flips sign of $p^a_i(e_I)$, i.e. $p^a_i(e_I^{-1})=-p^a_i(e_I)$. 

We may introduce the conventional LQC variables $C_i,P_i$ (often denoted by $c,p$ in LQC literature). We take into account two possible schemes for the relation between $(\theta_i,p_i)$ and $(C_i,P_i)$ 
\be
\text{Scheme I}:&& \theta_i=C_i\mu,\quad p_i=\frac{\mu^2}{a^2\b}P_i,\label{ansatzI}\\
\text{Scheme II}:&& \theta_i=C_i\mu,\quad p_i=\frac{P_i}{a^2\b}\frac{\sin^2(\mu C_i/2)}{ C_i^2/4}.\label{ansatzII}
\ee
The scheme II insert $A^a_j=C\delta^a_j$ and $E^j_a=P\delta^j_a/{2}$ in the expression of gauge covariant flux in Eq.\Ref{hpvari}, in which we choose $h(\rho_e(x))$ from $S_{e_I(v)}\cap e_I(v)$ to $x=(x^{I+1},x^{I+2})\in S_{e_I(v)}$ with $x^{I+1},x^{I+2}\in[-\mu/2,\mu/2]$ (the coordinate index of $x$ is evaluated mod 3) to be $h(\rho_e(x))=e^{C x^{I+1}\t^{I+1}/2}e^{C x^{I+2}\t^{I+2}/2}$ \cite{Liegener:2019zgw}, while the scheme I set $h(\rho_e(x))=1$ \cite{Dapor:2017rwv}. The constant $\mu>0$ is the coordinate length of edges in $\g$ and $\g^*$, and endows cubic lattices $\g, \g^*$ a flat fiducial geometry. 

$|\theta_i-\theta_{i-1}|\sim |p_i-p_{i-1}|\sim\Delta\t$ are infinitesimal by the argument below Lemma \ref{isolatezero} in the last section. In the continuum limit $\Delta\t\to0$, we may replace
\be
\theta_i\to\theta(\t),\quad p_i\to p(\t),\quad \text{or}\quad C_i\to C(\t),\quad P_i\to P(\t).
\ee
where $\t$ is the continuous time parameter.

We set the boundary data $g$ consistently:
\be
g(e_I(v))\equiv g(I)=e^{\lt(\theta-{i}p\rt)\frac{\t^I}{2}}=n_I\,e^{\lt(\theta-{i} p\rt)\frac{\t^3}{2}}\,n_I^\dagger.\label{finaldata}
\ee
The final condition in Eq.\Ref{bc} implies
\be
\theta_{N+1}=\theta, \quad p_{N+1}=p. 
\ee
Moreover, in order to be consistent with the 2nd boundary condition, we require the boundary data $g'$ can be written as Eq.\Ref{nen} up to a gauge transformation by $h$:
\be
g'{}^h(e_I(v))\equiv g'{}^h(I)=e^{\lt(\theta'-{i}p'\rt)\frac{\t^I}{2}}=n_I\,e^{\lt(\theta'-{i} p'\rt)\frac{\t^3}{2}}\,n_I^\dagger, 
\ee
then the initial condition in Eq.\Ref{bc} implies 
\be
\theta_{1}=\theta', \quad p_{1}=p'. 
\ee

Because the boundary data $g$ in Eq.\Ref{finaldata} respects the symmetry of the cubic graph $\g$, we have $U_{\varphi_2}\psi_g^t=\psi_g^t$ thus $U_{\varphi_2}\Psi_{[g]}^t=\Psi_{[g]}^t$ (Diffeomorphisms $U_\varphi$ commute with SU(2) gauge transformations) for all $\varphi_2\in GS_\g$ so the diffeomorphism average in Eq.\Ref{Aggdiff} is trivial: $A_{Diff;[g],[g']}=A_{[g],[g']}$.

By the ansatz, Eq.\Ref{closure0} reduces to the closure condition of a geometrical squared cube with all 6 areas equal to $p_1$:
\be
-\sum^3_{I=1, s(I)=v}p_1 \delta^a_I+\sum^3_{I=1, t(I)=v}p_1\delta^a_I=0,\label{closure1}
\ee 
which is automatically satisfied by the homogeneous and isotropic ansatz.


The ansatz gives the following simplifications
\be
z_{i+1,i}(e_I(v))&=&\mathrm{arccosh}\lt[\half\tr\lt(e^{-\lt[\lt(\theta_{i+1}-\theta_i\rt)+i(p_{i+1}+p_i)\rt]\frac{\t^3}{2}}\rt)\rt]=-i\frac{\theta_{i+1}-\theta_i}{2}+\frac{p_{i+1}+p_i}{2},\label{arccoshcosh} \\
\tr\lt[\t^a g_{i+1}(e_I(v))^\dagger g_i(e_I(v))\rt]&=&\tr\lt[\t^a e^{-\lt[\lt(\theta_{i+1}-\theta_i\rt)+i(p_{i+1}+p_i)\rt]\frac{\t_I}{2}}\rt]=2i\,\delta^a_I \sinh\lt[-i\frac{\theta_{i+1}-\theta_i}{2}+\frac{p_{i+1}+p_i}{2}\rt],
\ee
Infinitesimal $\theta_{i+1}-\theta_i$ and $p_i>0$ implies Eq.\Ref{arccoshcosh} and $[\cosh(z_{i+1,i})-1]^{1/2}[\cosh(z_{i+1,i})+1]^{1/2}=\sinh(z_{i+1,i})$. Effective equations \Ref{eoms1} and \Ref{eoms2} at the edge $e_I(v)$ reduces to
\be
\delta^a_I\lt[\frac{\theta_{i+1}-\theta_i}{\Delta\t}+i\,\frac{p_{i+1}-p_i}{\Delta\t}\rt]
&=&\frac{i\kappa }{a^2}\frac{\partial\,{\mathbf{H}[g_i^\eps]}}{\partial{\eps_i^a(e_I(v))}}\Bigg|_{\vec{\eps}=0}, \label{eomc1}\\
\delta^a_I\lt[\frac{\theta_{i}-\theta_{i-1}}{\Delta\t}-i\,\frac{p_{i}-p_{i-1}}{\Delta\t}\rt]
&=&-\frac{i\kappa}{a^2}\frac{\partial\,{\mathbf{H}[g_i^\eps]}}{\partial{\bar{\eps}_i^a(e_I(v))}}\Bigg|_{\vec{\eps}=0}.\label{eomc2} 
\ee
Here $g^\eps_i$ includes perturbations away from the homogeneous and isotropic ansatz:
\be
g^\eps_i(e_I(v))=e^{\lt(\theta_i-{i}p_i\rt)\frac{\t^I}{2}}e^{\eps_i^a(e_I(v))\t^a}.\label{perturb}
\ee
Notice that the right hand side of Eq.\Ref{eomc1}-\Ref{eomc2} are also independent of vertex $v$ as we shall see later in Section \ref{CEE}.

Eqs.\Ref{eomc1} and \Ref{eomc2} contain diagonals $a=I$ and off-diagonals $a\neq I$:
\begin{itemize}

\item The diagonals $a=I$ give time-evolution equations: in the approximation by the continuum limit,
\be
\lt[\frac{\rmd\theta}{\rmd\t}+i\,\frac{\rmd p}{\rmd\t}\rt]
=\frac{i\kappa }{a^2}\frac{\partial\,{\mathbf{H}[g^\eps]}}{\partial{\eps^I(e_I(v))}}\Bigg|_{\vec{\eps}=0}, \quad
\lt[\frac{\rmd\theta}{\rmd\t}-i\,\frac{\rmd p}{\rmd\t}\rt]
=-\frac{i\kappa }{a^2}\frac{\partial\,{\mathbf{H}[g^\eps]}}{\partial{\bar{\eps}^I(e_I(v))}}\Bigg|_{\vec{\eps}=0}.\label{effeqc2}
\ee 
where right-hand sides are derivatives of ``longitudinal perturbations'' $\eps^I(e_I(v))$, $\bar{\eps}^I(e_I(v))$.

\item The off-diagonals $a\neq I$ give constraint equations at every instant $\t$:
\be
\frac{\partial\,{\mathbf{H}[g^\eps]}}{\partial{\eps^a(e_I(v))}}\Bigg|_{\vec{\eps}=0}=0, \quad
\frac{\partial\,{\mathbf{H}[g^\eps]}}{\partial{\bar{\eps}^a(e_I(v))}}\Bigg|_{\vec{\eps}=0}=0,\quad a\neq I,\label{effeqc3}
\ee
where right-hand sides are derivatives of ``transverse perturbations'' $\eps^a(e_I(v))$, $\bar{\eps}^a(e_I(v))$ with $a\neq I$.

\end{itemize}

\subsection{Perturbations}

Effective equations \Ref{effeqc2} and \Ref{effeqc3} involve perturbations from the homogeneous and isotropic ansatz as in Eq.\Ref{perturb}. Since (as it becomes clear in Section \ref{CEE}) $\mathbf{H}[g]$ is conveniently expressed in terms of holonomy-flux variables $h(e),p^a(e)$, in computing $\mathbf{H}[g^\eps]$ and derivatives, we need to translate perturbations in holomorphic variables $g(e)$ as Eq.\Ref{perturb} to perturbations of $h(e),p^a(e)$.

\begin{Lemma}\label{lemmaPerturb}

Eq.\Ref{perturb} can be rewritten in the polar-decomposition form as Eq.\Ref{gthetap} where we extract perturbations of $p^a,\theta^a$:
\be
&&g^\eps(e_I(v))=e^{\lt(\theta-{i}p\rt)\frac{\t^I}{2}}e^{\eps^a(e_I(v))\t^a}=e^{-ip^a(e_I(v))\t^a/2}e^{\theta^a(e_I(v))\t^a/2},\label{perturbeqn}
\ee
where $p^a,\theta^a$ contains longitudinal perturbations $\delta p_\parallel,\delta \theta_\parallel$ and transverse perturbations $\delta p_\bot^a,\delta \theta_\bot^a$ with $a=I+1,I+2$ mod $3$ ,
 \be
 p^a(e_I(v))=\lt[p+\delta p_\parallel(e_I(v))\rt]\delta^a_I+\delta p_\bot^a(e_I(v)),\quad 
 \theta^a(e_I(v))=\lt[\theta+\delta \theta_\parallel(e_I(v))\rt]\delta^a_I+\delta \theta_\bot^a(e_I(v)).\label{deltaptheta}
\ee
$\delta p_\parallel,\delta \theta_\parallel$ and $\delta p_\bot^a,\delta \theta_\bot^a$ relates to $\eps^a$ up to $O(\eps^2)$ by 
\be
\delta p_\parallel(e_I(v))&=&i\lt[{\eps^I(e_I(v))-\bar{\eps}^I(e_I(v))}\rt],\quad 
\delta \theta_\parallel(e_I(v))={\eps^I(e_I(v))+\bar{\eps}^I(e_I(v))},\\
\left(\begin{array}{c}\delta p^{I+1}_\bot(e_I(v))\\ \delta p^{I+2}_\bot(e_I(v))\end{array}\right)&=&\frac{i p}{\sinh(p)}\left(\begin{array}{cc}\cos(\theta) & -\sin(\theta)  \\ \sin(\theta) & \cos(\theta) \end{array}\right) \left(\begin{array}{c}{\eps^{I+1}(e_I(v))-\bar{\eps}^{I+1}(e_I(v))}\\ {\eps^{I+2}(e_I(v))-\bar{\eps}^{I+2}(e_I(v))}\end{array}\right),\\
\left(\begin{array}{c}\delta \theta^{I+1}_\bot(e_I(v))\\ \delta \theta^{I+2}_\bot(e_I(v))\end{array}\right)&=&\frac{\theta/2}{\sin(\theta/2)}\Bigg[\left(\begin{array}{cc}\cos(\theta/2) & -\sin(\theta/2)  \\ \sin(\theta/2) & \cos(\theta/2) \end{array}\right) \left(\begin{array}{c}{\eps^{I+1}(e_I(v))+\bar{\eps}^{I+1}(e_I(v))}\\ {\eps^{I+2}(e_I(v))+\bar{\eps}^{I+2}(e_I(v))}\end{array}\right)\nonumber\\
&&+i\tanh(p/2)\left(\begin{array}{cc}\sin(\theta/2) & \cos(\theta/2)  \\ -\cos(\theta/2) & \sin(\theta/2) \end{array}\right)\left(\begin{array}{c}{\eps^{I+1}(e_I(v))-\bar{\eps}^{I+1}(e_I(v))}\\ {\eps^{I+2}(e_I(v))-\bar{\eps}^{I+2}(e_I(v))}\end{array}\right)\Bigg].\label{solvedeltaptheta}
\ee
The above linear transformation between $\eps^a$ and $\delta p,\delta\theta$ is non-degenerate.

\end{Lemma}

\textbf{Proof:} The above result may be obtained straight-forwardly by the polar decomposition of $e^{\lt(\theta-{i}p\rt)\frac{\t^I}{2}}e^{\eps^a(e_I(v))\t^a}\simeq e^{\lt(\theta-{i}p\rt)\frac{\t^I}{2}}[1+\eps^a(e_I(v))\t^a]$. However it can also be derived by inserting the ansatz Eq.\Ref{deltaptheta} into Eq.\Ref{perturbeqn} and solving $\delta p_{\parallel,\bot},\delta\theta_{\parallel,\bot}$ in terms of $\eps^a$. It amounts to solve the following equations by linearizing the right-hand side in $\delta p,\delta\theta$ (we ignore $O(\eps^2)$):
\be
\eps^a(e_I(v))=-\half\tr\lt[\t^a e^{-\lt(\theta-{i}p\rt)\frac{\t^I}{2}}e^{-ip^a(e_I(v))\t^a/2}e^{\theta^a(e_I(v))\t^a/2}\rt],\quad a=I,I+1,I+2\ \text{mod}\ 3
\ee
which give Eqs.\Ref{solvedeltaptheta} as the solution.

$\Box$

Applying the above result, the time-evolution equations \Ref{effeqc2} are equivalent to  
\be
\frac{\rmd\theta}{\rmd\t}=-\frac{\kappa }{a^2}\frac{\partial\,{\mathbf{H}[g^\eps]}}{\partial{\delta p_\parallel(e_I(v))}}\Bigg|_{\delta\theta=\delta p=0}, \quad
\frac{\rmd p}{\rmd\t}=\frac{\kappa }{a^2}\frac{\partial\,{\mathbf{H}[g^\eps]}}{\partial{\delta \theta_\parallel(e_I(v))}}\Bigg|_{\delta\theta=\delta p=0},\label{c1}
\ee
while the constraint equations \Ref{effeqc2} are equivalent to
\be
\frac{\partial\,{\mathbf{H}[g^\eps]}}{\partial{\delta p_\bot^a(e_I(v))}}\Bigg|_{\delta\theta=\delta p=0}=0, \quad
\frac{\partial\,{\mathbf{H}[g^\eps]}}{\partial{\delta \theta_\bot^a(e_I(v))}}\Bigg|_{\delta\theta=\delta p=0}=0,\quad a\neq I,\label{c2}
\ee

\section{Cosmology effective dynamics}\label{CEE}

In this section we apply specific definitions of $\hat{C}_v$ and $\hat{C}_{j,v}$ to the definition of $\hat{\bf H}$ in Eq.\Ref{physHam}, and insert the concrete expression of ${\bf H}[g]$ into the effective equations of cosmology \Ref{c1} and \Ref{c2}. 

\subsection{Euclidean Hamiltonian}

We firstly consider the simplest situation with Euclidean Hamiltonian, which corresponds to $\b=1$ and the Euclidean signature $s=1$, where the Hamiltonian constraint simplifies $C=-\frac{2}{\kappa\sqrt{\det(q)}}\tr\lt(F_{jk}\lt[E^j,E^k\rt]\rt)\equiv C_0$. The Euclidean Hamiltonian constraint $C_0$ and the diffeomorphism constraint $C_j$ can be quantization by \cite{Giesel:2007wn}:
\be
\hat{C}_{\mu,v}:=-\frac{4}{3i\b\kappa\ell_p^2/2}\sum_{s_1,s_2,s_3=\pm1}s_1s_2s_3\ \eps^{I_1I_2I_3}\ \mathrm{Tr}\Bigg(\t_\mu\ \hat{h}(\a_{v;I_1s_1,I_2s_2})\ \hat{h}(e_{v;I_3s_3})\Big[\hat{h}(e_{v;I_3s_3})^{-1},\hat{V}_v\Big]\ \Bigg)\label{C}
\ee
where $\mu=0,j$ and $\t_0$ is the identity matrix. We still keep $\beta$ in the derivation since it will be useful later for the derivation of Lorentzian Hamiltonians. In the above notation, $e_{v;Is}$ is an edge beginning at the vertex $v$ in positive $(s=1)$ or negative $(s=-1)$ $I$-th direction ($I=1,2,3$): $e_{v;I+}=e_I(v)$ and $e_{v;I-}=e_I(v-\hat{I})^{-1}$. $\a_{v;Is,Js'}$ is the minimal square loop along edges in $\g$ associated with $v$, $e_{v;Is}$ and $e_{v;Js'}$.  $\hat{V}_v$ is the volume operator at $v$:
\be
\hat{V}_v=\lt(\hat{Q}_v^2\rt)^{1/4},\quad \hat{Q}_v
=-i\lt(\frac{\b\ell_P^2}{4}\rt)^3\eps_{abc}\frac{R^a_{e_{v;1+}}-R^a_{e_{v;1-}}}{2}\frac{R^b_{e_{v;2+}}-R^a_{e_{v;2-}}}{2}\frac{R^c_{e_{v;3+}}-R^c_{e_{v;3-}}}{2}
\ee 
where $R_e^a$ are right invariant vector fields on SU(2).

For gravity-dust models, the Euclidean Hamiltonian operators are given by
\be
\hat{\bf H}=\sum_{v\in V(\g)}\lt[\hat{M}_-^\dagger(v) \hat{M}_-(v)\rt]^{1/4}\quad \hat{M}_-(v)=\hat{C}_{0,v}^{\ \dagger}\hat{C}_{0,v}-\frac{\a}{4}\hat{C}_{j,v}^{\ \dagger}\hat{C}_{j,v},\quad \a=\begin{cases}
1,&\text{Brown-Kucha\v{r} dust,}\\
0,&\text{Gaussian dust.}\label{physHamEu}
\end{cases}
\ee
Given that $\hat{M}_-^\dagger(v) \hat{M}_-(v)$ is self-adjoint, computing coherent state expectation values of $\hat{\bf H}$ reduce to computing expectation values of $\hat{M}_-^\dagger(v) \hat{M}_-(v)$ at the leading order in $t=\ell_P^2/a^2$ \cite{Thiemann:2000bx,Giesel:2006um}:
\be
\langle\tilde{\psi}^t_g\,|\,\hat{\bf H}\,|\,\tilde{\psi}^t_g\rangle=\sum_{v\in V(\g)}\langle\tilde{\psi}^t_g\,|\hat{M}_-^\dagger(v) \hat{M}_-(v)|\,\tilde{\psi}^t_g\rangle^{1/4}+O(t).
\ee
Expectation values $\langle\tilde{\psi}^t_g\,|\hat{M}_-^\dagger(v) \hat{M}_-(v)|\,\tilde{\psi}^t_g\rangle$ can be computed by the semiclassical perturbation expansion in $t$, in which the volume operator $\hat{V}_v$ can be replaced by the truncated power expansion at certain $2k+1$ (the Giesel-Thiemann volume operator):
\be
\hat{V}_v=\langle \hat{Q}_v\rangle^{2q}\lt[1+\sum_{n=1}^{2k+1}(-1)^{n+1}\frac{q(1-q)\cdots(n-1+q)}{n!}\lt(\frac{\hat{Q}_v^2}{\langle\hat{Q}_v\rangle^2}-1\rt)^n\rt],\quad q=1/4\label{expandvolume}
\ee
where $\langle \hat{Q}_v\rangle=\langle\psi^t_g|\hat{Q}_v|\psi^t_g\rangle$. The expansion allows to replace the non-polynomial operator $\hat{V}_v$ by the polynomial operator, where the error is bounded by $O(t^{k+1})$. By applying the semiclassical property of the coherent state in Eq.\Ref{hpbound}, one can show that $\langle\tilde{\psi}^t_g\,|\hat{M}_-^\dagger(v) \hat{M}_-(v)|\,\tilde{\psi}^t_g\rangle$ reproduces the (discrete) classical expression at the leading order in $t$:
\be
\langle\tilde{\psi}^t_g\,|\hat{M}_-^\dagger(v) \hat{M}_-(v)|\,\tilde{\psi}^t_g\rangle=\lt({C}_{0,v}^{2}-\frac{\a}{4}{C}_{j,v}^{2}\rt)^2[g]+O(t)
\ee
where ${C}_{\mu,v}[g]$ with $g(e)=e^{-ip^a(e)\t^a/2}h(e)$ is given by replacing all $\hat{h}(e)\to{h}(e)$ and $it \hat{R}^a_e/2\to p^a(e)$ and replacing commutators to classical Poisson bracket: $[\cdot,\cdot]\to \frac{1}{i\hbar}\{\cdot,\cdot\}$.

We plug in the homogeneous and isotropic ansatz of $g$ in Eq.\Ref{nen}, and obtain \cite{Dapor:2017rwv,Dapor:2017gdk}
\be
{C}_{0,v}[g]=\frac{4a}{\kappa}\sqrt{2\b p}\sin^2(\theta) 
,\quad {C}_{j,v}[g]=0,\quad V_v[g]=\frac{a^3 (\beta  p)^{3/2}}{2 \sqrt{2}}.
\ee 
Therefore,
\be
&&\langle\tilde{\psi}^t_g\,|\,\hat{\bf H}\,|\,\tilde{\psi}^t_g\rangle={\bf H}[\theta,p]+O(t),\quad {\bf H}[\theta,p]
=\frac{4a\,|V(\g)|}{\kappa}\sqrt{2\b p}\sin^2(\theta)\quad \text{for dust models},\label{EuHdust}
\ee
where $|V(\g)|$ is the total number of vertices in $\g$. Brown-Kucha\v{r} dust and Gaussian dust models result in the same ${\bf H}[g]$ since ${C}_{j,v}[g]=0$.

$\langle\tilde{\psi}^t_g\,|\,\hat{\bf H}\,|\,\tilde{\psi}^t_g\rangle$ in the gravity-scalar model can be computed similarly, by the replacement $\hat{M}_-\to \hat{C}_v \hat{V}_v$:
\be
&&\langle\tilde{\psi}^t_g\,|\,\hat{\bf H}\,|\,\tilde{\psi}^t_g\rangle={\bf H}[\theta,p]+O(t),\quad{\bf H}[\theta,p]=\frac{\sqrt{2}\,a^2  |V(\g)| }{\sqrt{\kappa}}\beta p\sqrt{ \sin^2(\theta )} \quad \text{for gravity-scalar}.\label{EuHscalar}
\ee

We would like to emphasize that, unlike the symmetry reduced models, these cosmological effective hamiltonians are obtained by imposing the cosmological ansarz after performing the Poisson brackets.

\begin{Theorem}\label{Euclideanresult}

For the Euclidean Hamiltonian of either Brown-Kucha\v{r} or Gaussian dust models, the cosmological evolution equations \Ref{c1} are equivalent to
\be
\frac{\rmd\theta}{\rmd\t}=-\frac{\partial}{\partial p}\lt[\frac{4 }{3a } \sqrt{2\beta p } \sin^2 (\theta )\rt], \quad
\frac{\rmd p}{\rmd\t}=
\frac{\partial}{\partial\theta}\lt[\frac{4 }{3 a } \sqrt{2\beta p } \sin^2 (\theta )\rt],\label{pthevodust11}
\ee
while the constraint equations \Ref{c2} are satisfied automatically.

For the Euclidean Hamiltonian of gravity-scalar model, the cosmological evolution equations \Ref{c1} are equivalent to
\be
\frac{\rmd\theta}{\rmd\t}=-\frac{\partial}{\partial p}\lt[\frac{\sqrt{2\kappa}\,  \beta p }{3  }\sqrt{ \sin^2(\theta )} \rt]
, \quad
\frac{\rmd p}{\rmd\t}=\frac{\partial}{\partial\theta}\lt[\frac{\sqrt{2\kappa}\,  \beta p }{3  }\sqrt{ \sin^2(\theta )} \rt],\label{pthevoscalar11}
\ee
while the constraint equations \Ref{c2} are again satisfied automatically. 

\end{Theorem}

\textbf{Proof:} At every vertex $v=(x,y,z)\in\Z^3$, we expand ${C}_{0,v}[g^\eps]$ in $\delta p,\delta\theta$ as in Lemma \ref{lemmaPerturb} (the computation uses Mathematica):
\be
&&\frac{4 a}{\kappa } \sqrt{2\beta p } \sin ^2(\theta )+C^{1}_{0,v}[\theta, p, \delta \theta, \delta p]+O(\delta^2)\label{expansionC}
\ee
where $C^{1}_{0,v}[\theta, p, \delta \theta, \delta p]$ contains linear terms of $\delta \theta$ and $\delta p$, whose explicit form can be found at \cite{Liugit2019}.
We neglect the quadratic order in $\delta\theta,\delta p$ because Eqs.\Ref{c1} and \Ref{c2} only involve 1st order derivatives. 

The diffeomorphism constraint ${C}_{j,v}[g]=0$ and ${C}_{j,v}[g^\eps]=O(\delta\theta,\delta p)$, therefore as far as ${C}_{0,v}[g]=\frac{4 a}{\kappa } \sqrt{2\beta p } \sin ^2(\theta )\neq 0$, 
\be
\text{Brown-Kucha\v{r}/Gaussian dust:}&&{\bf H}[g^\eps]=\sum_{v\in V(\g)}\lt|{C}_{0,v}[g^\eps]\rt|+O(\delta^2),\\
\text{Gravity-scalar:}&& {\bf H}[g^\eps]=\sum_{v\in V(\g)}\sqrt{\lt|V_v[g^\eps]{C}_{0,v}[g^\eps]\rt|}
\ee
The expansion of volume is given by
\be
V_v[g^\eps]&=&\frac{a^3 (\beta  p)^{3/2}}{2 \sqrt{2}}+\frac{a^3 \beta ^{3/2} \sqrt{p} }{8 \sqrt{2}}
   \Big[\delta p^1_\parallel(x-1,y,z)+\delta p^1_\parallel(x,y,z)+\delta p^2_\parallel(x,y-1,z)\nonumber\\
   &&+\delta p^2_\parallel(x,y,z)+\delta p^3_\parallel(x,y,z-1)+\delta p^3_\parallel(x,y,z)\Big]+O(\delta^2)
\ee
where the linear order only involve $\delta p^I_\parallel$.

When we sum over all cubic lattice vertices $v=(x,y,z)$ and impose periodic boundary conditions $x+L\sim x$, $y+L\sim y$, and $z+L\sim z$ (we have assumed $\g$ to be a lattice in $T^3$), all terms in Eq.\Ref{expansionC} linear to $\delta\theta^{a}_\bot$ cancels between ${C}_{0,v}[g^\eps]$ at different $v$'s. ${C}_{0,v}[g^\eps]$ in Eq.\Ref{expansionC} has no term linear to $\delta p^{a}_\bot$. Therefore the constraint equations \Ref{c2} are satisfied automatically in both gravity-dust and gravity-scalar models. 

The hamiltonian density at each vertex enrolls only finite terms of fluxes and holonomies at this vertex and neighboring vertices. In other words, a perturbation at certain vertex only appears in finite terms of hamiltonian densities at different vertices. The cancellation happens exactly among these finite hamiltonian densities (when there is no boundary) when one derives the EOM for certain phase space variable at a given vertex. So the result is the same in the case of an infinite space.

Moreover we check that for the Brown-Kucha\v{r}/Gaussian dust,
\be
\frac{\partial\,{\bf H}[g^\eps]}{\partial\delta\theta^I_\parallel(x,y,z)}\Bigg|_{\delta\theta=\delta p=0}&=&\frac{4 a}{3\kappa } \sqrt{2\beta p } \sin (2 \theta )=\frac{\partial}{\partial\theta}\lt[\frac{4 a}{3\kappa } \sqrt{2\beta p } \sin^2 (\theta )\rt],\nonumber\\
\frac{\partial\,{\bf H}[g^\eps]}{\partial\delta p^I_\parallel(x,y,z)}\Bigg|_{\delta\theta=\delta p=0}&=&\frac{2a}{3\kappa } \sqrt{\frac{2\beta}{p} } \sin^2 (\theta )=\frac{\partial}{\partial p}\lt[\frac{4 a}{3\kappa } \sqrt{2\beta p } \sin^2 (\theta )\rt].\label{derivativeHthetap}
\ee
are both independent of $(x,y,z)$ and $I$. Eqs.\Ref{c1} imply the following evolution equations
\be
\frac{\rmd\theta}{\rmd\t}=-\frac{\partial}{\partial p}\lt[\frac{4 }{3a } \sqrt{2\beta p } \sin^2 (\theta )\rt], \quad
\frac{\rmd p}{\rmd\t}=
\frac{\partial}{\partial\theta}\lt[\frac{4 }{3 a } \sqrt{2\beta p } \sin^2 (\theta )\rt].
\ee

On the other hand, for the gravity-scalar model,
\be
\frac{\partial\,{\bf H}[g^\eps]}{\partial\delta\theta^I_\parallel(x,y,z)}\Bigg|_{\delta\theta=\delta p=0}&=&\frac{\sqrt{2} a^2 \beta  p }{3 \sqrt{\kappa}} \cot (\theta ) \sqrt{ \sin^2(\theta )}=\frac{\partial}{\partial\theta}\lt[\frac{\sqrt{2} a^2 \beta p }{3 \sqrt{\kappa }}\sqrt{ \sin^2(\theta )} \rt]\nonumber\\
\frac{\partial\,{\bf H}[g^\eps]}{\partial\delta p^I_\parallel(x,y,z)}\Bigg|_{\delta\theta=\delta p=0}&=&\frac{\sqrt{2} a^2 \beta }{3 \sqrt{\kappa }} \sqrt{ \sin^2(\theta )}=\frac{\partial}{\partial p}\lt[\frac{\sqrt{2} a^2 \beta p }{3 \sqrt{\kappa }}\sqrt{ \sin^2(\theta )} \rt]
\ee
are both independent of $(x,y,z)$ and $I$. Eqs.\Ref{c1} imply the following evolution equations
\be
\frac{\rmd\theta}{\rmd\t}=-\frac{\partial}{\partial p}\lt[\frac{\sqrt{2\kappa}\,  \beta p }{3  }\sqrt{ \sin^2(\theta )} \rt]
, \quad
\frac{\rmd p}{\rmd\t}=\frac{\partial}{\partial\theta}\lt[\frac{\sqrt{2\kappa}\,  \beta p }{3  }\sqrt{ \sin^2(\theta )} \rt].
\ee

$\Box$

A careful reader may have noticed that Eqs.\Ref{pthevodust11} and \Ref{pthevoscalar11} closely relate to the classical Hamiltonian evolution given by cosmological Hamiltonian $\mathbf{H}[g]$ in Eqs.\Ref{EuHdust} and \Ref{EuHscalar}, namely Eqs.\Ref{pthevodust11} and \Ref{pthevoscalar11} is equivalent to 
\be
\frac{\rmd \theta}{\rmd \t}=-\frac{\kappa}{a^2}\frac{\partial}{\partial p}\lt[\frac{1}{3|V(\g)|}\mathbf{H}[\theta,p]\rt],\quad \frac{\rmd p}{\rmd\t}=\frac{\kappa}{a^2}\frac{\partial}{\partial\theta}\lt[\frac{1}{3|V(\g)|}\mathbf{H}[\theta,p]\rt].\label{pthevoHHH}
\ee
If we define a Poisson bracket $\{p,\theta\}=\frac{\kappa}{a^2}$ (this Poisson bracket is consistent with Eq.\Ref{ph}), the above equations can be written as Hamiltonian evolutions $\rmd f/d\t=\{f,\frac{1}{3}\mathbf{H}[\theta,p]/{|V(\g)|}\}$ where $f$ is $p$ or $\theta$, or any function of $p,\theta$. 

Eqs.\Ref{pthevodust11} and \Ref{pthevoscalar11} are derived from the full Hamiltonian and its variation ${\bf H}[g^\eps]$ (see Eq.\Ref{derivativeHthetap}), their relations to derivatives of cosmological Hamiltonian ${\bf H}[\theta,p]$ need some explanation: Firstly we notice that the full Hamiltonian ${\bf H}$ is invariant under the lattice translation $v\to v+\hat{I}$ and rotation $I\to I+1$ mod 3 (since the underlying smooth expression doesn't depend on choices of coordinates). Therefore when evaluating at the homogeneous and isotropic ansatz, 
\be
\frac{\partial\, {\bf H}[g^\eps]}{\partial\delta\theta_\parallel((e_I(v))}\Bigg|_{\delta\theta=\delta p=0}\quad \text{and}\quad \frac{\partial\,{\bf H}[g^\eps]}{\partial\delta p_\parallel(e_I(v))}\Bigg|_{\delta\theta=\delta p=0}
\ee
are independent of $I$ and $v$. On the other hand, derivatives of $\mathbf{H}[\theta,p]$ are global variations $\delta p_\parallel(e_I(v)),\delta \theta_\parallel(e_I(v))$ over all edges $e_I(v)$. Therefore
\be
\frac{\partial}{\partial p}\lt[\frac{1}{3|V(\g)|}\mathbf{H}[\theta,p]\rt]=\frac{1}{3|V(\g)|}\sum_{e\in E(\g)}\frac{\partial\,{\bf H}[g^\eps]}{\partial\delta p_\parallel(e)}\Bigg|_{\delta\theta=\delta p=0}=\frac{|E(\g)|}{3|V(\g)|}\frac{\partial\,{\bf H}[g^\eps]}{\partial\delta p_\parallel(e)}\Bigg|_{\delta\theta=\delta p=0}=\frac{\partial\,{\bf H}[g^\eps]}{\partial\delta p_\parallel(e)}\Bigg|_{\delta\theta=\delta p=0}\label{argument}
\ee
where $|E(\g)|=3|V(\g)|$ since every vertex in $\g$ is 6-valent. The relation for the derivative with respect to $\theta$ can be derived similarly. Inserting the above relation in Eq.\Ref{c1} reproduces the evolution equations \Ref{pthevoHHH}.

We may translate the time evolution equations to $(C,P)$ variables. For dust models, the schemes in Eqs.\Ref{ansatzI} and \Ref{ansatzII} imply
\be
\text{Scheme I}:&&\frac{\rmd P}{\rmd\t}=\frac{\partial}{\partial C}\mathbf{h}_{dust}(C,P),\quad \frac{\rmd C}{\rmd\t}=-\frac{\partial}{\partial P}\mathbf{h}_{dust}(C,P),\quad \mathbf{h}_{dust}(C,P)=\frac{4\b }{3 } \sqrt{2P }\, \frac{\sin^2 (C\mu )}{\mu^2},\label{evoI}\\
\text{Scheme II}:&&\frac{\rmd P}{\rmd \t}= \beta  C \,\cot \left(\frac{\mu  C}{2} \right) \lt(\frac{1}{3}  C \lt[5 \cos (\mu C)+1\rt]-\frac{2}{3} \frac{\sin (\mu  C)}{\mu}\rt)  \sqrt{\frac{P \lt[1-\cos (\mu  C)\rt]}{C^2}},\nonumber\\
&&\frac{\rmd C}{\rmd\t}= -\frac{\beta  \sin ^2(\mu C)}{3 \mu }\sqrt{\frac{C^2}{P\lt[1- \cos (\mu C)\rt]}},
\ee
where for the scheme I we can extract an ``effective cosmology Hamiltonian'' $\mathbf{h}_{dust}(C,P)$, which coincides with the standard effective Hamiltonian constraint in LQC (the $\mu_0$-scheme). Similarly for the gravity-scalar,
\be
\text{Scheme I}:&&\frac{\rmd P}{\rmd\t}=\frac{\partial}{\partial C}\mathbf{h}_{\phi}(C,P),\quad 
\frac{\rmd C}{\rmd\t}=-\frac{\partial}{\partial P}\mathbf{h}_{\phi}(C,P),\quad
\mathbf{h}_{\phi}(C,P)=\frac{ \sqrt{2\kappa}\,\beta  P}{3 \mu } \sqrt{\sin^2 ( \mu  C)}\\
\text{Scheme II}:&&\frac{\rmd P}{\rmd\t}=\frac{\sqrt{2\kappa} \beta  P}{3}\lt[2 \cot (\mu  C)+\csc (\mu  C)-\frac{2}{\mu  C}\rt]\sqrt{  \sin ^2(\mu  C)},\nonumber\\
&&\frac{\rmd C}{\rmd\t}=-\frac{ \sqrt{2\kappa}\,\beta  }{3 \mu } \sqrt{\sin^2 ( \mu  C)}.
\ee
where for the scheme I we can again extract an ``effective cosmology Hamiltonian'' $\mathbf{h}_{\phi}(C,P)$. However the scheme II cannot lead to Hamiltonian evolution equations in terms of $C,P$, because in contrast to the scheme I, $C,P$ are not anymore conjugate variables in the scheme II, by $\{p,\theta\}=\frac{\kappa}{a^2}$. This is different from the proposal in \cite{Liegener:2019zgw}. In our context the scheme II should be understood as change of variables and not affect the physics of effective dynamics. We come back to this point in Section \ref{BOUNCE}.

\subsection{Giesel-Thiemann's Lorentzian Hamiltonian}

In Lorentzian signature $s=-1$ and for an arbitrary $\b\in\R$, the Hamiltonian constraint $C$ in Eq.\Ref{Csigmatau} is quantized by Thiemann's method \cite{QSD} as the following
\be
\hat{C}_v&=&\hat{C}_{0,v}+\frac{1+\b^2}{2}\hat{C}_{L,v},\quad \hat{K}=\frac{i}{\hbar\b^2}\lt[\sum_{v\in V(\g)}\hat{C}_{0,v},\sum_{v\in V(\g)}V_v\rt]\nonumber\\
\hat{C}_{L,v}&=&\frac{16}{3\kappa\lt(i\b\ell_p^2/2\rt)^3}\sum_{s_1,s_2,s_3=\pm1}s_1s_2s_3\ \eps^{I_1I_2I_3}\times\nonumber\\
&&\times\quad \mathrm{Tr}\Bigg( \hat{h}(e_{v;I_1s_1})\Big[\hat{h}(e_{v;I_1s_1})^{-1},\hat{K}\Big]\ \hat{h}(e_{v;I_2s_2})\Big[\hat{h}(e_{v;I_2s_2})^{-1},\hat{K}\Big]\ \hat{h}(e_{v;I_3s_3})\Big[\hat{h}(e_{v;I_3s_3})^{-1},\hat{V}_v\Big]\ \Bigg).\label{HCO}
\ee

The cosmological coherent state expectation value $\langle\tilde{\psi}_g^t|\hat{\bf H}|\tilde{\psi}_g^t\rangle$ can be computed similarly as in the last subsection, by the replacement $\hat{C}_{0,v}\to\hat{C}_v$ and using the expansion Eq.\Ref{expandvolume} for volume operators 
\be
&&\langle\tilde{\psi}_g^t|\hat{\bf H}|\tilde{\psi}_g^t\rangle={\bf H}[\theta,p]+O(t),\nonumber\\
&&{\bf H}[\theta,p]=\begin{cases}
  -\frac{4a|V(\g)|}{\kappa}\sqrt{2\b p}\lt[\sin^2(\theta)-\frac{4}{9}\frac{1+\b^2}{\b^2}\sin^2(2\theta)\rt]
  ,& \text{for Brown-Kucha\v{r}/Gaussian dusts}\\
  \frac{\sqrt{2}\,a^2|V(\g)|}{\sqrt{\kappa}}{\b p}\sqrt{-\sin^2(\theta)+\frac{4}{9}\frac{1+\b^2}{\b^2}\sin^2(2\theta)}
  ,& \text{for gravity-scalar}.\label{thiemannHam}
\end{cases}
\ee

\begin{Theorem}\label{cosThiemann}

  For Giesel-Thiemann's's Lorentzian Hamiltonian of either Brown-Kucha\v{r}/Gaussian dust or gravity-scalar, the cosmological evolution equations \Ref{c1} are equivalent to
  \be
\frac{\rmd \theta}{\rmd \t}=-\frac{\kappa}{a^2}\frac{\partial}{\partial p}\lt[\frac{1}{3|V(\g)|}\mathbf{H}[\theta,p]\rt],\quad \frac{\rmd p}{\rmd\t}=\frac{\kappa}{a^2}\frac{\partial}{\partial\theta}\lt[\frac{1}{3|V(\g)|}\mathbf{H}[\theta,p]\rt],\label{evoThiemann}
  \ee
  where $\mathbf{H}[\theta,p]$ is given by Eq.\Ref{thiemannHam}. The constraint equations \Ref{c2} are satisfied automatically.
  
  \end{Theorem}

\textbf{Proof:} The constraint equations \Ref{c2} follows from a brute-force computation with Mathematica. We expand Giesel-Thiemann's Lorentzian Hamiltonian in terms of $\delta p_\parallel,\delta p_\bot$ and $\delta \theta_\parallel,\delta \theta_\bot$. This expansion is done in a CPU+GPU server, and uses the parallel computing environment of Mathematica with 30 parallel kernels. The computation lasts about 2 hours. The resulting explicit form of the expansion is shown in \cite{Liugit2019}. Firstly, when we sum over all cubic lattice vertices $v=(x,y,z)$ and impose periodic boundary conditions $x+L\sim x$, $y+L\sim y$, and $z+L\sim z$, all terms in ${C}_{L,v}[g^\eps]$ linear to $\delta\theta^{a}_\bot$ and $\delta p^{a}_\bot$ cancels between different $v$'s. The situation is similar to the Euclidean Hamiltonian constraint ${C}_{0,v}[g^\eps]$. Therefore, the constraint equations \Ref{c2} are satisfied automatically.

Secondly, expansions of ${C}_{L,v}[g^\eps]$ (in \cite{Liugit2019}) and ${C}_{0,v}[g^\eps]$ (in the proof of Theorem \ref{Euclideanresult}) show that 
\be
\frac{\partial\,{\bf H}[g^\eps]}{\partial\delta\theta^I_\parallel(x,y,z)}\Bigg|_{\delta\theta=\delta p=0}&=& - \frac{2a}{3 \kappa} \sqrt{\frac{{2 \beta}}{p}} \lt[\sin^2(\theta)-\frac{4}{9}\frac{1+\b^2}{\b^2}\sin^2(2\theta)\rt]=\frac{\partial}{\partial p}\lt[\frac{1}{3|V(\g)|}\mathbf{H}[\theta,p]\rt],\nonumber\\
\frac{\partial\,{\bf H}[g^\eps]}{\partial\delta p^I_\parallel(x,y,z)}\Bigg|_{\delta\theta=\delta p=0}&=&-\frac{4a}{3 \kappa}\sqrt{2\b p}\lt[\sin(2\theta)-\frac{8}{9}\frac{1+\b^2}{\b^2}\sin(4\theta)\rt]=\frac{\partial}{\partial \theta}\lt[\frac{1}{3|V(\g)|}\mathbf{H}[\theta,p]\rt].\label{derivativeLHthetap}
\ee
are both independent of $(x, y, z)$ and $I$. These results can also be obtained from the same argument as for Eq.\Ref{argument}. Eqs.\eqref{c1} and \eqref{derivativeLHthetap} gives Eq.\Ref{evoThiemann}.\\
$\Box$

The scheme I in Eq.\Ref{ansatzI} relates Eq.\Ref{evoThiemann} to the $\mu_0$-scheme effective dynamics proposed in \cite{Dapor:2017rwv,Yang:2009fp} (by a re-definition of $\b$ and a constant rescaling of the time parameter)
\be
&&\frac{\rmd P}{\rmd\t}=\frac{\partial}{\partial C}\mathbf{h}(C,P),\quad \frac{\rmd C}{\rmd\t}=-\frac{\partial}{\partial P}\mathbf{h}(C,P),\nonumber\\
 &&\mathbf{h}(C,P)=\begin{cases}
  -\frac{4\b }{3{\mu^2} } \sqrt{2P }\, \lt[{\sin^2 (C\mu )}-\frac{4}{9}\frac{1+\b^2}{\b^2}{\sin^2(2C\mu)} \rt]&  \text{for Brown-Kucha\v{r}/Gaussian dusts}\\
  \frac{ \sqrt{2\kappa}\,\beta  P}{3 \mu } \sqrt{-\sin^2 ( \mu  C)+\frac{4}{9}\frac{1+\b^2}{\b^2}{\sin^2(2C\mu)} }& \text{for gravity-scalar}.
 \end{cases}
\ee

\subsection{Alesci-Assanioussi-Lewandowski-Makinen's Hamiltonian}\label{LHWSCO}

There is a different proposal for the Hamiltonian operator by Alesci-Assanioussi-Lewandowski-Makinen (AALM) \cite{Assanioussi:2015gka,Alesci:2014aza}, based on a classically equivalent expression of the Hamiltonian constraint:
\be
C=-\frac{1}{\b^2} C_0-\frac{{1+\b^2}}{\kappa\b^2}\sqrt{\det(q)} \, {}^3\! \calr
\ee
where ${}^3\! \calr$ is the 3d scalar curvature. The operator $\hat{C}_{0,v}$ has been constructed above. The idea for constructing a scalar curvature operator is inspired by Regge calculus: $\int_\cs\rmd^3\sig\sqrt{\det(q)}\, {}^3\! \calr $ can be discretized on the dual cubic lattice $\g^*$ \cite{regge} (edges and 3-cells in $\g^*$ are denoted by $l$ and $\Box_v$, and every 3-cell $v^*$ covers a neighborhood at a unique $v\in V(\g)$)
\be
\int_\cs\rmd^3\sig\sqrt{\det(q)}\, {}^3\! \calr \simeq \sum_{l}L_l\lt[2\pi-\sum_{v,l\subset \Box_v}\theta_l(v)\rt]=\sum_{v}\sum_{l\subset \Box_v}L_l\lt[\frac{2\pi}{\a}-\theta_l(v)\rt],\quad \a=4
\ee 
where $L_l$ is the edge length, $\theta_l(v)$ is the dihedral angle between 2 faces of $v^*$ hinged by $l$, and $\a$ is the number of 3-cells sharing a given $l$. $\a=4$ since $\g^*$ is a cubic lattice. $\sum_{l\subset \Box_v}L_l[\frac{2\pi}{\a}-\theta_l(v)]$ is the local contribution at $v$ and discretizes $\int_{\Box_v}\rmd^3x\sqrt{\det(q)} \, {}^3\! R$. In the following, we restrict the construction in \cite{Alesci:2014aza} to the case of cubic graph. We refer to \cite{Alesci:2014aza} for detailed discussions about the scalar curvature operator, and generalization to arbitrary graphs. 

We rewrite $\sum_{l\subset \Box_v}L_l[\frac{2\pi}{\a}-\theta_l(v)]$ in terms of the LQG flux variables $p^a(e)$:
\be
{}^3\!\calr_v=\sum_{I\neq J}\sum_{s_1,s_2=\pm1}L_v(I,s_1;J,s_2) \lt(\frac{2\pi}{\a}-\pi+\arccos\lt[\frac{\vec{p}({e_{v;Is_1}})\cdot\vec{p}({e_{v;Js_2}})}{p({e_{v;Is_1}})p({e_{v;Js_2}})}\rt]\rt)
\ee
where $L_v(I,s_1;J,s_2)$ is the edge length in $\g^*$ determined by 2 cubic faces dual to $e_{v;Is_1},e_{v;Js_2}$ and is given by
\be
L_v(I,s_1;J,s_2)=\frac{1}{V_v}\sqrt{\eps^{abc}p_b(e_{v;Is_1})p_c(e_{v;Js_2})\eps^{ab'c'}p_{b'}(e_{v;Is_1})p_{c'}(e_{v;Js_2})}.\label{length}
\ee 
${\vec{p}({e_{v;Is_1}})}/{{p}({e_{v;Is_1}})}$ are viewed as the unit outward-pointing normal of the face dual to ${e_{v;Is_1}}$, so $\pi-\arccos\lt[\frac{\vec{p}({e_{v;Is_1}})\cdot\vec{p}({e_{v;Js_2}})}{p({e_{v;Is_1}})p({e_{v;Js_2}})}\rt]$ is the dihedral angle $\theta_l(v)$. Note that because $p^a(e)$ is covariant flux, we have
\be
p^{a}\left(e_{v ; I,-}\right)=\frac{1}{2} \operatorname{Tr}\left[\tau^{a} h\left(e_{v-\hat{I} ; I,+}\right)^{-1} p^{b}\left(e_{v-\hat{I} ; I,+}\right) \tau^{b} h\left(e_{v-\hat{I} ; I,+}\right)\right].
\ee

To quantize ${}^3\!R_v$, \cite{Alesci:2014aza} employs Bianchi's length operator in \cite{Bianchi:2008es} to quantize $L_v(I,s_1;J,s_2)$. Bianchi's length operator in \cite{Bianchi:2008es} replace $p^a(e)$ by $\hat{p}^a(e)=it R^a_e$, and ${V_v}^{-1}$ by $\widehat{V^{-1}}=\lim_{\epsilon\to0}(\hat{V}^{2}+\epsilon^6)^{-1}\hat{V}$ (the length operator in \cite{Ma:2010fy} involves the same $\widehat{V^{-1}}$). The Lorentzian Hamiltonian constraint operator $\hat{C}_v$ is given by
\be
\hat{C}_v=-\frac{1}{\b^2}\hat{C}_{0,v}-\frac{{1+\b^2}}{\kappa\b^2}{}^3\!\hat{\calr}_v.\label{CRconstraint}
\ee

Unfortunately the expansion method in \cite{Giesel:2006um} cannot be applied to computing the semiclassical expectation value $\langle\tilde{\psi}^t_g|\hat{\bf H}|\tilde{\psi}^t_g\rangle$ due to the inverse volume $\widehat{V^{-1}}$ \footnote{The method in \cite{Giesel:2006um} applies only for positive powers of $\hat{Q}_v$. However we may apply the following quantization of the inverse volume which has been used in coupling matter fields in canonical LQG \cite{Sahlmann:2002qj,Thiemann:1997rt,Han:2006iqa,Yang:2016kia}: Classically we have $
{V_v}^{-1}=
({\int_{\Box_v}\det(e)}/{V_v^{3/2}})^2$, Each factor ${\int_{\Box_v}\det(e)}/{V_v^{3/2}}\equiv q_v^{(1/2)}$ can be quantized by 
\be
\hat{q}_v^{(1/2)}&=&\frac{-4\times 2^3\times 2^3}{6(-i\b\ell_P^2/2)^3}\sum_{s_1,s_2,s_3=\pm1}s_1s_2s_3\eps^{IJK}\nonumber\\
&&\tr\lt(\hat{h}(e_{v;Is_1})\lt[\hat{h}(e_{v;Is_1})^{-1},\hat{V}^{1/2}_v\rt]\hat{h}(e_{v;Js_2})\lt[\hat{h}(e_{v;Js_2})^{-1},\hat{V}^{1/2}_v\rt]\hat{h}(e_{v;Ks_3})\lt[\hat{h}(e_{v;Ks_3})^{-1},\hat{V}^{1/2}_v\rt]\rt).
\ee
Therefore the length operator can be defined by a symmetric ordering:
\be
\hat{L}_v(I,s_1;J,s_2)&=&\hat{q}_v^{(1/2)}\sqrt{\hat{\bf X}_v(I,s_1;J,s_2)}\ \hat{q}_v^{(1/2)}.\label{lengthoperator}\\
\hat{\bf X}_v(I,s_1;J,s_2)&=&\eps^{abc}\hat{p}_b(e_{v;Is_1})\hat{p}_c(e_{v;Js_2})\eps^{ab'c'}\hat{p}_{b'}(e_{v;Is_1})\hat{p}_{c'}(e_{v;Js_2})\nonumber
\ee
We obtain the scalar curvature operator ${}^3\!\hat{\calr}_v$ by 
\be
{}^3\!\hat{\calr}_v=\sum_{I\neq J}\sum_{s_1,s_2=\pm1}\hat{L}_v(I,s_1;J,s_2) \lt(\frac{2\pi}{\a}-\pi+\arccos\lt[\frac{\hat{p}^a({e_{v;Is_1}})\hat{p}^a({e_{v;Js_2}})}{\hat{p}({e_{v;Is_1}})\hat{p}({e_{v;Js_2}})}\rt]\rt).
\ee
Inside the arccosine in ${}^3\!\hat{\calr}_v$, ${\hat{p}^a(e)}/{\hat{p}(e)}$ is a bounded operator whose eigenvalues belongs $[-1,1]$. Therefore we can expand the arccosine in power series which converges in operator norm:
\be
\arccos\lt[\frac{\hat{p}^a({e_{v;Is_1}})\hat{p}^a({e_{v;Js_2}})}{\hat{p}({e_{v;Is_1}})\hat{p}({e_{v;Js_2}})}\rt]=\frac{\pi}{2}-\sum_{n=0}^\infty\frac{(2n)!}{(2n+1)2^{2n}(n!)^2}\lt[\frac{\hat{p}^a({e_{v;Is_1}})\hat{p}^a({e_{v;Js_2}})}{\hat{p}({e_{v;Is_1}})\hat{p}({e_{v;Js_2}})}\rt]^{2n+1}.
\ee
When we compute the coherent state expectation value $\langle\tilde{\psi}^t_g|{}^3\!\hat{\calr}_v|\tilde{\psi}^t_g\rangle$, we expand $\hat{V}^{1/2}_v$ in $\hat{q}_v^{(1/2)}$ as in Eq.\Ref{expandvolume} with $q=1/8$, and apply the expansion to the square-root in Eq.\Ref{lengthoperator}, similar to the semiclassical perturbation theory in \cite{Giesel:2006um}. These expansions approximate square-roots by polynomial operators in $\hat{h}(e),\hat{p}^a(e)$, and reduce $\langle\tilde{\psi}^t_g|{}^3\!\hat{\calr}_v|\tilde{\psi}^t_g\rangle$ to expectation values of polynomials in $\hat{h}(e),\hat{p}^a(e)$ and ${\hat{p}^a(e)}/{\hat{p}(e)}$. $\langle\tilde{\psi}^t_g|{}^3\!\hat{\calr}_v|\tilde{\psi}^t_g\rangle$ reproduces correctly the classical ${}^3\!{\calr}_v$ if expectation values of monomials of $\hat{h}(e),\hat{p}^a(e)$ and ${\hat{p}^a(e)}/{\hat{p}(e)}$ reproduce the semiclassically the classical expression. Although expectation values of monomials of $\hat{h}(e),\hat{p}^a(e)$ has been shown to be semiclassically consistent, it seems to us nontrivial to check if they are still consistent when including ${\hat{p}^a(e)}/{\hat{p}(e)}$.}. There isn't a general proof of the semiclassical limit of $\langle\tilde{\psi}^t_g|{}^3\!\hat{\calr}_v|\tilde{\psi}^t_g\rangle$, although evidences are found numerically in \cite{Alesci:2014aza}. Our following discussion is based on the conjecture that $\langle\tilde{\psi}^t_g|\hat{\bf H}|\tilde{\psi}^t_g\rangle$ with $\hat{\bf H}$ defined with Eq.\Ref{CRconstraint} can reproduce the classical $\mathbf{H}[g]$ as $t\to0$.


At the homogeneous and isotropic ansatz, ${}^3\!\mathcal{R}_v[g]=0$ as $\a=4$, only the term with the Euclidean Hamiltonian constraint ${C}_{0,v}$ survives in $C_v$
\be
C_v[g]=-\frac{1}{\b^2}C_{0,v}[g]=-\frac{4a}{\b^2\kappa}\sqrt{2\b p}\, \sin^2(\theta).
\ee
Therefore, the physical Hamiltonian is only different from Eq.\Ref{EuHdust} or \Ref{EuHscalar} by an overall factor $1/\b^2$ or $1/\b$:
\be\label{eq:Hamwarsaw}
 {\bf H}[\theta,p]
=\begin{cases}
\frac{4a\,|V(\g)|}{\b^2\kappa}\sqrt{2\b p}\sin^2(\theta) & \text{for dust models}\\
\frac{\sqrt{2}\,a^2  |V(\g)| }{\sqrt{\kappa}} p\sqrt{ \sin^2(\theta )} & \text{for gravity-scalar}.
\end{cases}
\ee
The effective equations are almost the same as the case with Euclidean Hamiltonian ($s=1,\b=1$) in Theorem \ref{Euclideanresult}: 

\begin{Theorem}

For either Brown-Kucha\v{r} or Gaussian dust models, the cosmological evolution equations \Ref{c1} are equivalent to
\be
\frac{\rmd\theta}{\rmd\t}=-\frac{\partial}{\partial p}\lt[\frac{4 }{3\b^2a } \sqrt{2\beta p } \sin^2 (\theta )\rt], \quad
\frac{\rmd p}{\rmd\t}=
\frac{\partial}{\partial\theta}\lt[\frac{4 }{3 \b^2a } \sqrt{2\beta p } \sin^2 (\theta )\rt],\label{pthevodust}
\ee
while the constraint equations \Ref{c2} are satisfied automatically.

For the gravity-scalar model, the cosmological evolution equations \Ref{c1} are equivalent to
\be
\frac{\rmd\theta}{\rmd\t}=-\frac{\partial}{\partial p}\lt[\frac{\sqrt{2\kappa}\,  p }{3  }\sqrt{ \sin^2(\theta )} \rt]
, \quad
\frac{\rmd p}{\rmd\t}=\frac{\partial}{\partial\theta}\lt[\frac{\sqrt{2\kappa}\, p }{3  }\sqrt{ \sin^2(\theta )} \rt],\label{pthevoscalar}
\ee
while the constraint equations \Ref{c2} are again satisfied automatically. 

\end{Theorem}

\textbf{Proof:} As $\a=4$, the expansion of ${}^3\!\mathcal{R}_v$ has no term linear in $\delta p_\bot,\delta p_\parallel,\delta\theta_\bot,\delta\theta_\parallel$ at the homogeneous and isotropic ansatz:
\be
{}^3\!\mathcal{R}_v[g^\eps]=O(\delta^2).
\ee
Therefore computing derivatives of ${\bf H}[g^\eps]$ with respect to $\delta p_\bot,\delta p_\parallel,\delta\theta_\bot,\delta\theta_\parallel$ become the same as in proving Theorem \ref{Euclideanresult}, except for the factor $-1/\b^2$ in front of $C_{0,v}$.

$\Box$

When translating the time evolution equations to $(C,P)$ variables. For dust models, 
\be
\text{Scheme I}:&&\frac{\rmd P}{\rmd\t}=\frac{\partial}{\partial C}\mathbf{h}_{dust}(C,P),\quad \frac{\rmd C}{\rmd\t}=-\frac{\partial}{\partial P}\mathbf{h}_{dust}(C,P),\quad \mathbf{h}_{dust}(C,P)=\frac{4 }{3 \b} \sqrt{2P }\, \frac{\sin^2 (C\mu )}{\mu^2},\label{evoI}\\
\text{Scheme II}:&&\frac{\rmd P}{\rmd \t}= \frac{ C }{\b}\,\cot \left(\frac{\mu  C}{2} \right) \lt(\frac{1}{3}  C \lt[5 \cos (\mu C)+1\rt]-\frac{2}{3} \frac{\sin (\mu  C)}{\mu}\rt)  \sqrt{\frac{P \lt[1-\cos (\mu  C)\rt]}{C^2}},\nonumber\\
&&\frac{\rmd C}{\rmd\t}= -\frac{  \sin ^2(\mu C)}{3 \beta\mu }\sqrt{\frac{C^2}{P\lt[1- \cos (\mu C)\rt]}},
\ee
For the gravity-scalar,
\be
\text{Scheme I}:&&\frac{\rmd P}{\rmd\t}=\frac{\partial}{\partial C}\mathbf{h}_{\phi}(C,P),\quad 
\frac{\rmd C}{\rmd\t}=-\frac{\partial}{\partial P}\mathbf{h}_{\phi}(C,P),\quad
\mathbf{h}_{\phi}(C,P)=\frac{ \sqrt{2\kappa}\,  P}{3 \mu } \sqrt{\sin^2 ( \mu  C)}\\
\text{Scheme II}:&&\frac{\rmd P}{\rmd\t}=\frac{\sqrt{2\kappa}\,  P}{3}\lt[2 \cot (\mu  C)+\csc (\mu  C)-\frac{2}{\mu  C}\rt]\sqrt{  \sin ^2(\mu  C)},\nonumber\\
&&\frac{\rmd C}{\rmd\t}=-\frac{ \sqrt{2\kappa}  }{3 \mu } \sqrt{\sin^2 ( \mu  C)}.
\ee
The scheme I coincides with the standard effective dynamics in LQC (the $\mu_0$-scheme).

\section{Singularity resolution and bounce}\label{BOUNCE}

The cosmological effective dynamics derived from full LQG path integral coincide with the $\mu_0$-scheme in LQC effective dynamics. This coincidence is shown by the scheme I of the homogeneous and isotropic ansatz, while the scheme II is change of variables. When solving the effective equations, we may proceed with variable $\theta,p$ and ignore the choices of schemes I and II. It is useful to consider the following ($V,b$)-variables which directly relate to $(\theta,p)$,
\be
V=\frac{a^3(\b p)^{3/2}}{2\sqrt{2}},\quad b=\theta
\ee
where $V$ is the 3-volume at every vertex $v$. ($V,b$)-variables are also standard in the literature of LQC. 

In the following, we firstly study the effective dynamics in Eqs.\Ref{evoThiemann} from Giesel-Thiemann's Hamiltonian, while the other choice of Hamiltonian is discussed at the end of this section. Our derivation uses ($V,b$)-variables thus clearly independent of choices of scheme I and II. The following discussion demonstrates the resolution of big-bang singularity and unsymmetric bounce from the Brown-Kucha\v{r}/Gaussian dust models. The discussion closely relates to the $\mu_0$-scheme in LQC. The analysis for the gravity-scalar model is a re-derivation of results in \cite{Dapor:2017rwv}, thus is skipped here. 

Eqs.\Ref{evoThiemann} for the Brown-Kucha\v{r}/Gaussian dust models can be rewritten in terms of ($V,b$)-variables:
\be
\frac{\rmd V}{\rmd\t}&=&-\frac{2 V^{2/3} \left[9 \beta ^2 \sin (2 b)-8\left(\beta
   ^2+1\right) \sin (4 b)\right]}{9 \beta },\label{dustee11110}\\
   \frac{\rmd b}{\rmd\t}&=&-\frac{2 \sin ^2(b) \left[- \beta
   ^2+8 \left(\beta ^2+1\right) \cos (2 b)+8\right]}{27 \beta  V^{1/3}}.\label{dustee11111}
\ee
Moreover the physical Hamiltonian is conserved in the dust-time evolution:
\be
\frac{\mathbf{H}}{|V(\g)|}=\frac{8 {V}^{1/3} \sin ^2(b) \left[- \beta ^2+8 \left(\beta ^2+1\right) \cos (2 b)+8\right]}{9 \beta ^2 \kappa }=\rho V\equiv\ce,\label{conservationlaw}
\ee
where $\rho>0$ is the energy density of physical dust. The conservation law can be solved for $b$, assuming $b\in[-\pi/2,\pi/2]$:
\be
b&=&\arcsin\lt( \pm\frac{1}{4 \sqrt{2}}\sqrt{\frac{7 \beta ^2+16-\sqrt{\left(7 \beta ^2+16\right)^2-72 \beta ^2
   \left(\beta ^2+1\right) \kappa  \rho  V^{2/3}}}{\beta ^2+1}}\rt),\label{physol}\\
b&=&\arcsin\lt( \pm\frac{1}{4 \sqrt{2}}\sqrt{\frac{7
   \beta ^2+16+\sqrt{\left(7 \beta ^2+16\right)^2-72 \beta ^2 \left(\beta ^2+1\right) \kappa  \rho 
   V^{2/3}}}{\beta ^2+1}}\rt).\label{unphysol}
\ee
Inserting solutions to the EOM of $\rmd V/\rmd\t$ gives
\be
\frac{\rmd V}{\text{d$\tau $}}&=& \pm\frac{V^{2/3}}{12\sqrt{2} \left(\beta ^2+1\right)} \sqrt{\left(7 \beta ^2+16\right)^2-72 \beta ^2
   \left(\beta ^2+1\right) \kappa  \rho  V^{2/3}} \nonumber\\
&&\sqrt{{7 \beta ^2}+4 \left(\beta ^2+1\right)
   \kappa  \rho  V^{2/3}- \sqrt{\left(7 \beta ^2+16\right)^2-72 \beta ^2 \left(\beta
   ^2+1\right) \kappa  \rho  V^{2/3}}+16},\label{physV}\\
\frac{\rmd V}{\text{d$\tau $}}&=& \mp\frac{V^{2/3}}{12 \sqrt{2} \left(\beta ^2+1\right)} \sqrt{\left(7 \beta ^2+16\right)^2-72 \beta ^2
\left(\beta ^2+1\right) \kappa  \rho  V^{2/3}} \nonumber\\
&&\sqrt{{7 \beta ^2}+4 \left(\beta ^2+1\right)
\kappa  \rho  V^{2/3} +\sqrt{\left(7 \beta ^2+16\right)^2-72 \beta ^2 \left(\beta
^2+1\right) \kappa  \rho  V^{2/3}}+16}.\label{unphysV}
\ee
When replacing $V$ by $\fa^3$ where $\fa$ is the scale factor we obtain
\be
\lt(\frac{\rmd{\fa}/\rmd\t}{\fa}\rt)^2&=& \frac{1}{3^4 2^5 \fa^2 \left(\beta ^2+1\right)^2}\left[\left(7 \beta ^2+16\right)^2-72 \fa^2 \beta ^2
\left(\beta ^2+1\right) \kappa  \rho   \right]\nonumber\\
 &&  \left[{7 \beta ^2}+4 \fa^2 \left(\beta ^2+1\right)
 \kappa  \rho   - \sqrt{\left(7 \beta ^2+16\right)^2-72 \fa^2 \beta ^2 \left(\beta
 ^2+1\right) \kappa  \rho  }+16\right],\label{mF1}\\
\lt(\frac{\rmd{\fa}/\rmd\t}{\fa}\rt)^2&=&\frac{1}{3^4 2^5 \fa^2 \left(\beta ^2+1\right)^2}\left[\left(7 \beta ^2+16\right)^2-72 \fa^2 \beta ^2
\left(\beta ^2+1\right) \kappa  \rho   \right]\nonumber\\
 &&  \left[{7 \beta ^2}+4 \fa^2 \left(\beta ^2+1\right)
 \kappa  \rho   + \sqrt{\left(7 \beta ^2+16\right)^2-72 \fa^2 \beta ^2 \left(\beta
 ^2+1\right) \kappa  \rho  }+16\right].\label{mF2}
\ee

We expand Eqs.\Ref{mF1} and \Ref{mF2} at low density $\rho\ll1$ or $\fa\gg1$ (the low dust density reduces the dust models to pure gravity):
\be
\text{Eq}.\Ref{mF1}\quad&\Rightarrow&\quad\lt(\frac{\rmd{\fa}/\rmd\t}{\fa}\rt)^2=\frac{2(16+7\b^2)}{81}\kappa\,\rho+\frac{1}{\fa^2}\sum_{n=1}^\infty c_n(\b)\,\kappa^{n+1}\rho^{n+1}\fa^{2(n+1)},\label{mFRW}\\
\text{Eq}.\Ref{mF2}\quad&\Rightarrow&\quad\lt(\frac{\rmd{\fa}/\rmd\t}{\fa}\rt)^2=\frac{\left(7 \beta ^2+16\right)^3}{1296 \left(\beta ^2+1\right)^2\fa^2}+\frac{\left(-19 \beta ^2+8\right) \left(7 \beta ^2+16\right) \kappa  \rho  }{324 \left(\beta
   ^2+1\right)}\nonumber\\
 &&\quad \quad \quad \quad \quad \quad \quad  +\frac{1}{\fa^2}\sum_{n=1}^\infty c'_n(\b)\,\kappa^{n+1}\rho^{n+1}\fa^{2(n+1)}.\label{mFRW2}
\ee
where $c_n(\b)$ are expansion coefficients. Since $\ce=\rho \fa^3$ is conserved, $\rho^{n+1}\fa^{2n}=(\rho \fa^3)^{n+1}/\fa^{n+3}\ll (\rho \fa^3)/\fa^{3}$ at $n>0$ becomes negligible as $\fa$ large. Eq.\Ref{mFRW} truncated at the leading order reduces to the Friedmann equation (up a constant rescaling of $\t$) without cosmological constant.

When going back in time toward early universe, the scale factor $\fa$ and volume $V$ shrink, and the density $\rho$ grows, until we approach $\rmd V/\rmd\t=0$, which happens at the following finite critical density:
\be
\rho_c=\frac{\ce}{V_c}, \quad V_c=\lt(\frac{6\sqrt{2}\, \beta  }{7 \beta ^2+16}\rt)^6(\beta ^2+1)^3 \kappa ^3 \ce^3.
\ee
$\rmd V/\rmd\t=0$ at nonzero $V_c$ and finite $\rho_c$ indicates that the big-bang singularity is replaced by a big bounce. However this result shares the same issue as the $\mu_0$-scheme in LQC: $\rho_c$ depends on the conserved energy $\ce$, so here $\rho_c$ might be small if $\ce$ was large. We will come back to this issue in Section \ref{Onmubar}.

At the bounce where we set $\t=0$, $\sin(b(0))$ takes the following value:
\be
\text{Eq.}\Ref{physol}\quad&\Rightarrow&\quad \sin(b(0))=\pm\frac{1}{4}\sqrt{\frac{{{7 \beta ^2+16}}}{{2({\beta ^2+1})}}}\nonumber\\
\text{Eq.}\Ref{unphysol}\quad&\Rightarrow&\quad \sin(b(0))=\pm\frac{1}{4}\sqrt{\frac{{{7 \beta ^2+16}}}{{2({\beta ^2+1})}}}.\label{sinbplusminus}
\ee
The continuity of $\sin(b)$ requires that at the bounce $\rmd V/\rmd\t=0$, the plus (minus) branch in Eq.\Ref{physV} has to be connected to the minus (plus) branch in Eq.\Ref{unphysV}, i.e. the solution mixes between Eqs.\Ref{mF1} and \Ref{mF2}. 

\begin{figure}[h]
   \begin{center}
   \includegraphics[width = 0.8\textwidth]{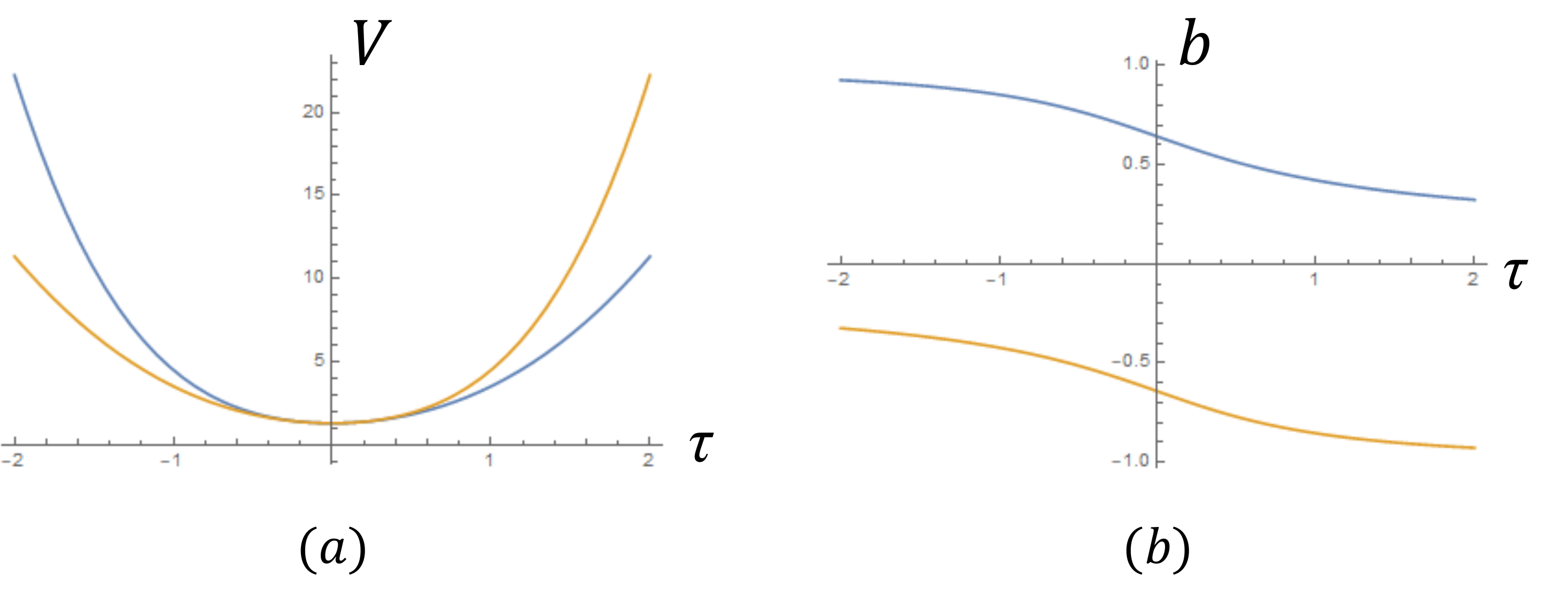}
   \end{center}
   \caption{($a$) Plots of two solutions $V(\t)$ of effective equation at $\ce=4$, $\b=1$, $\kappa=1$. Each solution exhibits an unsymmetric bounce at $\t=0$. The big-bang singularity is resolved by the bounce with nonzero volume $V_c=1.29092$. The solution represented by the blue curve reduces asymptotically to a solution to the Friedmann equation or Eq.\Ref{mFRW} as $\t\to\infty$ and reduces to Eq.\Ref{mFRW2} as $\t\to-\infty$. The solution represented by the orange curve is given by a time refection. $(b)$ Plots of two corresponding $b(\t)$.}
   \label{plotvt}
 \end{figure}

Fixing $V_c$, The cosmological effective equation \Ref{dustee11110} and \Ref{dustee11111} admit 2 independent solutions $V(\t)$ related by a time-reflection (2 solutions of $V(\t)$ corresponding to $b(\t)$ with different $b(0)$). Each solution resolves the singularity and gives an unsymmetric bounce (see FIG.\ref{plotvt}).

We can even obtain analytically solutions of $b(\t)$, which is going to be useful for computing semiclassical amplitude in the next section. From the \eqref{conservationlaw},
\be\label{veq1}
  V =\frac{9^3 \ce^3 \beta ^6 \kappa^3 }{8^3 \sin ^6(b) \left[- \beta ^2+8 \left(\beta ^2+1\right) \cos (2 b)+8\right]^3 }
\ee
substitute into \eqref{dustee11111}, we then have
\be\label{beq1}
 \frac{\rmd b}{\rmd\t}&=&-\frac{16 \sin ^4(b) \left[- \beta
    ^2+8 \left(\beta ^2+1\right) \cos (2 b)+8\right]^2}{243 \beta^3 \kappa }
\ee
whose sulution is given by 
\be\label{eq:bsol1}
   b(\tau) = f^{-1}\lt(c_1+\frac{16 \tau }{9 \beta ^3 \ce \kappa }\rt)\label{bfinv}
\ee
where $c_1$ is a integration constant and $x=f^{-1}(y)$ is the inverse function of $f(x) =y$ with $f(x)$ given by
\be\label{eq:ff}
f(x) =& \frac{18 \left(55 \beta ^2+64\right) \cot (x)}{\left(7 \beta ^2+16\right)^3}-\frac{3072 \left(\beta ^2+1\right)^3 \sin (2 x)}{\beta ^2 \left(7 \beta ^2+16\right)^3 \left(-\beta ^2+8 \left(\beta ^2+1\right) \cos (2 x)+8\right)} \nonumber\\
&+\frac{9 \cot (x) \csc ^2(x)}{\left(7 \beta ^2+16\right)^2}-\frac{256 \left(19 \beta ^2-8\right) \left(\beta ^2+1\right)^2 }{\beta ^3 \left(7 \beta ^2+16\right)^{7/2}} \tanh ^{-1}\left(\frac{3 \beta  \tan (x)}{\sqrt{7 \beta ^2+16}}\right) 
\ee

Solution of Eq.\Ref{beq1} is unique once fixing the inital condition. With a suitable $c_1$,
Eq.\Ref{bfinv} corresponds to the solution of the ``+'' branch in Eq.\Ref{sinbplusminus} with $\sin(b(0))=\frac{1}{4}\sqrt{\frac{7\b^2+16}{2(\b^2+1)}}$. The ``-'' branch corresponds to a different solution with $\sin(b(0))=-\frac{1}{4}\sqrt{\frac{7\b^2+16}{2(\b^2+1)}}$:
\be\label{eq:bsol2}
b(\tau) = -f^{-1}\lt(c_1-\frac{16 \tau }{9 \beta ^3 \ce \kappa }\rt),
\ee
by observing that, \eqref{beq1} is invariant under simultaneously $b \to -b$ and $\t\to-\t$. $c_1$ is fixed to 
 \be
 	c_1 = \frac{6 \beta  \sqrt{25 \beta ^2+16} \left(149 \beta ^4+112 \beta ^2-64\right)-256 \left(\beta 	^2+1\right)^2 \left(19 \beta ^2-8\right) \tanh ^{-1}\left(\frac{3 \beta }{\sqrt{25 \beta ^2+16}}\right)}{\beta ^3 \left(7 \beta ^2+16\right)^{7/2}}.\label{c1=}
 \ee

\begin{figure}[h]
   \begin{center}
   \includegraphics[width = 0.8\textwidth]{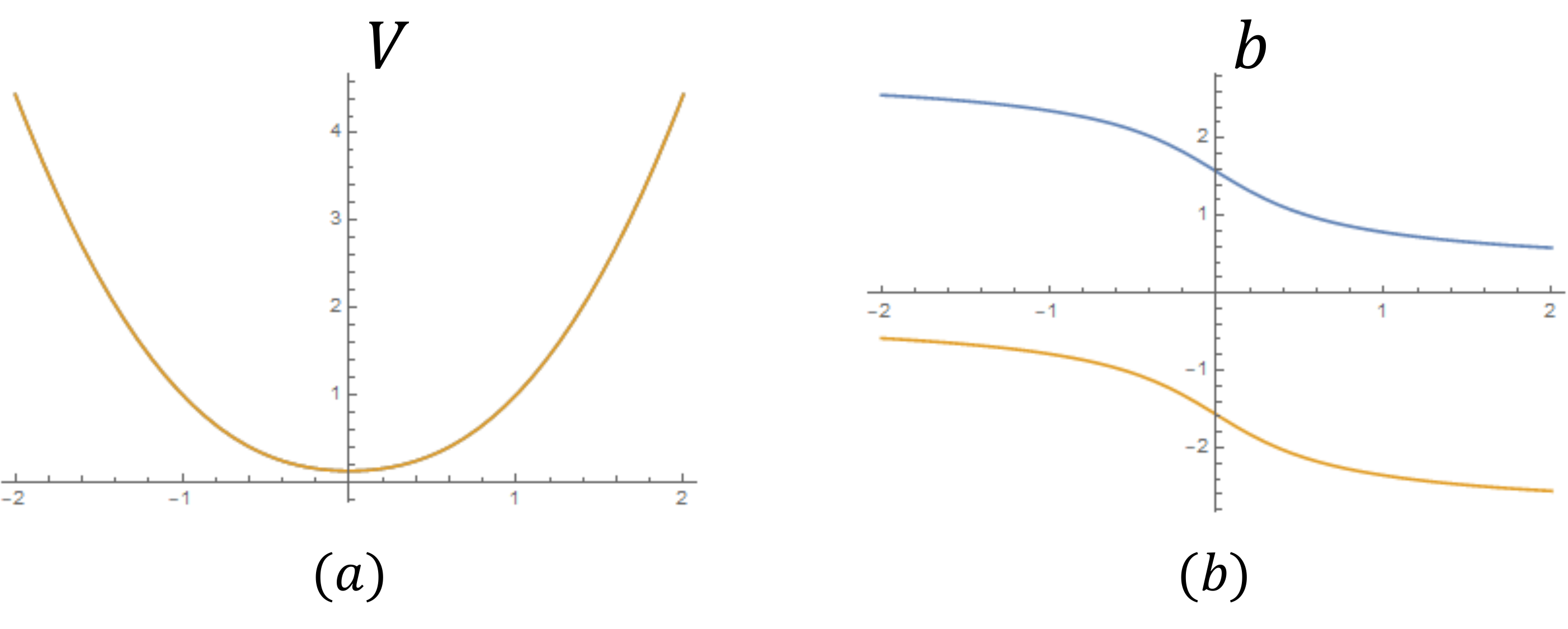}
   \end{center}
   \caption{($a$) Plots of two solutions $V(\t)$ of effective equation from the AALM Hamiltonian, at $\ce=4$, $\b=1$, $\kappa=1$. Two solutions $V(\t)$ coincide and exhibit a symmetric bounce at $\t=0$. The big-bang singularity is resolved by the bounce with nonzero volume $V_c=0.125$. Both solutions reduce asymptotically to a solution to the Friedmann equation as $\t\to\pm\infty$. $(b)$ Plots of two corresponding $b(\t)$.}
   \label{plotvb1}
 \end{figure}

Let's briefly discuss the solutions from the AALM Hamiltonian given in \eqref{eq:Hamwarsaw}. The equations of motion in this case are given by 
\be
\frac{\rmd V}{\rmd\t}&=&\frac{2 V^{2/3} \sin (2 b) }{ \beta },\label{dustee111100}\\
   \frac{\rmd b}{\rmd\t}&=&-\frac{2 \sin ^2(b)  }{3 \beta  V^{1/3}}.\label{dustee111110}
\ee
and the conserved physical Hamiltonian in the dust-time evolution:
\be
\frac{\mathbf{H}}{|V(\g)|}=\frac{8 {V}^{1/3} \sin ^2(b) }{\beta ^2 \kappa }=\rho V\equiv\ce.\label{conservationlaw0}
\ee
All above discussions of solutions can be repeated analogously, so we skip the details. Two solutions $b(\tau)$ are still given by \eqref{bfinv} and \eqref{eq:bsol2}, but with $f(x)$ given by:
\be\label{eq:ff1}
f(x) =&\frac{1}{9} \cot (x) \lt[ 2 +  \csc ^2(x) \rt].
\ee
where $c_1$ is now fixed to $c_1 = 0$ with initial condition $b(0) = \pm \frac{\pi}{2}$.
In this case the two solutions $V(\tau)$ become coincide, leading to the symmetric bounce as in the standard LQC. The critical values of $b$ and $V$ at the bounce are given by 
\be
 	b(0) = \pm \frac{\pi}{2}, \qquad V_c = \frac{\mathcal{E}^3 \beta^6 \kappa^3 }{512 }
\ee
Solutions $V(\t),b(\t)$ are plotted in FIG.\ref{plotvb1}

\section{Semiclassical amplitude}\label{Semiclassical amplitude}

Given the initial and final states $\Psi^t_{[g]}, \Psi^t_{[g']}$ in the amplitude $A_{[g],[g']}$ in Eq.\Ref{Agg}. The semiclassical approximation of $A_{[g],[g']}$ can be obtained by evaluating the integrand in the path integral Eq.\Ref{integral} at the solution of EOMs, i.e. up to prefactors including the integration measure and 1-loop determinant (determinant of Hessian),
\be
\frac{A_{[g],[g']}}{||\psi^t_{g}||\,||{\psi}^t_{g'}||}\sim e^{S[g,h]/t}\Big|_\text{solution},\label{onshellA}
\ee
provided that there exists a unique solution satisfying the initial and final conditions. If the initial and final condition do not fix uniquely a solution, the above gives the contribution to the semiclassica amplitude from one solution, while $A_{[g],[g']}$ is approximated by summing over all solutions.

The solution in FIG.\ref{plotvt} or \ref{plotvb1} (with fixed initial and final conditions) is unique within the homogeneous and isotropic sector. The uniqueness of the solution in full theory will be studied in \cite{Han:2019unique}.

We evaluate the on-shell action as follows: In the action \eqref{action}, applying the homogeneous and isotropic ansatz Eq.\Ref{nen} to $K[g_{i+1},g_{i}]$ gives
\be
K[g_{i+1},g_{i}]=& - \frac{3 |V(\gamma)| }{4} \Big[(p_{i+1} - p_{i})^2 + (\theta_{i+1} - \theta_{i})^2 + 2 {i} (p_{i}+p_{i+1})(\theta_{i+1} - \theta_{i})  \Big] \nonumber\\
=&- \frac{3 |V(\gamma)| }{4} \Delta \tau \lt[ 2 {i} (p_{i}+p_{i+1})\frac{(\theta_{i+1} - \theta_{i})}{\Delta \tau}  +\Big( \frac{(p_{i+1} - p_{i})^2}{\Delta \tau^2} + \frac{(\theta_{i+1} - \theta_{i})^2}{\Delta \tau^2} \Big) \Delta \tau  \rt] 
\ee
In approximation $\Delta\to0$, the action becomes
\be
S = - i |V(\gamma)| \int d \tau \Big( 3 p \dot{\theta} + \frac{ \kappa}{a^2} \frac{\bf{H}}{|V(\gamma)|} \Big) = - i \frac{|V(\gamma)|}{a^2 \beta} \int d \tau \Big( {6 V^{\frac{2}{3}} \dot{b}} + { \kappa} \beta \frac{\bf{H}}{|V(\gamma)|} \Big) 
\ee
where $ \dot{\theta} = {\rmd \theta(\tau)}/{\rmd \tau}$. $\frac{\bf{H}}{|V(\gamma)|}$ is given by \eqref{conservationlaw}.
The equation of motion implies
\be
   \frac{{\bf H}}{|V(\gamma)|} = - \frac{12 V^{\frac{2}{3}}}{\beta \kappa } \frac{db}{d \tau}.
\ee
The action can be computed explicitly
\begin{align}
   S =&  i \frac{6 |V(\gamma)|}{a^2 \beta} \int_0^T d \,\tau \; \dot{b} V^{\frac{2}{3}} 
=  i \frac{6 |V(\gamma)|}{ a^2 \beta} \int_{b(0)}^{b(T)} d \, b \; V(b)^{\frac{2}{3}} \\
  =& i \frac{6 |V(\gamma)|}{ a^2 \beta} \int_{b(0)}^{b(T)} d \, {b} \;  \frac{81 \beta ^4 \mathcal{E}^2 \kappa ^2 }{64 \sin ^4({b})\left[\beta ^2-8 \left(\beta ^2+1\right) \cos (2 {b})-8\right]^2} \\
  =&- i \frac{9   |V(\gamma)|  \beta ^3 \mathcal{E}^2 \kappa ^2 }{32 a^2}  f \big(b(\tau) \big) \Bigg|_{b(0)}^{b(T)}
\end{align}
where $f(x)$ is given in \eqref{eq:ff}. Substitute the solution of $b(\tau)$  into the action, we obtain $S$ is linear to $T$ 
\begin{align}
 	S =-  i \frac{\kappa |V(\gamma)|    }{2 a^2} \,\mathcal{E} T  \label{onshellS}
\end{align}
The above result is from the Giesel-Thiemann's Hamiltonian. But interestingly the AALM Hamiltonian gives exactly the same expression as Eq.\Ref{onshellS}.


\section{Outlook: continuum limit, quantum effects, and $\bar{\mu}$-scheme}\label{Onmubar}

Although the effective equation derived from full LQG demonstrates the bounce resolving the big-bang singularity, the result has the same issue as the $\mu_0$-scheme in LQC, i.e. the critical density $\rho_c$ and critical volume $V_c$ depends on the conserved energy $\ce$. It might seem that for large $\ce$, the bounce could happen at arbitrarily low density or large volume. 

Our expectation is that this issue should be resolved by taking into account the continuum limit and quantum effects in the path integral formula Eq.\Ref{integral}. The role played by the continuum limit is suggested by Eq.\Ref{conservationlaw}, where $\ce$ is the energy per lattice vertex, given by the total energy ${\bf H}$ divided by the total number of lattice sites $|V(\g)|$. Heuristically, the continuum limit of the lattice $\g$, $|V(\g)|\to\infty$, prevents large $\ce$ (thus prevents low $\rho_c$) for any finite total energy ${\bf H}$. But this argument is heuristic because Eq.\Ref{conservationlaw} with $|V(\g)|\to\infty$ would lead to $\ce\to0$ so $V_c\to 0$, which contradicted the minimal area and volume gaps in LQG. We might introduce an effective cut-off $\ce_{min}$ or $|V(\g)|_{max}$ to regulate $V_c$ from zero. But more systematically, when applying the continuum limit to the cosmological effective dynamics, we should take into account the quantum effects which is important in the deep Planck regime, and presumably, modifies Eq.\Ref{conservationlaw}.

The LQC $\bar{\mu}$-scheme \cite{Ashtekar:2006wn,Assanioussi:2019iye} suggests the following ad hoc modification in the Hamiltonian ${\bf H}$ in e.g. Eq.\Ref{conservationlaw},
\be
V^{1/3}\ \to\ V,
\ee
and modify the Hamiltonian evolution equations Eqs.\Ref{dustee11110} and \Ref{dustee11111} accordingly\footnote{The corresponding modification of ${\bf H}$ in Eq.\Ref{thiemannHam} is 
\be
{\bf H}\to-\frac{4a^3|V(\g)|}{\kappa\,\Delta}(2\b p)^{3/2}\lt[\sin^2(\theta)-\frac{4}{9}\frac{1+\b^2}{\b^2}\sin^2(2\theta)\rt],
\ee
then the modified evolution equations can be derived from Eq.\Ref{evoThiemann}}. In order to balance the dimension in Eq.\Ref{conservationlaw}, we have to add an area scale $\Delta$ in the denominator, i.e. we make the following modification
\be
\kappa\ \to\ \kappa\Delta\label{modkappa}
\ee
in the denominator of Eq.\Ref{conservationlaw}. A similar computation as in Section \ref{BOUNCE} shows that this ad hoc modification leads to the critical density 
\be
\rho_c=\frac{\left(7 \beta ^2+16\right)^2}{72 \beta ^2 \left(\beta ^2+1\right)   \kappa \Delta}
\ee
independent of the conserved energy $\ce$. $\rho_c$ is Planckian if $\Delta$ is Planckian. The LQC $\bar{\mu}$ scheme takes $\Delta$ to be the minimal LQG area eigenvalue.

We anticipate that the above modification should be explained by taking into account quantum effects in the path integral Eq.\Ref{integral} \footnote{There is a recent work in \cite{Alesci:2016rmn} where average over graphs is taken into count in computing the expectation value of Hamiltonian to relate the $\bar{\mu}$-scheme.}. The $\bar{\mu}$-scheme should relate to the \emph{quantum} effective EOM, which may be derived from the full \emph{quantum} effective action of our path integral. As a preliminary idea, we know that a typical implication of quantum effects is the running of coupling constant with respect to energy scales. In the path integral of LQG, $\kappa$ is the only coupling constant, while the energy scale relates to the length scale $\mu$ in e.g. Eq.\Ref{ansatzI}, thus relates to the number of lattice sites $|V(\g)|$ (we may let $\mu=1/|V(\g)|$ in certain macroscopic unit). Thus the continuum limit $|V(\g)|\to\infty$ should run $\kappa$ away from its infrared value. The running of $\kappa$ seems to relate to the above ad hoc modification in Eq.\Ref{modkappa}. The following conjectured $\b$-function of $\kappa$ may lead to Eq.\Ref{modkappa} with $\Delta\sim\ell^2_P$ when $\mu\to\ell_P$:
\be
\b(\kappa)=\mu\frac{\rmd}{\rmd\mu}\kappa=4\kappa. 
\ee
$\kappa$ scaling as $\mu^4$ might force $V^{1/3}\ \to\ V$ in the effective Hamiltonian as above, since the scaling dimension of the Hamiltonian shouldn't change by quantum effects. 

The above argument depends on the conjectured running of $\kappa$, and assumes the effective Hamiltonian taking into account the quantum effects depends on $b$ in the same manner as in Eq.\Ref{conservationlaw}. At the moment, we still don't have good justifications of these assumptions from full LQG perspective. A systematic study requires to understand the quantum effective action of the path integral.

In addition to the above idea about $\bar{\mu}$-scheme, there are several other perspectives in which our work should be developed further:
\begin{itemize}
\item The effective equations \Ref{eoms1}, \Ref{eoms2}, and \Ref{closure0} are valid for full LQG, therefore should be applied to other situation such as black holes, gravitational waves, etc. It is especially interesting to study the effective dynamics with non-homogeneous space and its time evolution.  

\item The effective equations of full LQG is formulated to be suitable for numerical simulation. It is interesting to develop numerical techniques to solve the effective equations for complicated dynamical situations, and make contact with numerical relativity.  

\item In the path integral method the derivation do not requires a dynamically stable coherent states. However this strategy may relate to the proposal of stable coherent states, because the saddle point approximation of path integral shows that the transition amplitude between two coherent states are not suppressed (namely time evolution of one coherent state has a large overlap with the other) only when they are related by equations of motion. In this sense our proof is related to the question whether these coherent states are dynamically stable. It is useful to make the relation more precise in the future.

\item Section \ref{LHWSCO} suggests that the AALM Hamiltonian with scalar curvature operator should relate to a simpler effective Hamiltonian. There are also some other recent results based on the simplicity of this Hamiltonian (see e.g. \cite{Kisielowski:2018oiv,Assanioussi:2017tql,Zhang:2019dgi}. However as is mentioned in Section \ref{LHWSCO}, the semiclassical limit of the scalar curvature operator has not yet been checked in the literature for the most general situation. Especially its expectation values with respect to coherent states used in this work has not been well studied yet. It is useful to spend some future work on the semiclassical limit of scalar curvature operator and the corresponding Hamiltonian.

\end{itemize}


\section*{Acknowledgements}

The authors acknowledge Kristina Giesel, Andrea Dapor, Klaus Liegener, Yongge Ma, and Xiangdong Zhang for useful communications, and acknowledge Hui Luo and Ling-Yan Hung for sharing computational resources. This work receives support from the National Science Foundation through grants PHY-1602867 and PHY-1912278. 


\appendix

\section{Variations of $S[g,h]$}\label{VariationofS}

\subsection{Variations $g_{i}(e)\protect\mapsto g_{i}^{\varepsilon}(e)=g_{i}(e)e^{\varepsilon_{i}^{a}(e)\tau^{a}}\quad i=1,\cdots,N+1$}

First of all, the variation of $x_{i+1,i}$ gives 
\begin{eqnarray*}
x_{i+1,i}^{\varepsilon}(e)&=&\frac{1}{2}\mathrm{Tr}\left[g_{i+1}^{\varepsilon\dagger}(e)g_{i}^{\varepsilon}(e)\right]=\frac{1}{2}\mathrm{Tr}\left[e^{-\varepsilon_{i+1}^{a}(e)^{*}\tau^{a}}g_{i+1}^{\dagger}(e)g_{i}(e)e^{\varepsilon_{i}^{a}(e)\tau^{a}}\right]\\
 & = & \frac{1}{2}\mathrm{Tr}\left[\left(1-\varepsilon_{i+1}^{a}(e)^{*}\tau^{a}\right)g_{i+1}^{\dagger}(e)g_{i}(e)\left(1+\varepsilon_{i}^{a}(e)\tau^{a}\right)\right]\\
 & = & \frac{1}{2}\mathrm{Tr}\left[g_{i+1}^{\dagger}(e)g_{i}(e)\right]+\frac{\varepsilon_{i}^{a}(e)}{2}\mathrm{Tr}\left[\tau^{a}g_{i+1}^{\dagger}(e)g_{i}(e)\right]-\frac{\varepsilon_{i+1}^{a}(e)^{*}}{2}\mathrm{Tr}\left[\tau^{a}g_{i+1}^{\dagger}(e)g_{i}(e)\right]+O(\varepsilon^{2}).
\end{eqnarray*}
Since
\[
z_{i+1,i}^{\varepsilon}(e)=\mathrm{arccosh}\left(x_{i+1,i}^{\varepsilon}(e)\right),\quad
\frac{\partial z_{i+1,i}^{\varepsilon}(e)}{\partial x_{i+1,i}^{\varepsilon}(e)}=\frac{1}{\sqrt{x_{i+1,i}^{\varepsilon}(e)-1}\sqrt{x_{i+1,i}^{\varepsilon}(e)+1}},
\]
we obtain that 
\begin{align*}
&\delta_\eps\lt( z_{i+1,i}^{\varepsilon}(e)^{2}\rt)  =\varepsilon_{i}^{a}(e)\frac{\partial z_{i+1,i}^{\varepsilon}(e)^{2}}{\partial\varepsilon_{i}^{a}(e)}|_{\varepsilon=0}+\varepsilon_{i+1}^{a}(e)^{*}\frac{\partial z_{i+1,i}^{\varepsilon}(e)^{2}}{\partial\varepsilon_{i+1}^{a}(e)^{*}}|_{\varepsilon=0}\\
 & =2\varepsilon_{i}^{a}(e)z_{i+1,i}(e)\frac{\partial z_{i+1,i}^{\varepsilon}(e)}{\partial x_{i+1,i}^{\varepsilon}(e)}\frac{\partial x_{i+1,i}^{\varepsilon}(e)}{\partial\varepsilon_{i}^{a}(e)}|_{\varepsilon=0}+2\varepsilon_{i+1}^{a}(e)^{*}z_{i+1,i}(e)\frac{\partial z_{i+1,i}^{\varepsilon}(e)}{\partial x_{i+1,i}^{\varepsilon}(e)}\frac{\partial x_{i+1,i}^{\varepsilon}(e)}{\partial\varepsilon_{i+1}^{a}(e)^{*}}|_{\varepsilon=0}\\
 & =\frac{2\varepsilon_{i}^{a}(e)z_{i+1,i}(e)}{\sqrt{x_{i+1,i}(e)-1}\sqrt{x_{i+1,i}(e)+1}}\frac{1}{2}\mathrm{Tr}\left[\tau^{a}g_{i+1}^{\dagger}(e)g_{i}(e)\right]-\frac{2\varepsilon_{i+1}^{a}(e)^{*}z_{i+1,i}(e)}{\sqrt{x_{i+1,i}(e)-1}\sqrt{x_{i+1,i}(e)+1}}\frac{1}{2}\mathrm{Tr}\left[\tau^{a}g_{i+1}^{\dagger}(e)g_{i}(e)\right]\\
 & =\frac{\varepsilon_{i}^{a}(e)z_{i+1,i}(e)}{\sqrt{x_{i+1,i}(e)-1}\sqrt{x_{i+1,i}(e)+1}}\mathrm{Tr}\left[\tau^{a}g_{i+1}^{\dagger}(e)g_{i}(e)\right]-\frac{\varepsilon_{i+1}^{a}(e)^{*}z_{i+1,i}(e)}{\sqrt{x_{i+1,i}(e)-1}\sqrt{x_{i+1,i}(e)+1}}\mathrm{Tr}\left[\tau^{a}g_{i+1}^{\dagger}(e)g_{i}(e)\right].
\end{align*}
Similarly
\be
\delta_\eps \lt(p_{i+1}(e)^{2}\rt)&=&\frac{\varepsilon_{i+1}^{a}(e)p_{i+1}(e)}{\sinh\left(p_{i+1}(e)\right)}\mathrm{Tr}\left[\tau^{a}g_{i+1}^{\dagger}(e)g_{i+1}(e)\right]-\frac{\varepsilon_{i+1}^{a}(e)^{*}p_{i+1}(e)}{\sinh\left(p_{i+1}(e)\right)}\mathrm{Tr}\left[\tau^{a}g_{i+1}^{\dagger}(e)g_{i+1}(e)\right],\nonumber\\
\delta_\eps \lt(p_{i}(e)^{2}\rt)&=&\frac{\varepsilon_{i}^{a}(e)p_{i}(e)}{\sinh\left(p_{i}(e)\right)}\mathrm{Tr}\left[\tau^{a}g_{i}^{\dagger}(e)g_{i}(e)\right]-\frac{\varepsilon_{i}^{a}(e)^{*}p_{i}(e)}{\sinh\left(p_{i}(e)\right)}\mathrm{Tr}\left[\tau^{a}g_{i}^{\dagger}(e)g_{i}(e)\right],\nonumber
\ee
since $p_{i+1}(e)>0$
\[
\sqrt{\cosh\left(p_{i+1}(e)\right)-1}\sqrt{\cosh\left(p_{i+1}(e)\right)+1}=\sqrt{\cosh\left(p_{i+1}(e)\right)^{2}-1}=\sqrt{\sinh\left(p_{i+1}(e)\right)^{2}}=\sinh\left(p_{i+1}(e)\right).
\]
We apply the above relations to $\delta S$
\begin{eqnarray*}
\delta_\eps S[g,h] & = & \sum_{i=0}^{N+1}\sum_{e\in E(\gamma)}\left[\delta_\eps\lt( z_{i+1,i}(e)^{2}\rt)-\frac{1}{2}\delta_\eps\lt( p_{i+1}(e)^{2}\rt)-\frac{1}{2}\delta_\eps\lt( p_{i}(e)^{2}\rt)\right]-\frac{i\kappa}{a^{2}}\sum_{i=1}^{N}\Delta\tau\left[\delta_\eps\frac{\langle\psi_{g_{i+1}}^{t}|\hat{\mathbf{H}}|\psi_{g_{i}}^{t}\rangle}{\langle\psi_{g_{i+1}}^{t}|\psi_{g_{i}}^{t}\rangle}\right]\\
 & = & \sum_{e\in E(\gamma)}\sum_{i=0}^{N+1}\Bigg(\frac{\varepsilon_{i}^{a}(e)z_{i+1,i}(e)}{\sqrt{x_{i+1,i}(e)-1}\sqrt{x_{i+1,i}(e)+1}}\mathrm{Tr}\left[\tau^{a}g_{i+1}^{\dagger}(e)g_{i}(e)\right]-\frac{\varepsilon_{i+1}^{a}(e)^{*}z_{i+1,i}(e)}{\sqrt{x_{i+1,i}(e)-1}\sqrt{x_{i+1,i}(e)+1}}\mathrm{Tr}\left[\tau^{a}g_{i+1}^{\dagger}(e)g_{i}(e)\right]\\
 &  & -\frac{1}{2}\frac{\varepsilon_{i+1}^{a}(e)p_{i+1}(e)}{\sinh\left(p_{i+1}(e)\right)}\mathrm{Tr}\left[\tau^{a}g_{i+1}^{\dagger}(e)g_{i+1}(e)\right]+\frac{1}{2}\frac{\varepsilon_{i+1}^{a}(e)^{*}p_{i+1}(e)}{\sinh\left(p_{i+1}(e)\right)}\mathrm{Tr}\left[\tau^{a}g_{i+1}^{\dagger}(e)g_{i+1}(e)\right]\\
 &  & -\frac{1}{2}\frac{\varepsilon_{i}^{a}(e)p_{i}(e)}{\sinh\left(p_{i}(e)\right)}\mathrm{Tr}\left[\tau^{a}g_{i}^{\dagger}(e)g_{i}(e)\right]+\frac{1}{2}\frac{\varepsilon_{i}^{a}(e)^{*}p_{i}(e)}{\sinh\left(p_{i}(e)\right)}\mathrm{Tr}\left[\tau^{a}g_{i}^{\dagger}(e)g_{i}(e)\right]\Bigg)\\
 &  & -\frac{i\kappa}{a^{2}}\sum_{i=1}^{N}\Delta\tau\left[\sum_{e\in E(\gamma)}\varepsilon_{i}^{a}(e)\frac{\partial}{\partial\varepsilon_{i}^{a}(e)}\frac{\langle\psi_{g_{i+1}^{\varepsilon}}^{t}|\hat{\mathbf{H}}|\psi_{g_{i}^{\varepsilon}}^{t}\rangle}{\langle\psi_{g_{i+1}^{\varepsilon}}^{t}|\psi_{g_{i}^{\varepsilon}}^{t}\rangle}|_{\varepsilon=0}+\sum_{e\in E(\gamma)}\varepsilon_{i+1}^{a}(e)^{*}\frac{\partial}{\partial\varepsilon_{i+1}^{a}(e)^{*}}\frac{\langle\psi_{g_{i+1}^{\varepsilon}}^{t}|\hat{\mathbf{H}}|\psi_{g_{i}^{\varepsilon}}^{t}\rangle}{\langle\psi_{g_{i+1}^{\varepsilon}}^{t}|\psi_{g_{i}^{\varepsilon}}^{t}\rangle}|_{\varepsilon=0}\right]\\
 & = & \sum_{e\in E(\gamma)}\sum_{i=1}^{N+1}\varepsilon_{i}^{a}(e)\Bigg(\frac{z_{i+1,i}(e)}{\sqrt{x_{i+1,i}(e)-1}\sqrt{x_{i+1,i}(e)+1}}\mathrm{Tr}\left[\tau^{a}g_{i+1}^{\dagger}(e)g_{i}(e)\right]-\frac{p_{i}(e)}{\sinh\left(p_{i}(e)\right)}\mathrm{Tr}\left[\tau^{a}g_{i}^{\dagger}(e)g_{i}(e)\right]\Bigg)\\
 &  & +\sum_{e\in E(\gamma)}\sum_{i=1}^{N+1}\varepsilon_{i}^{a}(e)^{*}\Bigg(-\frac{z_{i,i-1}(e)}{\sqrt{x_{i,i-1}(e)-1}\sqrt{x_{i,i-1}(e)+1}}\mathrm{Tr}\left[\tau^{a}g_{i}^{\dagger}(e)g_{i-1}(e)\right]+\frac{p_{i}(e)}{\sinh\left(p_{i}(e)\right)}\mathrm{Tr}\left[\tau^{a}g_{i}^{\dagger}(e)g_{i}(e)\right]\Bigg)\\
 &  & -\frac{i\kappa\Delta\tau}{a^{2}}\sum_{e\in E(\gamma)}\left[\sum_{i=1}^{N}\varepsilon_{i}^{a}(e)\frac{\partial}{\partial\varepsilon_{i}^{a}(e)}\frac{\langle\psi_{g_{i+1}^{\varepsilon}}^{t}|\hat{\mathbf{H}}|\psi_{g_{i}^{\varepsilon}}^{t}\rangle}{\langle\psi_{g_{i+1}^{\varepsilon}}^{t}|\psi_{g_{i}^{\varepsilon}}^{t}\rangle}|_{\varepsilon=0}+\sum_{i=2}^{N+1}\varepsilon_{i}^{a}(e)^{*}\frac{\partial}{\partial\varepsilon_{i}^{a}(e)^{*}}\frac{\langle\psi_{g_{i}^{\varepsilon}}^{t}|\hat{\mathbf{H}}|\psi_{g_{i-1}^{\varepsilon}}^{t}\rangle}{\langle\psi_{g_{i}^{\varepsilon}}^{t}|\psi_{g_{i-1}^{\varepsilon}}^{t}\rangle}|_{\varepsilon=0}\right]
\end{eqnarray*}
where $\varepsilon_{0}^{a}(e)=\varepsilon_{N+2}^{a}(e)=0$ in the 2nd step.

We can extract the following EOMs: \\
For $i=1,\cdots,N$,
\[
\frac{\delta S}{\delta\varepsilon_{i}^{a}(e)}=0:\quad\frac{z_{i+1,i}(e)\mathrm{Tr}\left[\tau^{a}g_{i+1}^{\dagger}(e)g_{i}(e)\right]}{\sqrt{x_{i+1,i}(e)-1}\sqrt{x_{i+1,i}(e)+1}}-\frac{p_{i}(e)\mathrm{Tr}\left[\tau^{a}g_{i}^{\dagger}(e)g_{i}(e)\right]}{\sinh\left(p_{i}(e)\right)}=\frac{i\kappa\Delta\tau}{a^{2}}\frac{\partial}{\partial\varepsilon_{i}^{a}(e)}\frac{\langle\psi_{g_{i+1}^{\varepsilon}}^{t}|\hat{\mathbf{H}}|\psi_{g_{i}^{\varepsilon}}^{t}\rangle}{\langle\psi_{g_{i+1}^{\varepsilon}}^{t}|\psi_{g_{i}^{\varepsilon}}^{t}\rangle}\Bigg|_{\varepsilon=0}
\]
For $i=N+1$
\[
\frac{\delta S}{\delta\varepsilon_{N+1}^{a}(e)}=0:\quad\frac{z_{N+2,N+1}(e)\mathrm{Tr}\left[\tau^{a}g_{N+2}^{\dagger}(e)g_{N+1}(e)\right]}{\sqrt{x_{N+2,N+1}(e)-1}\sqrt{x_{N+2,N+1}(e)+1}}-\frac{p_{N+1}(e)\mathrm{Tr}\left[\tau^{a}g_{N+1}^{\dagger}(e)g_{N+1}(e)\right]}{\sinh\left(p_{N+1}(e)\right)}=0.
\]
For $i=2,\cdots,N+1$,
\[
\frac{\delta S}{\delta\varepsilon_{i}^{a}(e)^{*}}=0:\quad\frac{z_{i,i-1}(e)\mathrm{Tr}\left[\tau^{a}g_{i}^{\dagger}(e)g_{i-1}(e)\right]}{\sqrt{x_{i,i-1}(e)-1}\sqrt{x_{i,i-1}(e)+1}}-\frac{p_{i}(e)\mathrm{Tr}\left[\tau^{a}g_{i}^{\dagger}(e)g_{i}(e)\right]}{\sinh\left(p_{i}(e)\right)}=-\frac{i\kappa\Delta\tau}{a^{2}}\frac{\partial}{\partial\varepsilon_{i}^{a}(e)^{*}}\frac{\langle\psi_{g_{i}^{\varepsilon}}^{t}|\hat{\mathbf{H}}|\psi_{g_{i-1}^{\varepsilon}}^{t}\rangle}{\langle\psi_{g_{i}^{\varepsilon}}^{t}|\psi_{g_{i-1}^{\varepsilon}}^{t}\rangle}\Bigg|_{\varepsilon=0}.
\]
For $i=1$
\[
\frac{\delta S}{\delta\varepsilon_{1}^{a}(e)^{*}}=0:\quad\frac{z_{10}(e)\mathrm{Tr}\left[\tau^{a}g_{1}^{\dagger}(e)g_{0}(e)\right]}{\sqrt{x_{10}(e)-1}\sqrt{x_{10}(e)+1}}-\frac{p_{1}(e)\mathrm{Tr}\left[\tau^{a}g_{1}^{\dagger}(e)g_{1}(e)\right]}{\sinh\left(p_{1}(e)\right)}=0.
\]

\subsection{Variation $h_{v}^{\eta}=h_{v}e^{\eta_{v}^{a}\tau^{a}}\text{ with }\eta_{v}^{a}\in\mathbb{R}$}

Firstly the variation of $g'^h$ gives
\begin{align*}
g_{0}(e) & =g'^{h^{\eta}}(e)=h_{s(e)}^{\eta,-1}g'(e)h_{t(e)}^{\eta}=e^{-\eta_{s(e)}^{a}\tau^{a}}h_{s(e)}^{-1}g'(e)h_{t(e)}e^{\eta_{t(e)}^{a}\tau^{a}}\\
 & =\left(1-\eta_{s(e)}^{a}\tau^{a}\right)h_{s(e)}^{-1}g'(e)h_{t(e)}\left(1+\eta_{t(e)}^{a}\tau^{a}\right)+O\left(\eta^{2}\right)\\
 & =\left(1-\eta_{s(e)}^{a}\tau^{a}\right)g'^{h}(e)\left(1+\eta_{t(e)}^{a}\tau^{a}\right)+O\left(\eta^{2}\right)
\end{align*}
The variation of $x_{10}(e)$ can be computed
\begin{align*}
x_{10}^{\eta}(e) & =\frac{1}{2}\mathrm{Tr}\left[g_{1}^{\dagger}(e)g_{0}^{\eta}(e)\right]=\frac{1}{2}\mathrm{Tr}\left[g_{1}^{\dagger}(e)g'^{h^{\eta}}(e)\right]\\
 & =\frac{1}{2}\mathrm{Tr}\left[g_{1}^{\dagger}(e)\left(1-\eta_{s(e)}^{a}\tau^{a}\right)g'^{h}(e)\left(1+\eta_{t(e)}^{a}\tau^{a}\right)\right]+O\left(\eta^{2}\right)\\
 & =\frac{1}{2}\mathrm{Tr}\left[g_{1}^{\dagger}(e)g'^{h}(e)\right]+\frac{\eta_{t(e)}^{a}}{2}\mathrm{Tr}\left[\tau^{a}g_{1}^{\dagger}(e)g'^{h}(e)\right]-\frac{\eta_{s(e)}^{a}}{2}\mathrm{Tr}\left[g_{1}^{\dagger}(e)\tau^{a}g'^{h}(e)\right]+O\left(\eta^{2}\right).
\end{align*}
Moreover,
\begin{align*}
x_{0}^{\eta}(e) & \equiv\frac{1}{2}\mathrm{Tr}\left[g_{0}^{\eta\dagger}(e)g_{0}^{\eta}(e)\right]=\frac{1}{2}\mathrm{Tr}\left[g'^{h^{\eta}}(e)^{\dagger}g'^{h^{\eta}}(e)\right]\\
 & =\frac{1}{2}\mathrm{Tr}\left[\left(e^{-\eta_{t(e)}^{a}\tau^{a}}h_{t(e)}^{\dagger}g'(e)^{\dagger}h_{s(e)}e^{\eta_{s(e)}^{a}\tau^{a}}\right)\left(e^{-\eta_{s(e)}^{a}\tau^{a}}h_{s(e)}^{-1}g'(e)h_{t(e)}e^{\eta_{t(e)}^{a}\tau^{a}}\right)\right]\\
 & =\frac{1}{2}\mathrm{Tr}\left[\left(h_{t(e)}^{\dagger}g'(e)^{\dagger}h_{s(e)}\right)\left(h_{s(e)}^{-1}g'(e)h_{t(e)}\right)\right]\\
 & =x_{0}(e).
\end{align*}
Therefore we have
\[
\delta_{\eta}x_{10}(e)=\frac{\eta_{t(e)}^{a}}{2}\mathrm{Tr}\left[\tau^{a}g_{1}^{\dagger}(e)g'^{h}(e)\right]-\frac{\eta_{s(e)}^{a}}{2}\mathrm{Tr}\left[g_{1}^{\dagger}(e)\tau^{a}g'^{h}(e)\right]
\]
\[
\delta_{\eta}x_{0}(e)=0\quad\Rightarrow\quad\delta_{\eta}p_{0}(e)=0.
\]
Then we compute $\delta_{\eta}\lt(z_{10}(e)^{2}\rt)$:
\begin{align*}
\delta_{\eta}\lt(z_{10}(e)^{2}\rt) & =2z_{10}(e)\delta_{\eta}z_{10}(e)=2z_{10}(e)\frac{\partial z_{10}(e)}{\partial x_{10}(e)}\delta_{\eta}x_{10}(e)\\
 & =\frac{2z_{10}(e)}{\sqrt{x_{10}(e)-1}\sqrt{x_{10}(e)+1}}\left(\frac{\eta_{t(e)}^{a}}{2}\mathrm{Tr}\left[\tau^{a}g_{1}^{\dagger}(e)g'^{h}(e)\right]-\frac{\eta_{s(e)}^{a}}{2}\mathrm{Tr}\left[g_{1}^{\dagger}(e)\tau^{a}g'^{h}(e)\right]\right)\\
 & =\frac{z_{10}(e)}{\sqrt{x_{10}(e)-1}\sqrt{x_{10}(e)+1}}\left(\eta_{t(e)}^{a}\mathrm{Tr}\left[\tau^{a}g_{1}^{\dagger}(e)g'^{h}(e)\right]-\eta_{s(e)}^{a}\mathrm{Tr}\left[g_{1}^{\dagger}(e)\tau^{a}g'^{h}(e)\right]\right).
\end{align*}
Applying the above results to the variation of $S[g,h]$ gives
\begin{align*}
\delta_{\eta}S[g,h] & =\sum_{e\in E(\gamma)}\left[\delta_{\eta}z_{10}(e)^{2}\right]\\
 & =\sum_{e\in E(\gamma)}\frac{z_{10}(e)}{\sqrt{x_{10}(e)-1}\sqrt{x_{10}(e)+1}}\left(\eta_{t(e)}^{a}\mathrm{Tr}\left[\tau^{a}g_{1}^{\dagger}(e)g'^{h}(e)\right]-\eta_{s(e)}^{a}\mathrm{Tr}\left[g_{1}^{\dagger}(e)\tau^{a}g'^{h}(e)\right]\right),
\end{align*}
which leads to the following variation equation:
\[
\frac{\delta S[g,h]}{\delta\eta_{v}^{a}}=\sum_{e,t(e)=v}\frac{z_{10}(e)\mathrm{Tr}\left[\tau^{a}g_{1}^{\dagger}(e)g'^{h}(e)\right]}{\sqrt{x_{10}(e)-1}\sqrt{x_{10}(e)+1}}-\sum_{e,s(e)=v}\frac{z_{10}(e)\mathrm{Tr}\left[g_{1}^{\dagger}(e)\tau^{a}g'^{h}(e)\right]}{\sqrt{x_{10}(e)-1}\sqrt{x_{10}(e)+1}}=0.
\]

When we impose the initial condition $g_{1}=g'^{h}$, we obtain
\[
z_{10}(e)=p_{1}(e),\quad x_{10}(e)=\cosh\left(p_{1}(e)\right),\quad\sqrt{x_{10}(e)-1}\sqrt{x_{10}(e)+1}=\sqrt{\cosh\left(p_{1}(e)\right)^{2}-1}=\sinh\left(p_{1}(e)\right),
\]
and the following simplifications:
\begin{align*}
\mathrm{Tr}\left[\tau^{a}g_{1}^{\dagger}(e)g'^{h}(e)\right] & =\mathrm{Tr}\left[\tau^{a}g_{1}^{\dagger}(e)g_{1}(e)\right]=\mathrm{Tr}\left[\tau^{a}e^{-\theta^{a}\tau^{a}/2}e^{-ip^{a}\tau^{a}/2}e^{-ip^{a}\tau^{a}/2}e^{\theta^{a}\tau^{a}/2}\right]\\
 & =\mathrm{Tr}\left[e^{\theta^{a}\tau^{a}/2}\tau^{a}e^{-\theta^{a}\tau^{a}/2}e^{-ip^{a}\tau^{a}}\right]=\Lambda_{\ b}^{a}(\theta)\mathrm{Tr}\left[\tau^{b}e^{-ip^{c}\tau^{c}}\right]\\
 & =\Lambda_{\ b}^{a}(\theta)\mathrm{Tr}\left[\tau^{b}\left(\cosh(p)-i\frac{p^{c}\tau^{c}}{p}\sinh(p)\right)\right]\\
 & =-i\frac{p^{c}}{p}\sinh(p)\Lambda_{\ b}^{a}(\theta)\mathrm{Tr}\left[\tau^{b}\tau^{c}\right]\\
 & =2i\frac{\sinh(p)}{p}\Lambda_{\ c}^{a}(\theta)p^{c},
\end{align*}
where $e^{\theta^{a}\tau^{a}/2}\tau^{a}e^{-\theta^{a}\tau^{a}/2}=\Lambda_{\ b}^{a}\mathrm{(\theta)}\tau^{b}$. Similarly,
\begin{align*}
\mathrm{Tr}\left[g_{1}^{\dagger}(e)\tau^{a}g'^{h}(e)\right] & =\mathrm{Tr}\left[g_{1}^{\dagger}(e)\tau^{a}g_{1}(e)\right]=\mathrm{Tr}\left[e^{-\theta^{a}\tau^{a}/2}e^{-ip^{a}\tau^{a}/2}\tau^{a}e^{-ip^{a}\tau^{a}/2}e^{\theta^{a}\tau^{a}/2}\right]\\
 & =\mathrm{Tr}\left[\tau^{a}e^{-ip^{a}\tau^{a}}\right]=\mathrm{Tr}\left[\tau^{a}\left(\cosh(p)-i\frac{p^{c}\tau^{c}}{p}\sinh(p)\right)\right]\\
 & =-i\frac{p^{c}}{p}\sinh(p)\mathrm{Tr}\left[\tau^{a}\tau^{c}\right]\\
 & =2i\frac{\sinh(p)}{p}p^{a}.
\end{align*}
The variation equation reduces to
\begin{align*}
0=\frac{\delta S[g,h]}{\delta\eta_{v}^{a}} & =\sum_{e,t(e)=v}\frac{p_{1}(e)2i\frac{\sinh(p(e))}{p(e)}\Lambda_{\ c}^{a}(\theta)p^{c}(e)}{\sinh\left(p_{1}(e)\right)}-\sum_{e,s(e)=v}\frac{p_{1}(e)2i\frac{\sinh(p(e))}{p(e)}p^{a}(e)}{\sinh\left(p_{1}(e)\right)}\\
 & =\sum_{e,t(e)=v}2i\Lambda_{\ c}^{a}(\theta)p^{c}(e)-\sum_{e,s(e)=v}2ip^{a}(e).
\end{align*}

\section{Derivation in Lemma 4.1}\label{DL4.1}

\subsection{$D_{1}^{a}(g_{i},g_{j})$}

We apply $g_{i}=g_{j}\left[1+\Delta\phi^{a}\tau^{a}+O\left(\Delta\phi^{2}\right)\right]$ to
\begin{eqnarray*}
D_{1}^{a}(g_{i},g_{j}) & = & \frac{z_{ij}}{\sqrt{x_{ij}-1}\sqrt{x_{ij}+1}}\mathrm{Tr}\left[\tau^{a}g_{i}^{\dagger}g_{j}\right]-\frac{p_{j}}{\sinh\left(p_{j}\right)}\mathrm{Tr}\left[\tau^{a}g_{j}^{\dagger}g_{j}\right].
\end{eqnarray*}
We use the following relations: Firstly,
\begin{align*}
x_{ij} & =\frac{1}{2}\mathrm{Tr}\left[g_{i}^{\dagger}g_{j}\right]=\frac{1}{2}\mathrm{Tr}\left[\left[1+\Delta\phi^{a}\tau^{a}+O\left(\Delta\phi^{2}\right)\right]^{\dagger}g_{j}^{\dagger}g_{j}\right]\\
 & =\frac{1}{2}\mathrm{Tr}\left[\left[1-\Delta\phi^{a*}\tau^{a}\right]g_{j}^{\dagger}g_{j}\right]+O\left(\Delta\phi^{2}\right)\\
 & =\frac{1}{2}\mathrm{Tr}\left[g_{j}^{\dagger}g_{j}\right]-\frac{1}{2}\Delta\phi^{a*}\mathrm{Tr}\left[\tau^{a}g_{j}^{\dagger}g_{j}\right]+O\left(\Delta\phi^{2}\right)\\
 & =\cosh(p_{j})-i\Delta\phi^{a*}\frac{\sinh(p_{j})}{p_{j}}\Lambda_{\ c}^{a}(\theta_{j})p_{j}^{c}++O\left(\Delta\phi^{2}\right).
\end{align*}
Secondly, since $z_{ij}=\mathrm{arccosh}\left(x_{ij}\right)$,
\begin{eqnarray*}
\delta z_{ij}&=&\frac{\partial z_{ij}}{\partial x_{ij}}\Big|_{g_{i}=g_{j}}\delta x_{ij}=\frac{\delta x_{ij}}{\sqrt{x_{ij}-1}\sqrt{x_{ij}+1}}\Big|_{g_{i}=g_{j}}\\
&=&\frac{1}{\sinh(p_{j})}\left[-i\Delta\phi^{a*}\frac{\sinh(p_{j})}{p_{j}}\Lambda_{\ c}^{a}(\theta_{j})p_{j}^{c}\right]=-i\Delta\phi^{a*}\Lambda_{\ c}^{a}(\theta_{j})\frac{p_{j}^{c}}{p_{j}}.
\end{eqnarray*}
Thirdly,
\begin{eqnarray*}
\mathrm{Tr}\left[\tau^{a}g_{i}^{\dagger}g_{j}\right] & = & \mathrm{Tr}\left[\tau^{a}\left[1-\Delta\phi^{a*}\tau^{a}\right]g_{j}^{\dagger}g_{j}\right]\\
 & = & \mathrm{Tr}\left[\tau^{a}g_{j}^{\dagger}g_{j}\right]-\Delta\phi^{b*}\mathrm{Tr}\left[\tau^{a}\tau^{b}g_{j}^{\dagger}g_{j}\right]\\
 & = & \mathrm{Tr}\left[\tau^{a}g_{j}^{\dagger}g_{j}\right]-\Delta\phi^{b*}\mathrm{Tr}\left[\tau^{a}\tau^{b}e^{-\theta_{j}^{a}\tau^{a}/2}e^{-ip_{j}^{a}\tau^{a}}e^{\theta_{j}^{a}\tau^{a}/2}\right]\\
 & = & \mathrm{Tr}\left[\tau^{a}g_{j}^{\dagger}g_{j}\right]-\Delta\phi^{b*}\mathrm{Tr}\left[e^{\theta_{j}^{a}\tau^{a}/2}\tau^{a}e^{-\theta_{j}^{a}\tau^{a}/2}e^{\theta_{j}^{a}\tau^{a}/2}\tau^{b}e^{-\theta_{j}^{a}\tau^{a}/2}e^{-ip_{j}^{a}\tau^{a}}\right]\\
 & = & \mathrm{Tr}\left[\tau^{a}g_{j}^{\dagger}g_{j}\right]-\Delta\phi^{b*}\Lambda_{\ c}^{a}(\theta_{j})\Lambda_{\ d}^{b}(\theta_{j})\mathrm{Tr}\left[\tau^{c}\tau^{d}\left(\cosh(p_{j})-i\frac{p_{j}^{e}\tau^{e}}{p_{j}}\sinh(p_{j})\right)\right]\\
 & = & \mathrm{Tr}\left[\tau^{a}g_{j}^{\dagger}g_{j}\right]-\Delta\phi^{b*}\Lambda_{\ c}^{a}(\theta_{j})\Lambda_{\ d}^{b}(\theta_{j})\left(-2\delta^{cd}\cosh(p_{j})+2i\frac{p_{j}^{e}}{p_{j}}\sinh(p_{j})\varepsilon^{cde}\right).
\end{eqnarray*}
where we have used $\mathrm{Tr}\left[\tau^{b}\tau^{c}\right]=-\mathrm{Tr}\left[\sigma^{b}\sigma^{c}\right]=-2\delta^{bc}$ and $\mathrm{Tr}\left[\tau^{a}\tau^{b}\tau^{c}\right]=i\mathrm{Tr}\left[\sigma^{a}\sigma^{b}\sigma^{c}\right]=i(2i\varepsilon^{abc})=-2\varepsilon^{abc}$.

Applying above relations to $D_{1}^{a}(g_{i},g_{j})$ gives 
\begin{align*}
&D_{1}^{a}(g_{i},g_{j}) \\
 & =\frac{p_{j}-i\Delta\phi^{a*}\Lambda_{\ c}^{a}(\theta_{j})\frac{p_{j}^{c}}{p_{j}}}{\sqrt{\left[\cosh(p_{j})-1\right]\left[1-i\Delta\phi^{a*}\frac{\sinh(p_{j})}{p_{j}\left[\cosh(p_{j})-1\right]}\Lambda_{\ c}^{a}(\theta_{j})p_{j}^{c}\right]}\sqrt{\left[\cosh(p_{j})+1\right]\left[1-i\Delta\phi^{a*}\frac{\sinh(p_{j})}{p_{j}\left[\cosh(p_{j})+1\right]}\Lambda_{\ c}^{a}(\theta_{j})p_{j}^{c}\right]}}\\
 & \left(\mathrm{Tr}\left[\tau^{a}g_{j}^{\dagger}g_{j}\right]-\Delta\phi^{b*}\Lambda_{\ c}^{a}(\theta_{j})\Lambda_{\ d}^{b}(\theta_{j})\left(-2\delta^{cd}\cosh(p_{j})+2i\frac{p_{j}^{e}}{p_{j}}\sinh(p_{j})\varepsilon^{cde}\right)\right) -\frac{p_{j}}{\sinh\left(p_{j}\right)}\mathrm{Tr}\left[\tau^{a}g_{j}^{\dagger}g_{j}\right]\\
 & =\frac{\left[p_{j}-i\Delta\phi^{a*}\Lambda_{\ c}^{a}(\theta_{j})\frac{p_{j}^{c}}{p_{j}}\right]\left[1+i\Delta\phi^{a*}\frac{\sinh(p_{j})}{2p_{j}\left[\cosh(p_{j})-1\right]}\Lambda_{\ c}^{a}(\theta_{j})p_{j}^{c}\right]\left[1+i\Delta\phi^{a*}\frac{\sinh(p_{j})}{2p_{j}\left[\cosh(p_{j})+1\right]}\Lambda_{\ c}^{a}(\theta_{j})p_{j}^{c}\right]}{\sinh(p_{j})}\\
 & \left(\mathrm{Tr}\left[\tau^{a}g_{j}^{\dagger}g_{j}\right]-\Delta\phi^{b*}\Lambda_{\ c}^{a}(\theta_{j})\Lambda_{\ d}^{b}(\theta_{j})\left(-2\delta^{cd}\cosh(p_{j})+2i\frac{p_{j}^{e}}{p_{j}}\sinh(p_{j})\varepsilon^{cde}\right)\right)-\frac{p_{j}}{\sinh\left(p_{j}\right)}\mathrm{Tr}\left[\tau^{a}g_{j}^{\dagger}g_{j}\right]\\
 & =-i\Delta\phi^{a*}\Lambda_{\ c}^{a}(\theta_{j})p_{j}^{c}\left(\frac{1}{\sinh(p_{j})p_{j}}-\frac{1}{2\left[\cosh(p_{j})-1\right]}-\frac{1}{2\left[\cosh(p_{j})+1\right]}\right)2i\frac{\sinh(p_{j})}{p_{j}}\Lambda_{\ c}^{a}(\theta_{j})p_{j}^{c}\\
 & -\Delta\phi^{b*}\Lambda_{\ c}^{a}(\theta_{j})\Lambda_{\ d}^{b}(\theta_{j})\left(-2\delta^{cd}\cosh(p_{j})+2i\frac{p_{j}^{e}}{p_{j}}\sinh(p_{j})\varepsilon^{cde}\right)\frac{p_{j}}{\sinh(p_{j})}\\
 & =2\Delta\phi^{b*}\Lambda_{\ c}^{a}(\theta_{j})\Lambda_{\ d}^{b}(\theta_{j})\left(\frac{p_{j}^{c}}{p_{j}}\frac{p_{j}^{d}}{p_{j}}-\frac{p_{j}^{c}}{p_{j}}\frac{p_{j}^{d}}{p_{j}}\frac{\cosh(p_{j})p_{j}}{\sinh(p_{j})}+\delta^{cd}\frac{\cosh(p_{j})p_{j}}{\sinh(p_{j})}-ip_{j}^{e}\varepsilon^{cde}\right).
\end{align*}
\\

\subsection{$D_{2}^{a}(g_{i},g_{j})$}

We apply $g_{j}=g_{i}\left[1-\Delta\phi^{a}\tau^{a}+O\left(\Delta\phi^{2}\right)\right]$ to
\[
D_{2}^{a}(g_{i},g_{j})=\frac{z_{ij}}{\sqrt{x_{ij}-1}\sqrt{x_{ij}+1}}\mathrm{Tr}\left[\tau^{a}g_{i}^{\dagger}g_{j}\right]-\frac{p_{i}}{\sinh\left(p_{i}\right)}\mathrm{Tr}\left[\tau^{a}g_{i}^{\dagger}g_{i}\right].
\]
We use the following relations: Firstly,
\begin{align*}
x_{ij} & =\frac{1}{2}\mathrm{Tr}\left[g_{i}^{\dagger}g_{j}\right]=\frac{1}{2}\mathrm{Tr}\left[g_{i}^{\dagger}g_{i}\left[1-\Delta\phi^{a}\tau^{a}\right]\right]+O\left(\Delta\phi^{2}\right)\\
 & =\frac{1}{2}\mathrm{Tr}\left[g_{i}^{\dagger}g_{i}\right]-\frac{1}{2}\Delta\phi^{a}\mathrm{Tr}\left[\tau^{a}g_{i}^{\dagger}g_{i}\right]+O\left(\Delta\phi^{2}\right)\\
 & =\cosh(p_{i})-i\Delta\phi^{a}\frac{\sinh(p_{i})}{p_{i}}\Lambda_{\ c}^{a}(\theta_{i})p_{i}^{c}+O\left(\Delta\phi^{2}\right).
\end{align*}
Secondly, 
\[
\delta z_{ij}=\frac{1}{\sinh(p_{j})}\left[-i\Delta\phi^{a}\frac{\sinh(p_{i})}{p_{i}}\Lambda_{\ c}^{a}(\theta_{i})p_{i}^{c}\right]=-i\Delta\phi^{a}\Lambda_{\ c}^{a}(\theta_{i})\frac{p_{i}^{c}}{p_{i}}.
\]
Thridly, 
\begin{eqnarray*}
\mathrm{Tr}\left[\tau^{a}g_{i}^{\dagger}g_{j}\right] & = & \mathrm{Tr}\left[\left[1-\Delta\phi^{b}\tau^{b}\right]\tau^{a}g_{i}^{\dagger}g_{i}\right]\\
 & = & \mathrm{Tr}\left[\tau^{a}g_{i}^{\dagger}g_{i}\right]-\Delta\phi^{b}\mathrm{Tr}\left[\tau^{b}\tau^{a}g_{i}^{\dagger}g_{i}\right]\\
 & = & \mathrm{Tr}\left[\tau^{a}g_{i}^{\dagger}g_{i}\right]-\Delta\phi^{b}\mathrm{Tr}\left[\tau^{b}\tau^{a}e^{-\theta_{i}^{a}\tau^{a}/2}e^{-ip_{i}^{a}\tau^{a}}e^{\theta_{i}^{a}\tau^{a}/2}\right]\\
 & = & \mathrm{Tr}\left[\tau^{a}g_{i}^{\dagger}g_{i}\right]-\Delta\phi^{b}\Lambda_{\ d}^{b}(\theta_{i})\Lambda_{\ c}^{a}(\theta_{i})\mathrm{Tr}\left[\tau^{d}\tau^{c}\left(\cosh(p_{i})-i\frac{p_{i}^{e}\tau^{e}}{p_{i}}\sinh(p_{i})\right)\right]\\
 & = & \mathrm{Tr}\left[\tau^{a}g_{i}^{\dagger}g_{i}\right]-\Delta\phi^{b}\Lambda_{\ c}^{a}(\theta_{j})\Lambda_{\ d}^{b}(\theta_{j})\mathrm{Tr}\left[\tau^{d}\tau^{c}\left(\cosh(p_{i})-i\frac{p_{i}^{e}\tau^{e}}{p_{i}}\sinh(p_{i})\right)\right]\\
 & = & \mathrm{Tr}\left[\tau^{a}g_{i}^{\dagger}g_{i}\right]-\Delta\phi^{b}\Lambda_{\ c}^{a}(\theta_{j})\Lambda_{\ d}^{b}(\theta_{j})\left(-2\delta^{cd}\cosh(p_{i})-2i\frac{p_{i}^{e}}{p_{i}}\sinh(p_{i})\varepsilon^{cde}\right).
\end{eqnarray*}
Applying above relations to $D_{2}^{a}(g_{i},g_{j})$ gives
\begin{align*}
&D_{2}^{a}(g_{i},g_{j}) \\ 
 & =\frac{p_{i}-i\Delta\phi^{a}\Lambda_{\ c}^{a}(\theta_{i})\frac{p_{i}^{c}}{p_{i}}}{\sqrt{\left[\cosh(p_{i})-1\right]\left[1-i\Delta\phi^{a}\frac{\sinh(p_{i})}{p_{i}\left[\cosh(p_{i})-1\right]}\Lambda_{\ c}^{a}(\theta_{i})p_{i}^{c}\right]}\sqrt{\left[\cosh(p_{i})+1\right]\left[1-i\Delta\phi^{a}\frac{\sinh(p_{i})}{p_{i}\left[\cosh(p_{i})+1\right]}\Lambda_{\ c}^{a}(\theta_{i})p_{i}^{c}\right]}}\\
 & \left[\mathrm{Tr}\left[\tau^{a}g_{i}^{\dagger}g_{i}\right]-\Delta\phi^{b*}\Lambda_{\ c}^{a}(\theta_{j})\Lambda_{\ d}^{b}(\theta_{j})\left(-2\delta^{cd}\cosh(p_{i})-2i\frac{p_{i}^{e}}{p_{i}}\sinh(p_{i})\varepsilon^{cde}\right)\right]-\frac{p_{i}}{\sinh\left(p_{i}\right)}\mathrm{Tr}\left[\tau^{a}g_{i}^{\dagger}g_{i}\right]\\
 & =\frac{\left[p_{i}-i\Delta\phi^{b}\Lambda_{\ d}^{b}(\theta_{i})\frac{p_{i}^{d}}{p_{i}}\right]\left[1+i\Delta\phi^{b}\frac{\sinh(p_{i})}{2p_{i}\left[\cosh(p_{i})-1\right]}\Lambda_{\ d}^{b}(\theta_{i})p_{i}^{d}\right]\left[1+i\Delta\phi^{b}\frac{\sinh(p_{i})}{2p_{i}\left[\cosh(p_{i})+1\right]}\Lambda_{\ d}^{b}(\theta_{i})p_{i}^{d}\right]}{\sqrt{\left[\cosh(p_{i})-1\right]}\sqrt{\left[\cosh(p_{i})+1\right]}}\\
 & \left[\mathrm{Tr}\left[\tau^{a}g_{i}^{\dagger}g_{i}\right]-\Delta\phi^{b*}\Lambda_{\ c}^{a}(\theta_{j})\Lambda_{\ d}^{b}(\theta_{j})\left(-2\delta^{cd}\cosh(p_{i})-2i\frac{p_{i}^{e}}{p_{i}}\sinh(p_{i})\varepsilon^{cde}\right)\right]-\frac{p_{i}}{\sinh\left(p_{i}\right)}\mathrm{Tr}\left[\tau^{a}g_{i}^{\dagger}g_{i}\right]\\
 & =2\Delta\phi^{b}\frac{\sinh(p_{i})}{p_{i}}\Lambda_{\ c}^{a}(\theta_{i})\Lambda_{\ d}^{b}(\theta_{i})p_{i}^{c}p_{i}^{d}\left(\frac{1}{\sinh(p_{i})p_{i}}-\frac{\cosh(p_{i})}{\sinh(p_{i})^{2}}\right)\\
 & +2\Delta\phi^{b}\Lambda_{\ c}^{a}(\theta_{j})\Lambda_{\ d}^{b}(\theta_{j})\left(\delta^{cd}\cosh(p_{i})+i\frac{p_{i}^{e}}{p_{i}}\sinh(p_{i})\varepsilon^{cde}\right)\frac{p_{i}}{\sinh(p_{i})}\\
 & =2\Delta\phi^{b}\Lambda_{\ c}^{a}(\theta_{i})\Lambda_{\ d}^{b}(\theta_{i})\left(\frac{p_{i}^{c}}{p_{i}}\frac{p_{i}^{d}}{p_{i}}-\frac{p_{i}\cosh(p_{i})}{\sinh(p_{i})}\frac{p_{i}^{c}}{p_{i}}\frac{p_{i}^{d}}{p_{i}}+\delta^{cd}\frac{p_{i}\cosh(p_{i})}{\sinh(p_{i})}+ip_{i}^{e}\varepsilon^{cde}\right).
\end{align*}

\bibliographystyle{jhep}

\bibliography{muxin}

\providecommand{\href}[2]{#2}\begingroup\raggedright\begin{thebibliography}{10}

\bibitem{book}
T.~Thiemann, {\em Modern Canonical Quantum General Relativity}.
\newblock Cambridge University Press, 2007.

\bibitem{review}
M.~Han, W.~Huang, and Y.~Ma, {\it {Fundamental structure of loop quantum
  gravity}},  {\em Int.J.Mod.Phys.} {\bf D16} (2007) 1397--1474,
  [\href{http://arxiv.org/abs/gr-qc/0509064}{{\tt gr-qc/0509064}}].

\bibitem{review1}
A.~Ashtekar and J.~Lewandowski, {\it {Background independent quantum gravity: A
  Status report}},  {\em Class.Quant.Grav.} {\bf 21} (2004) R53,
  [\href{http://arxiv.org/abs/gr-qc/0404018}{{\tt gr-qc/0404018}}].

\bibitem{Bojowald:2001xe}
M.~Bojowald, {\it {Absence of singularity in loop quantum cosmology}},  {\em
  Phys. Rev. Lett.} {\bf 86} (2001) 5227--5230,
  [\href{http://arxiv.org/abs/gr-qc/0102069}{{\tt gr-qc/0102069}}].

\bibitem{Ashtekar:2006wn}
A.~Ashtekar, T.~Pawlowski, and P.~Singh, {\it {Quantum Nature of the Big Bang:
  Improved dynamics}},  {\em Phys. Rev.} {\bf D74} (2006) 084003,
  [\href{http://arxiv.org/abs/gr-qc/0607039}{{\tt gr-qc/0607039}}].

\bibitem{Ashtekar:2006rx}
A.~Ashtekar, T.~Pawlowski, and P.~Singh, {\it {Quantum nature of the big
  bang}},  {\em Phys. Rev. Lett.} {\bf 96} (2006) 141301,
  [\href{http://arxiv.org/abs/gr-qc/0602086}{{\tt gr-qc/0602086}}].

\bibitem{Singh:2009mz}
P.~Singh, {\it {Are loop quantum cosmos never singular?}},  {\em Class. Quant.
  Grav.} {\bf 26} (2009) 125005, [\href{http://arxiv.org/abs/0901.2750}{{\tt
  arXiv:0901.2750}}].

\bibitem{Assanioussi:2019iye}
M.~Assanioussi, A.~Dapor, K.~Liegener, and T.~Pawlowski, {\it {Emergent de
  Sitter epoch of the Loop Quantum Cosmos: a detailed analysis}},
  \href{http://arxiv.org/abs/1906.05315}{{\tt arXiv:1906.05315}}.

\bibitem{Ashtekar:2018cay}
A.~Ashtekar, J.~Olmedo, and P.~Singh, {\it {Quantum extension of the Kruskal
  spacetime}},  {\em Phys. Rev.} {\bf D98} (2018), no.~12 126003,
  [\href{http://arxiv.org/abs/1806.02406}{{\tt arXiv:1806.02406}}].

\bibitem{Ashtekar:2018lag}
A.~Ashtekar, J.~Olmedo, and P.~Singh, {\it {Quantum Transfiguration of Kruskal
  Black Holes}},  {\em Phys. Rev. Lett.} {\bf 121} (2018), no.~24 241301,
  [\href{http://arxiv.org/abs/1806.00648}{{\tt arXiv:1806.00648}}].

\bibitem{Assanioussi:2019twp}
M.~Assanioussi, A.~Dapor, and K.~Liegener, {\it {Perspectives on the dynamics
  in loop effective black hole interior}},
  \href{http://arxiv.org/abs/1908.05756}{{\tt arXiv:1908.05756}}.

\bibitem{Gambini:2013hna}
R.~Gambini, J.~Olmedo, and J.~Pullin, {\it {Quantum black holes in Loop Quantum
  Gravity}},  {\em Class. Quant. Grav.} {\bf 31} (2014) 095009,
  [\href{http://arxiv.org/abs/1310.5996}{{\tt arXiv:1310.5996}}].

\bibitem{BenAchour:2018khr}
J.~Ben~Achour, F.~Lamy, H.~Liu, and K.~Noui, {\it {Polymer Schwarzschild black
  hole: An effective metric}},  {\em EPL} {\bf 123} (2018), no.~2 20006,
  [\href{http://arxiv.org/abs/1803.01152}{{\tt arXiv:1803.01152}}].

\bibitem{Bojowald:2018xxu}
M.~Bojowald, S.~Brahma, and D.-h. Yeom, {\it {Effective line elements and
  black-hole models in canonical loop quantum gravity}},  {\em Phys. Rev.} {\bf
  D98} (2018), no.~4 046015, [\href{http://arxiv.org/abs/1803.01119}{{\tt
  arXiv:1803.01119}}].

\bibitem{Bodendorfer:2019cyv}
N.~Bodendorfer, F.~M. Mele, and J.~M{\"u}nch, {\it {Effective Quantum Extended
  Spacetime of Polymer Schwarzschild Black Hole}},  {\em Class. Quant. Grav.}
  {\bf 36} (2019), no.~19 195015, [\href{http://arxiv.org/abs/1902.04542}{{\tt
  arXiv:1902.04542}}].

\bibitem{Rovelli:2014cta}
C.~Rovelli and F.~Vidotto, {\it {Planck stars}},  {\em Int. J. Mod. Phys.} {\bf
  D23} (2014), no.~12 1442026, [\href{http://arxiv.org/abs/1401.6562}{{\tt
  arXiv:1401.6562}}].

\bibitem{Han:2016fgh}
M.~Han and M.~Zhang, {\it {Spinfoams near a classical curvature singularity}},
  {\em Phys. Rev.} {\bf D94} (2016), no.~10 104075,
  [\href{http://arxiv.org/abs/1606.02826}{{\tt arXiv:1606.02826}}].

\bibitem{Bojowald:2006da}
M.~Bojowald, {\it {Loop quantum cosmology}},  {\em Living Rev. Rel.} {\bf 8}
  (2005) 11, [\href{http://arxiv.org/abs/gr-qc/0601085}{{\tt gr-qc/0601085}}].

\bibitem{Ashtekar:2008zu}
A.~Ashtekar, {\it {Loop Quantum Cosmology: An Overview}},  {\em Gen. Rel.
  Grav.} {\bf 41} (2009) 707--741, [\href{http://arxiv.org/abs/0812.0177}{{\tt
  arXiv:0812.0177}}].

\bibitem{Agullo:2016tjh}
I.~Agullo and P.~Singh, {\it {Loop Quantum Cosmology}},  in {\em Loop Quantum
  Gravity: The First 30 Years} (A.~Ashtekar and J.~Pullin, eds.), pp.~183--240.
\newblock WSP, 2017.
\newblock \href{http://arxiv.org/abs/1612.01236}{{\tt arXiv:1612.01236}}.

\bibitem{Taveras:2008ke}
V.~Taveras, {\it {Corrections to the Friedmann Equations from LQG for a
  Universe with a Free Scalar Field}},  {\em Phys. Rev.} {\bf D78} (2008)
  064072, [\href{http://arxiv.org/abs/0807.3325}{{\tt arXiv:0807.3325}}].

\bibitem{Bojowald:2018sgf}
M.~Bojowald, {\it {The BKL scenario, infrared renormalization, and quantum
  cosmology}},  {\em JCAP} {\bf 1901} (2019), no.~01 026,
  [\href{http://arxiv.org/abs/1810.00238}{{\tt arXiv:1810.00238}}].

\bibitem{Alesci:2013xd}
E.~Alesci and F.~Cianfrani, {\it {Quantum-Reduced Loop Gravity: Cosmology}},
  {\em Phys. Rev.} {\bf D87} (2013), no.~8 083521,
  [\href{http://arxiv.org/abs/1301.2245}{{\tt arXiv:1301.2245}}].

\bibitem{Bodendorfer:2014vea}
N.~Bodendorfer, {\it {Quantum reduction to Bianchi I models in loop quantum
  gravity}},  {\em Phys. Rev.} {\bf D91} (2015), no.~8 081502,
  [\href{http://arxiv.org/abs/1410.5608}{{\tt arXiv:1410.5608}}].

\bibitem{Bodendorfer:2015hwl}
N.~Bodendorfer, {\it {An embedding of loop quantum cosmology in $(b,v)$
  variables into a full theory context}},  {\em Class. Quant. Grav.} {\bf 33}
  (2016), no.~12 125014, [\href{http://arxiv.org/abs/1512.00713}{{\tt
  arXiv:1512.00713}}].

\bibitem{Alesci:2016rmn}
E.~Alesci and F.~Cianfrani, {\it {Improved regularization from Quantum Reduced
  Loop Gravity}},  \href{http://arxiv.org/abs/1604.02375}{{\tt
  arXiv:1604.02375}}.

\bibitem{Dapor:2017rwv}
A.~Dapor and K.~Liegener, {\it {Cosmological Effective Hamiltonian from full
  Loop Quantum Gravity Dynamics}},  {\em Phys. Lett.} {\bf B785} (2018)
  506--510, [\href{http://arxiv.org/abs/1706.09833}{{\tt arXiv:1706.09833}}].

\bibitem{Engle:2007zz}
J.~Engle, {\it {Relating loop quantum cosmology to loop quantum gravity:
  Symmetric sectors and embeddings}},  {\em Class. Quant. Grav.} {\bf 24}
  (2007) 5777--5802, [\href{http://arxiv.org/abs/gr-qc/0701132}{{\tt
  gr-qc/0701132}}].

\bibitem{2016arXiv160105531H}
M.~{Hanusch}, {\it {Invariant Connections and Symmetry Reduction in Loop
  Quantum Gravity}},  {\em arXiv e-prints} (Jan, 2016) arXiv:1601.05531,
  [\href{http://arxiv.org/abs/1601.05531}{{\tt arXiv:1601.05531}}].

\bibitem{Fleischhack:2010zt}
C.~Fleischhack, {\it {Loop Quantization and Symmetry: Configuration Spaces}},
  {\em Commun. Math. Phys.} {\bf 360} (2018), no.~2 481--521,
  [\href{http://arxiv.org/abs/1010.0449}{{\tt arXiv:1010.0449}}].

\bibitem{Rovelli:2008aa}
C.~Rovelli and F.~Vidotto, {\it {Stepping out of Homogeneity in Loop Quantum
  Cosmology}},  {\em Class. Quant. Grav.} {\bf 25} (2008) 225024,
  [\href{http://arxiv.org/abs/0805.4585}{{\tt arXiv:0805.4585}}].

\bibitem{Giesel:2007wn}
K.~Giesel and T.~Thiemann, {\it {Algebraic quantum gravity (AQG). IV. Reduced
  phase space quantisation of loop quantum gravity}},  {\em Class. Quant.
  Grav.} {\bf 27} (2010) 175009, [\href{http://arxiv.org/abs/0711.0119}{{\tt
  arXiv:0711.0119}}].

\bibitem{Giesel:2012rb}
K.~Giesel and T.~Thiemann, {\it {Scalar Material Reference Systems and Loop
  Quantum Gravity}},  {\em Class. Quant. Grav.} {\bf 32} (2015) 135015,
  [\href{http://arxiv.org/abs/1206.3807}{{\tt arXiv:1206.3807}}].

\bibitem{Brown:1994py}
J.~D. Brown and K.~V. Kuchar, {\it {Dust as a standard of space and time in
  canonical quantum gravity}},  {\em Phys. Rev.} {\bf D51} (1995) 5600--5629,
  [\href{http://arxiv.org/abs/gr-qc/9409001}{{\tt gr-qc/9409001}}].

\bibitem{Kuchar:1990vy}
K.~V. Kuchar and C.~G. Torre, {\it {Gaussian reference fluid and interpretation
  of quantum geometrodynamics}},  {\em Phys. Rev.} {\bf D43} (1991) 419--441.

\bibitem{Domagala:2010bm}
M.~Domagala, K.~Giesel, W.~Kaminski, and J.~Lewandowski, {\it {Gravity
  quantized: Loop Quantum Gravity with a Scalar Field}},  {\em Phys. Rev.} {\bf
  D82} (2010) 104038, [\href{http://arxiv.org/abs/1009.2445}{{\tt
  arXiv:1009.2445}}].

\bibitem{Rovelli:1993bm}
C.~Rovelli and L.~Smolin, {\it {The Physical Hamiltonian in nonperturbative
  quantum gravity}},  {\em Phys. Rev. Lett.} {\bf 72} (1994) 446--449,
  [\href{http://arxiv.org/abs/gr-qc/9308002}{{\tt gr-qc/9308002}}].

\bibitem{QSD}
T.~Thiemann, {\it {Quantum spin dynamics (QSD)}},  {\em Class. Quant. Grav.}
  {\bf 15} (1998) 839--873, [\href{http://arxiv.org/abs/gr-qc/9606089}{{\tt
  gr-qc/9606089}}].

\bibitem{Alesci:2015wla}
E.~Alesci, M.~Assanioussi, J.~Lewandowski, and I.~Makinen, {\it {Hamiltonian
  operator for loop quantum gravity coupled to a scalar field}},  {\em Phys.
  Rev.} {\bf D91} (2015), no.~12 124067,
  [\href{http://arxiv.org/abs/1504.02068}{{\tt arXiv:1504.02068}}].

\bibitem{Assanioussi:2015gka}
M.~Assanioussi, J.~Lewandowski, and I.~Makinen, {\it {New scalar constraint
  operator for loop quantum gravity}},  {\em Phys. Rev.} {\bf D92} (2015),
  no.~4 044042, [\href{http://arxiv.org/abs/1506.00299}{{\tt
  arXiv:1506.00299}}].

\bibitem{Alesci:2014aza}
E.~Alesci, M.~Assanioussi, and J.~Lewandowski, {\it {Curvature operator for
  loop quantum gravity}},  {\em Phys. Rev.} {\bf D89} (2014), no.~12 124017,
  [\href{http://arxiv.org/abs/1403.3190}{{\tt arXiv:1403.3190}}].

\bibitem{Thiemann:2000bw}
T.~Thiemann, {\it {Gauge field theory coherent states (GCS): 1. General
  properties}},  {\em Class. Quant. Grav.} {\bf 18} (2001) 2025--2064,
  [\href{http://arxiv.org/abs/hep-th/0005233}{{\tt hep-th/0005233}}].

\bibitem{Thiemann:2000ca}
T.~Thiemann and O.~Winkler, {\it {Gauge field theory coherent states (GCS). 2.
  Peakedness properties}},  {\em Class. Quant. Grav.} {\bf 18} (2001)
  2561--2636, [\href{http://arxiv.org/abs/hep-th/0005237}{{\tt
  hep-th/0005237}}].

\bibitem{Thiemann:2000bx}
T.~Thiemann and O.~Winkler, {\it {Gauge field theory coherent states (GCS): 3.
  Ehrenfest theorems}},  {\em Class. Quant. Grav.} {\bf 18} (2001) 4629--4682,
  [\href{http://arxiv.org/abs/hep-th/0005234}{{\tt hep-th/0005234}}].

\bibitem{link}
M.~Han and T.~Thiemann, {\it {On the Relation between Operator Constraint --,
  Master Constraint --, Reduced Phase Space --, and Path Integral
  Quantisation}},  {\em Class.Quant.Grav.} {\bf 27} (2010) 225019,
  [\href{http://arxiv.org/abs/0911.3428}{{\tt arXiv:0911.3428}}].

\bibitem{Han:2009bb}
M.~Han, {\it {Canonical Path-Integral Measures for Holst and Plebanski Gravity.
  II. Gauge Invariance and Physical Inner Product}},  {\em Class. Quant. Grav.}
  {\bf 27} (2010) 245015, [\href{http://arxiv.org/abs/0911.3436}{{\tt
  arXiv:0911.3436}}].

\bibitem{Kisielowski:2018oiv}
M.~Kisielowski and J.~Lewandowski, {\it {Spin-foam model for gravity coupled to
  massless scalar field}},  {\em Class. Quant. Grav.} {\bf 36} (2019), no.~7
  075006, [\href{http://arxiv.org/abs/1807.06098}{{\tt arXiv:1807.06098}}].

\bibitem{Ashtekar:2009dn}
A.~Ashtekar, M.~Campiglia, and A.~Henderson, {\it {Loop Quantum Cosmology and
  Spin Foams}},  {\em Phys. Lett.} {\bf B681} (2009) 347--352,
  [\href{http://arxiv.org/abs/0909.4221}{{\tt arXiv:0909.4221}}].

\bibitem{Henderson:2010qd}
A.~Henderson, C.~Rovelli, F.~Vidotto, and E.~Wilson-Ewing, {\it {Local spinfoam
  expansion in loop quantum cosmology}},  {\em Class. Quant. Grav.} {\bf 28}
  (2011) 025003, [\href{http://arxiv.org/abs/1010.0502}{{\tt
  arXiv:1010.0502}}].

\bibitem{Craig:2016iuw}
D.~Craig and P.~Singh, {\it {The Vertex Expansion in the Consistent Histories
  Formulation of Spin Foam Loop Quantum Cosmology}},  in {\em {Proceedings,
  14th Marcel Grossmann Meeting on Recent Developments in Theoretical and
  Experimental General Relativity, Astrophysics, and Relativistic Field
  Theories (MG14) (In 4 Volumes): Rome, Italy, July 12-18, 2015}}, vol.~4,
  pp.~4065--4070, 2017.
\newblock \href{http://arxiv.org/abs/1603.09671}{{\tt arXiv:1603.09671}}.

\bibitem{Craig:2016lxm}
D.~A. Craig and P.~Singh, {\it {Cosmological dynamics in spin-foam loop quantum
  cosmology: challenges and prospects}},  {\em Class. Quant. Grav.} {\bf 34}
  (2017), no.~7 074001, [\href{http://arxiv.org/abs/1612.08800}{{\tt
  arXiv:1612.08800}}].

\bibitem{Qin:2012xh}
L.~Qin and Y.~Ma, {\it {Coherent State Functional Integral in Loop Quantum
  Cosmology: Alternative Dynamics}},  {\em Mod. Phys. Lett.} {\bf A27} (2012)
  1250078, [\href{http://arxiv.org/abs/1206.1128}{{\tt arXiv:1206.1128}}].

\bibitem{Yang:2009fp}
J.~Yang, Y.~Ding, and Y.~Ma, {\it {Alternative quantization of the Hamiltonian
  in loop quantum cosmology II: Including the Lorentz term}},  {\em Phys.
  Lett.} {\bf B682} (2009) 1--7, [\href{http://arxiv.org/abs/0904.4379}{{\tt
  arXiv:0904.4379}}].

\bibitem{Dittrich:2014ala}
B.~Dittrich, {\it {The continuum limit of loop quantum gravity - a framework
  for solving the theory}},  in {\em Loop Quantum Gravity: The First 30 Years}
  (A.~Ashtekar and J.~Pullin, eds.), pp.~153--179.
\newblock 2017.
\newblock \href{http://arxiv.org/abs/1409.1450}{{\tt arXiv:1409.1450}}.

\bibitem{Bahr:2016hwc}
B.~Bahr and S.~Steinhaus, {\it {Numerical evidence for a phase transition in 4d
  spin foam quantum gravity}},  {\em Phys. Rev. Lett.} {\bf 117} (2016), no.~14
  141302, [\href{http://arxiv.org/abs/1605.07649}{{\tt arXiv:1605.07649}}].

\bibitem{Lang:2017beo}
T.~Lang, K.~Liegener, and T.~Thiemann, {\it {Hamiltonian renormalisation I:
  derivation from Osterwalder?Schrader reconstruction}},  {\em Class. Quant.
  Grav.} {\bf 35} (2018), no.~24 245011,
  [\href{http://arxiv.org/abs/1711.05685}{{\tt arXiv:1711.05685}}].

\bibitem{Han:2018fmu}
M.~Han, Z.~Huang, and A.~Zipfel, {\it {Emergent 4-dimensional linearized
  gravity from spin foam model}},  \href{http://arxiv.org/abs/1812.02110}{{\tt
  arXiv:1812.02110}}.

\bibitem{Husain:2011tk}
V.~Husain and T.~Pawlowski, {\it {Time and a physical Hamiltonian for quantum
  gravity}},  {\em Phys. Rev. Lett.} {\bf 108} (2012) 141301,
  [\href{http://arxiv.org/abs/1108.1145}{{\tt arXiv:1108.1145}}].

\bibitem{Giesel:2007wi}
K.~Giesel, S.~Hofmann, T.~Thiemann, and O.~Winkler, {\it {Manifestly
  Gauge-Invariant General Relativistic Perturbation Theory. I. Foundations}},
  {\em Class. Quant. Grav.} {\bf 27} (2010) 055005,
  [\href{http://arxiv.org/abs/0711.0115}{{\tt arXiv:0711.0115}}].

\bibitem{Ashtekar:1995zh}
A.~Ashtekar, J.~Lewandowski, D.~Marolf, J.~Mourao, and T.~Thiemann, {\it
  {Quantization of diffeomorphism invariant theories of connections with local
  degrees of freedom}},  {\em J. Math. Phys.} {\bf 36} (1995) 6456--6493,
  [\href{http://arxiv.org/abs/gr-qc/9504018}{{\tt gr-qc/9504018}}].

\bibitem{Giesel:2016gxq}
K.~Giesel and A.~Vetter, {\it {Reduced loop quantization with four Klein-Gordon
  scalar fields as reference matter}},  {\em Class. Quant. Grav.} {\bf 36}
  (2019), no.~14 145002, [\href{http://arxiv.org/abs/1610.07422}{{\tt
  arXiv:1610.07422}}].

\bibitem{Sahlmann:2001nv}
H.~Sahlmann, T.~Thiemann, and O.~Winkler, {\it {Coherent states for canonical
  quantum general relativity and the infinite tensor product extension}},  {\em
  Nucl. Phys.} {\bf B606} (2001) 401--440,
  [\href{http://arxiv.org/abs/gr-qc/0102038}{{\tt gr-qc/0102038}}].

\bibitem{Bianchi:2009ky}
E.~Bianchi, E.~Magliaro, and C.~Perini, {\it {Coherent spin-networks}},  {\em
  Phys. Rev.} {\bf D82} (2010) 024012,
  [\href{http://arxiv.org/abs/0912.4054}{{\tt arXiv:0912.4054}}].

\bibitem{Giesel:2006um}
K.~Giesel and T.~Thiemann, {\it {Algebraic quantum gravity (AQG). III.
  Semiclassical perturbation theory}},  {\em Class. Quant. Grav.} {\bf 24}
  (2007) 2565--2588, [\href{http://arxiv.org/abs/gr-qc/0607101}{{\tt
  gr-qc/0607101}}].

\bibitem{Liegener:2019zgw}
K.~Liegener and P.~Singh, {\it {Gauge invariant bounce from quantum geometry}},
   \href{http://arxiv.org/abs/1906.02759}{{\tt arXiv:1906.02759}}.

\bibitem{Dapor:2017gdk}
A.~Dapor and K.~Liegener, {\it {Cosmological coherent state expectation values
  in loop quantum gravity I. Isotropic kinematics}},  {\em Class. Quant. Grav.}
  {\bf 35} (2018), no.~13 135011, [\href{http://arxiv.org/abs/1710.04015}{{\tt
  arXiv:1710.04015}}].

\bibitem{Liugit2019}
M.~Han and H.~Liu.
  \url{https://github.com/LQG-Florida-Atlantic-University/Hamiltonian}, 2019.

\bibitem{regge}
T.~Regge, {\it {General relativity without coordinates}},  {\em Nuovo Cim.}
  {\bf 19} (1961) 558--571.

\bibitem{Bianchi:2008es}
E.~Bianchi, {\it {The Length operator in Loop Quantum Gravity}},  {\em Nucl.
  Phys.} {\bf B807} (2009) 591--624,
  [\href{http://arxiv.org/abs/0806.4710}{{\tt arXiv:0806.4710}}].

\bibitem{Ma:2010fy}
Y.~Ma, C.~Soo, and J.~Yang, {\it {New length operator for loop quantum
  gravity}},  {\em Phys. Rev.} {\bf D81} (2010) 124026,
  [\href{http://arxiv.org/abs/1004.1063}{{\tt arXiv:1004.1063}}].

\bibitem{Sahlmann:2002qj}
H.~Sahlmann and T.~Thiemann, {\it {Towards the QFT on curved space-time limit
  of QGR. 1. A General scheme}},  {\em Class. Quant. Grav.} {\bf 23} (2006)
  867--908, [\href{http://arxiv.org/abs/gr-qc/0207030}{{\tt gr-qc/0207030}}].

\bibitem{Thiemann:1997rt}
T.~Thiemann, {\it {QSD 5: Quantum gravity as the natural regulator of matter
  quantum field theories}},  {\em Class. Quant. Grav.} {\bf 15} (1998)
  1281--1314, [\href{http://arxiv.org/abs/gr-qc/9705019}{{\tt gr-qc/9705019}}].

\bibitem{Han:2006iqa}
M.~Han and Y.~Ma, {\it {Dynamics of scalar field in polymer-like
  representation}},  {\em Class. Quant. Grav.} {\bf 23} (2006) 2741--2760,
  [\href{http://arxiv.org/abs/gr-qc/0602101}{{\tt gr-qc/0602101}}].

\bibitem{Yang:2016kia}
J.~Yang and Y.~Ma, {\it {New volume and inverse volume operators for loop
  quantum gravity}},  {\em Phys. Rev.} {\bf D94} (2016), no.~4 044003,
  [\href{http://arxiv.org/abs/1602.08688}{{\tt arXiv:1602.08688}}].

\bibitem{Han:2019unique}
M.~Han and H.~Liu, {\it {The uniqueness of cosmological solution in full loop
  quantum gravity}},  {\em in preparation}.

\bibitem{Assanioussi:2017tql}
M.~Assanioussi, J.~Lewandowski, and I.~Makinen, {\it {Time evolution in
  deparametrized models of loop quantum gravity}},  {\em Phys. Rev.} {\bf D96}
  (2017), no.~2 024043, [\href{http://arxiv.org/abs/1702.01688}{{\tt
  arXiv:1702.01688}}].

\bibitem{Zhang:2019dgi}
C.~Zhang, J.~Lewandowski, H.~Li, and Y.~Ma, {\it {Bouncing evolution in a model
  of loop quantum gravity}},  {\em Phys. Rev.} {\bf D99} (2019), no.~12 124012,
  [\href{http://arxiv.org/abs/1904.07046}{{\tt arXiv:1904.07046}}].

\end{thebibliography}\endgroup

\end{document}